\documentclass[a4paper,11pt]{article}
\pdfoutput=1 

\usepackage{jheppub} 

\usepackage{latexsym,amssymb,amstext,amsmath}
\usepackage{amsfonts}
\usepackage{enumerate}

\usepackage{url}

\usepackage{graphicx}

\usepackage{color}
\usepackage{comment}

\usepackage{amsthm}

\usepackage{esint}

\usepackage{hyperref}

\newtheorem{theorem}{Theorem}[section]

\newtheorem{proposition}[theorem]{Proposition}

\def\be {\begin{equation}}
\def\ee {\end{equation}}
\def\bea {\begin{eqnarray}}
\def\eea {\end{eqnarray}}

\begin{document}
%
\begin{titlepage}

\begin{center}
 {\LARGE\bfseries 
Rademacher expansion of a Siegel modular form \\
\vskip 2mm
 for ${\cal N}= 4$ counting} 
 \\[10mm]

\textbf{Gabriel Lopes Cardoso, Suresh Nampuri and Mart\'i Rossell\'o}

\vskip 6mm
{\em  Center for Mathematical Analysis, Geometry and Dynamical Systems,\\
  Department of Mathematics, 
  Instituto Superior T\'ecnico,\\ Universidade de Lisboa,
  Av. Rovisco Pais, 1049-001 Lisboa, Portugal}\\
\vskip 3mm

{\tt 
gabriel.lopes.cardoso@tecnico.ulisboa.pt, 
nampuri@gmail.com,}\\
{\tt martirossello@tecnico.ulisboa.pt}
\end{center}

\vskip .2in
\begin{center} {\bf ABSTRACT } \end{center}
\begin{quotation}

\noindent 
The degeneracies of $1/4$ BPS states with unit torsion in heterotic string theory compactified on a six-torus are given in terms of the 
Fourier coefficients of the reciprocal of the Igusa cusp Siegel modular form $\Phi_{10}$ of weight $10$. We use the symplectic symmetries of the latter
to construct a fine-grained Rademacher type expansion which expresses 
these BPS degeneracies as a regularized sum over residues of the poles of $1/\Phi_{10}$. The construction uses two distinct ${\rm SL}(2, \mathbb{Z})$
subgroups of ${\rm Sp}(2, \mathbb{Z})$ which encode multiplier systems, Kloosterman sums and Eichler integrals appearing therein.
Additionally, it shows how the polar data are explicitly built from the Fourier coefficients of $1/\eta^{24}$ by means of a 
continued
fraction structure.

\vskip 3mm
\noindent

\end{quotation}
\vfill
\end{titlepage}

\tableofcontents

\section{Introduction}
Automorphic forms have played an uncannily effective role in the counting of BPS states in superstring theory wherein they arise as generating functions of the degeneracies of said states. More precisely, the integral degeneracies in these cases arise as coefficients of a Fourier expansion of these automorphic forms.   Prominent examples include modular forms, Jacobi forms and Siegel modular forms in the context of BPS state counting in toroidal compactifications of type II/heterotic string theory to four dimensions, 
namely \cite{Dabholkar:1989jt,Dabholkar:2004yr,Maldacena:1999bp,Dijkgraaf:1996it,Shih:2005uc}
\bea
\label{1}
\frac{1}{2} \textrm {BPS in } {\cal N}=4: \qquad \frac{1}{\eta^{24}(\sigma)} &=& \sum_{n \geq -1} d(n) \, q^n,\,\,\,\,\,\,\, q = e^{2\pi i \sigma}\nonumber\\
\frac{1}{8} \textrm{BPS in } {\cal N}=8 :  \qquad \frac{\vartheta_1^2(\sigma,v)}{\eta^6(\sigma)} &=& 
\sum_{\ell \in \mathbb{Z}/2\mathbb{Z} } \vartheta_{1, \ell} ( \sigma, v) \sum_{\substack{\Delta \geq -1 \\ \Delta = - \ell^2 \, \text{mod} \, 4 }}
  c_{\ell} (\Delta) \,
q^{\Delta/4} \\
\frac{1}{4} \textrm{BPS in } {\cal N}=4:  \qquad \frac{1}{\Phi_{10}(\rho, \sigma, v)}&=& 
\sum_{\substack{m,n,\ell \in \mathbb{Z}  \\  m, n \geq -1 }} 
g(m,n,\ell)  \, p^m q^n y^{\ell},\,\,\,\,\,\, p = e^{2 \pi i \rho} \;,\; y = e^{2 \pi i v}. \nonumber
\eea
Modular and Jacobi forms, as shown in the first two examples, have automorphic properties under $\mathrm{SL}(2,\mathbb{Z})$ and are defined in 
the $\sigma$ upper half plane. 
Both of these forms have a finite number of Fourier coefficients associated with a negative power of $q$. 
 These coefficients are referred to as a polar coefficients. 
  The modular symmetries of these functions are powerful enough to constrain the Fourier coefficients so that each non-polar coefficient can be exactly expressed in terms of the polar ones in a Rademacher expansion \cite{rade2}. 
  As an illustrative example, we present the well-known Rademacher expansion for $1/\eta^{24}$, 
  which has been used to extract asymptotic degeneracies for $1/2$ BPS states in toroidally compactified heterotic string theory \cite{Dabholkar:2005dt},
    \begin{equation}
	d(n) = \sum_{\gamma = 1}^{+\infty} \, d(-1) \, 
	\frac{2\pi}{\gamma \,n^{13/2}} {\rm Kl} (n,-1,\gamma) I_{13}\left( \frac{4\pi \sqrt{n}}{\gamma} \right).
\end{equation}
We see that the polar coefficient $d(-1)$ is sufficient to reconstruct the modular form.

In this note we will focus on the third of the above examples, the reciprocal of the Igusa cusp Siegel modular form $\Phi_{10}$ of weight 10.
 It is defined on the Siegel upper half plane, with each of its Fourier coefficients defined in terms of three integers corresponding to the three variables 
 defining the Siegel upper half plane. 
Further, 
 each of its polar terms satisfies $\Delta < 0$, where $\Delta = 4 m n - \ell^2$, and is determined in terms of the coefficients of 
 $1/\eta^{24}$ and the continued fraction representation of $\ell /2m$ \cite{Sen:2011mh,Chowdhury:2019mnb, LopesCardoso:2020pmp}.   
Using the automorphic properties of $\Phi_{10}$ with respect to $\mathrm{Sp}(2, \mathbb{Z})$\footnote{This group is also referred to as 
$\mathrm{Sp}(4, \mathbb{Z})$.}, 
we will demonstrate that this symmetry group constraints the Fourier coefficients of $1/\Phi_{10}$ even more powerfully than in the 
$\mathrm{SL}(2, \mathbb{Z})$ case, resulting in a fine-grained Rademacher expansion that not only reconstructs each element of 
the  infinite set of its Fourier coefficients that satisfy $\Delta >0$ from the  finite polar data, whose elements satisfy $\Delta < 0$,
but also encodes the continued fraction structure underlying the polar terms. Thus, the result of this paper can be summarised as follows:
\newline

\noindent
{\it We use the 
 $\mathrm{Sp}(2,\mathbb{Z})$ symmetries of $1/\Phi_{10}$ to construct a fine-grained Rademacher expansion which expresses its Fourier coefficients
 as a regularized sum over residues of its poles. 
 The construction uses two distinct ${\rm SL}(2, \mathbb{Z})$
subgroups of ${\rm Sp}(2, \mathbb{Z})$ which encode multiplier systems, Kloosterman sums and Eichler integrals appearing therein.
Additionally, it shows how the polar data are explicitly built from the Fourier coefficients of $1/\eta^{24}$ by means of a 
continued
fraction structure.
}
 \\
 
The Rademacher expansion for $d(m,n,\ell)$ was derived in \cite{Ferrari:2017msn} by viewing $d(m,n,\ell)$ as Fourier coefficients of a Mock Jacobi form
 $\psi_m^F$ \cite{Zwegers:2008zna,Dabholkar:2012nd} using the mixed Mock Rademacher expansion developed in \cite{Bringmann:2010sd}. However,
 just as in the case of modular forms, the Rademacher expansion in \cite{Ferrari:2017msn} does not contain information about the explicit values of the polar coefficients of $\psi_m^F$, as computed in  \cite{Chowdhury:2019mnb,LopesCardoso:2020pmp}.

\section{Set up of the calculation and outline of the paper}

In order to define a Rademacher expansion for the reciprocal of the Igusa cusp  $\Phi_{10}$ of weight $10$, we adopt the following sequential program elucidated below:
\begin{enumerate}
\item Identify polar data in $1/\Phi_{10}$.\newline
In contrast to the modular and Jacobi forms displayed above, the Siegel modular form $\Phi_{10}$ has a countably infinite set of zeroes defined in the Siegel upper half plane by the loci \cite{borcherds},
\bea
D(n_2, n_1,m_1,m_2, j): \quad n_2 (\rho \sigma - v^2) + j v + n_1 \sigma - m_1 \rho + m_2 &=& 0 \\
m_1, n_1, m_2, n_2 \in \mathbb{Z}, \; j \in 2 \mathbb{Z} +1, \;\;\; \;\;
m_1 n_1 + m_2 n_2 &=& \frac{1-j^2}{4}. \nonumber
\eea
It has been shown \cite{Sen:2011mh,Chowdhury:2019mnb, LopesCardoso:2020pmp} that for $\Delta < 0$, 
\begin{equation}
	d(m, n, \ell) \equiv (-1)^{\ell+1} \, g(m, n, \ell)
\end{equation}
 can be constructed from the residues of $1/\Phi_{10}$ corresponding to the linear poles ($n_2=0$),
where $g(m, n, \ell)$ are the Fourier coefficients of $1/\Phi_{10}$ (cf. \eqref{1}).
{\it  Therefore, the polar data in  $1/\Phi_{10}$ are determined by its linear pole residues. }

Viewing non-polar data as the complementary information to its polar counterpart, one is motivated to identify the non-polar data in $1/\Phi_{10}$ as the contribution to $g(m,n,\ell)$ with $\Delta \geq 0$.
Generically, these Fourier coefficients receive contributions from both linear and quadratic poles.

\item Extracting non-polar data from $d(m,n,\ell)$  with $\Delta \geq 0$.\newline
From \eqref{1}, we can define, 
\bea 
	d(m, n, \ell) = (-1)^{\ell+1} \,  \int_0^1 d \sigma_1 \int_0^1 d  v_1 \int_0^1 d \rho_1 \,   \frac{1}{\Phi_{10} (\rho, \sigma, v)} \, e^{-2  \pi i \left(  m \rho +  n \sigma + \ell v \right) } \;,
\label{degg}
\eea

where $\rho= \rho_1 + i \rho_2, \sigma = \sigma_1 + i \sigma_2, v = v_1 + i v_2$, and 
where the imaginary parts of $\rho$, $\sigma$ and $v$ are held fixed in the Siegel upper half plane, which is defined by the conditions
$\rho_2 > 0, \, \sigma_2 > 0$ and $\rho_2 \sigma_2 - v_2^2 > 0$.
The $n_2 =0$ poles of $1/\Phi_{10}$ correspond to co-dimension one surfaces in the Siegel upper half plane.
Therefore, in order to identify the non-polar data in terms of the Fourier coefficients, we first need to isolate a chamber in the 
Siegel upper half plane where we perform a Fourier expansion.
Following \cite{Sen:2011mh} we will define this region, which is referred to as the ${\cal R}$-chamber, by
\bea
\frac{\rho_2}{\sigma_2} \gg 1\;\;\;,\;\;  \frac{v_2}{\sigma_2} = -\frac{ \ell}{2m} \;\;\;,\;\;\;  \frac{\ell}{2m} \in [0, 1) .
\label{rch1}
\eea
 We will write down the Rademacher expansion for the Fourier coefficients with $\Delta \geq 0$  in the ${\cal R}$-chamber.

\item Setting up the Rademacher expansion.
\newline
Viewing \eqref{degg} as a $\rho$ integral, we define: 
\bea
\label{2}
&& d(m,n,\ell)|_{\Delta \geq 0} = \\
&& (-1)^{\ell+1}  \sum_{\substack{D \\ n_2 \neq 0 }}
 \left(\int_{\Gamma_{\sigma}(D) } d\sigma \int_{\Gamma_v (D)} d  v \,  {\rm Res} \left(  \frac{1}{\Phi_{10} (\rho, \sigma, v)} \, e^{-2  \pi i \left(  m \rho +  n \sigma + \ell v \right) }\right) \right)\Big\vert_{\rm reg}, \nonumber
\eea
where `reg' refers to  a regularized sum of residues,
\begin{enumerate}
\item ${\rm Res} f $ is the residue of $f$ at an $n_2 \neq 0$ pole $D$ in the $\rho$ plane;

\item  $\Gamma_v(D) $ is the integration contour of $v$ with $v_1 \in [0,1)$, subject to the condition $\frac{v_2}{\sigma_2}=-\frac{\ell}{2m}$.
Further, the condition that 
$D$ lies in the Siegel upper half plane constraints the range of $v_1$;

\item  $\Gamma_{\sigma}(D)$ lies in the projection of $D$ in the $\sigma$ upper half plane with the restriction that $\sigma_1 \in [0,1)$.

\end{enumerate}

This defines our starting point. Our objective is to express the right hand side of \eqref{2} in terms of the polar data.
This implies that the sum over residues of quadratic poles must be rewritten as one over linear pole contributions.
\end{enumerate}

In section \ref{sec:spspr} we use the $\mathrm{Sp}(2, \mathbb{Z})$ symmetries to map each quadratic pole to the simplest linear pole $v=0$. We will reparametrize the five numbers defining the quadratic poles by entries of the $\mathrm{Sp}(2, \mathbb{Z})$ matrix that performs this map and consequently rewrite the summation on the right hand side of \eqref{2} as a sum over these matrix entries. In 
sections \ref{sec:vsint} and \ref{sec:sigint} we explicitly evaluate the summands by performing the $v$ and $\sigma$ integrals respectively. We isolate 
multiplier systems, Kloosterman sums and error functions giving rise to Eichler integrals and to the continued fraction structure. These 
elements serve as building blocks for the Rademacher expansion. We identify a symmetry that is crucial to obtain the Rademacher expansion
by enabling massive cancellations between various terms.
The resulting expansion
 is 
 given by
\eqref{crbs} in the case when $\Delta > 0$. Additionally, we write down the Rademacher expansion for the case when $\Delta = m = \ell =0$ in  \eqref{imm-dd-ml0}.
 In section \ref{sec:conc}
 we conclude with comments on the implications of this expansion for defining exact quantum entropy functions from a gravity path integral. In the appendices we review and discuss various useful relations and calculations.

\section{$\mathrm{Sp}(2, \mathbb{Z})$ symmetries, poles and residues of $1/\Phi_{10}$ \label{sec:spspr}}
In this section, we map each quadratic pole of  $1/\Phi_{10}$ to the simplest linear pole $v=0$. In order to so, 
we will use the $\mathrm{Sp}(2, \mathbb{Z})$ symmetries of $\Phi_{10}$ to reparametrize the five numbers defining the quadratic poles by entries of the $\mathrm{Sp}(2, \mathbb{Z})$ matrix that performs this map. Subsequently, we  rewrite the summation on the right hand side of \eqref{2} as a sum over 
the entries of these mapping matrices.

\subsection{$\mathrm{Sp}(2, \mathbb{Z})$ symmetries of $\Phi_{10}$ }

The Igusa cusp form
$\Phi_{10}$ transforms as follows under $\mathrm{Sp}(2, \mathbb{Z})$ transformations, 
\begin{equation}
	\Phi_{10}\left((A\Omega + B)(C\Omega+D)^{-1}\right) = \det(C\Omega+D)^{10} \, \Phi_{10}(\Omega) \;,
\end{equation}
where
\begin{equation}
	\Omega = 	\begin{pmatrix}
\rho & v \\
v & \sigma
\end{pmatrix} \;\;\;,\;\;\; \begin{pmatrix}
	A & B\\
	C & D
\end{pmatrix} \in \mathrm{Sp}(2, \mathbb{Z}) \;.
\end{equation}
The elements of $\mathrm{Sp}(2, \mathbb{Z})$  satisfy
\begin{equation}
	\begin{pmatrix}
	A & B\\
	C & D
\end{pmatrix}^T\begin{pmatrix}
	0 & I_2\\
	-I_2 & 0
\end{pmatrix}\begin{pmatrix}
	A & B\\
	C & D
\end{pmatrix} = \begin{pmatrix}
	0 & I_2\\
	-I_2 & 0
\end{pmatrix}, \hspace{5mm} I_2 = \begin{pmatrix}
	1 & 0\\
	0 & 1
\end{pmatrix}.
\end{equation}
$\Phi_{10}$ is invariant under S-duality transformations, which consist of $\mathrm{SL}(2,\mathbb{Z})$ transformations
\begin{equation}
\begin{pmatrix}
a & b \\
c & d
\end{pmatrix} \;\;\;,\;\;\;  ad - b c = 1 
\end{equation}
that operate on $\Omega$ through the $\mathrm{Sp}(2, \mathbb{Z})$ transformations
\begin{equation}
	\begin{pmatrix}
a & b \\
c & d
\end{pmatrix}_S = \begin{pmatrix}
	a & -b & 0 & 0 \\
	-c & d & 0 & 0 \\
	0 & 0 & d & c \\
	0 & 0 & b & a
\end{pmatrix} \;.
\end{equation}
Thus, under S-duality, we infer the transformation laws
\begin{align}
\rho' &= a^2 \rho + b^2 \sigma - 2 ab v \nonumber\\
\sigma ' &= c^2 \rho + d^2 \sigma - 2 c d v \nonumber\\
v' &= - a c \rho - bd \sigma + (a d + b c ) v \;.
\label{S-dual}
\end{align}
They leave $\Phi_{10}$ invariant,
\begin{equation}
	\Phi_{10}(\rho',\sigma',v') = \Phi_{10}(\rho,\sigma,v).
\end{equation}

Next, we consider a different set of $\mathrm{Sp}(2,\mathbb{Z})$ transformations, which we denote by 
$\mathrm{SL}(2, \mathbb{Z})_{\sigma}$. We denote its group elements by
\bea
\begin{pmatrix}
\alpha & \beta \\
\gamma& \delta
\end{pmatrix} \;\;\;,\;\;\; \alpha \delta - \beta \gamma = 1 .
\eea
They operate on $\Omega$ through the $\mathrm{Sp}(2, \mathbb{Z})$ transformations
\begin{equation}
\begin{pmatrix}
\alpha & \beta \\
\gamma& \delta
\end{pmatrix}_\sigma = 
\begin{pmatrix}
	1 &  0 & 0 & 0 \\
	0 & \alpha & 0 & \beta \\
	0 & 0 & 1 & 0 \\
	0 & \gamma & 0 & \delta 
\end{pmatrix} \;,
\label{sp2zsig}
\end{equation}
which results in the transformation laws
\bea
\rho' &=& \rho - \frac{\gamma v^2}{\gamma \sigma + \delta} \nonumber\\
\sigma' &=& \frac{\alpha \sigma + \beta}{\gamma \sigma + \delta} \nonumber\\
v' &=& \frac{v}{\gamma \sigma + \delta} \;.
\label{gtr}
\eea
Under this set of transformations, 
$\Phi_{10}$ transforms as
\bea
\Phi_{10} (\rho', \sigma', v') = \left( \det \left( C \Omega + D \right) \right)^{10} \, \Phi_{10} (\rho, \sigma, v) = \left( \gamma \sigma + \delta \right)^{10} 
\, \Phi_{10} (\rho, \sigma, v).
\eea

Furthermore, $\Phi_{10}$ is also invariant under the integer shifts 
\begin{equation}
	\Phi_{10}(\rho+\lambda, \sigma+\mu, v+\nu) = \Phi_{10}(\rho,\sigma,v), \hspace{5mm} \lambda,\mu,\nu \in\mathbb{Z} \;,
\end{equation}
which implies that it possesses a Fourier expansion.

Performing first an $\mathrm{SL}(2,\mathbb{Z})_\sigma$ transformation, then an S-duality transformation, and finally an  integer shift 
of $v$ by $-\Sigma$, $\Phi_{10}$ changes as
\begin{equation}
	{\Phi_{10}(\rho',\sigma',v')} = (\gamma\sigma + \delta)^{10}{\Phi_{10}(\rho,\sigma,v)} \;,
\end{equation}
where
\bea
\rho' & = & a^2\left(\rho-\frac{\gamma v^2}{\gamma\sigma+\delta} \right)+b^2\left(\frac{\alpha\sigma+\beta}{\gamma\sigma+\delta}\right)-2ab \frac{v}{\gamma\sigma+\delta} \nonumber\\
\sigma' & =&c^2\left(\rho-\frac{\gamma v^2}{\gamma\sigma+\delta} \right)+d^2\left(\frac{\alpha\sigma+\beta}{\gamma\sigma+\delta}\right)-2cd\frac{v}{\gamma\sigma+\delta}  \nonumber\\
v' & =  & -ac\left(\rho-\frac{\gamma v^2}{\gamma\sigma+\delta} \right)-bd\left(\frac{\alpha\sigma+\beta}{\gamma\sigma+\delta}\right)+(ad+bc)\frac{v}{\gamma\sigma+\delta}-\Sigma.
\label{rsvp}
\eea
The reciprocal of $\Phi_{10}$ transforms accordingly.

\subsection{Reparametrization of the poles of $1/\Phi_{10}$}

The Siegel modular form $\Phi_{10}^{-1}(\rho,\sigma,v)$ has the following behaviour near $v=0$,
\begin{equation}
	\frac{1}{\Phi_{10}(\rho,\sigma,v)} = -\frac{1}{4\pi^2}\frac{1}{v^2} \frac{1}{\eta^{24} (\rho)}\frac{1}{ \eta^{24}(\sigma)}+\mathcal{O}(v^0).
\end{equation}
Therefore, it has poles at all the $\mathrm{Sp}(2,\mathbb{Z})$ images of the divisor $v=0$ in the Siegel upper half plane.
The location of the poles is determined by
\begin{equation}
n_2 ( \rho \sigma - v^2) + j v + n_1 \sigma - m_1 \rho + m_2 =0 \;.
\label{n2p}
\end{equation}
The poles are labelled by five integers, $(m_1,m_2,j,n_1,n_2)$ that satisfy the constraint
\begin{equation}
	 m_1 n_1 + m_2 n_2 =  \frac14 \left( 1 - j^2 \right).
	 \label{conmnj}
\end{equation}
Since \eqref{n2p} and \eqref{conmnj} are invariant under  $(m_1,m_2,j,n_1,n_2) \mapsto  (-m_1,- m_2,- j,- n_1,- n_2)$, we 
 may restrict to $n_2 \geq 0$ \cite{Banerjee:2008ky}. 

A specific parametrization of these poles was given in \cite{Murthy:2009dq}. We will use an equivalent parametrization, as follows.

{\proposition 
Any pole \eqref{n2p} with $n_2 \geq 1$ can be reparametrized in terms of nine integers as 
\bea
n_2 &=& - a c \gamma \;, \nonumber\\
j &=& a d + b c \;, \nonumber\\
n_1 &=& - b d \alpha - \gamma \Sigma 
\; \nonumber\\
m_1 &=& a c \delta \; \nonumber\\
m_2 &=& - b d \beta - \delta  \Sigma  \;,
\label{set1set2enl}
\eea
where eight of the integers can be arranged into two $\mathrm{SL}(2,\mathbb{Z})$ matrices,
\begin{equation}
	\Gamma = \begin{pmatrix}
		\alpha & \beta \\
		\gamma & \delta 
	\end{pmatrix}\in \mathrm{SL}(2,\mathbb{Z}), \hspace{5mm} 
	 G= \begin{pmatrix}
		a & b \\
		c & d
	\end{pmatrix} \in \mathrm{SL}(2,\mathbb{Z}),
	\label{ggm}
\end{equation}
with entries that satisfy $a>0,\, c<0, \, \gamma>0,\, \alpha\in\mathbb{Z}/\gamma\mathbb{Z}$, while
$\Sigma\in \mathbb{Z}$.

Moreover, any pole \eqref{n2p} with $n_2 \geq 1$ can be mapped to the pole $v=0$ by the following sequence of 
$\mathrm{Sp}(2,\mathbb{Z})$ transformations, namely first an 
$\mathrm{SL}(2,\mathbb{Z})_\sigma$ transformation, then an S-duality transformation, and finally an integer shift of $v$.}

\begin{proof}

By performing the $\mathrm{SL}(2,\mathbb{Z})_{\sigma}$ transformation \eqref{sp2zsig}, a generic pole \eqref{n2p} with $n_2 \geq 1$ gets
mapped to 
\begin{equation}
	n_2 ( \rho' \sigma' - v'^2) + j v' + n_1 \sigma' - m_1 \rho' + m_2 =0 \;,
\end{equation}
which equals
\begin{equation}
	(\rho \sigma - v^2)(n_2\alpha-m_1\gamma) + j v + \sigma(n_1\alpha+m_2\gamma) -\rho(m_1\delta-n_2\beta) + n_1\beta+m_2\delta =0
	\label{polesig}
\end{equation}
in the original variables. Therefore, this transformation maps
\begin{equation}
	n_2 \mapsto n_2\alpha-m_1\gamma, \;\; n_1 \mapsto n_1\alpha+m_2\gamma, \;\; m_1 \mapsto m_1\delta-n_2\beta, \;\; m_2 \mapsto n_1\beta+m_2\delta,
\end{equation}
leaving $j$ invariant. Let
\begin{equation}
	r = \gcd(n_2,-m_1) >0 \;,
\end{equation}
so that
\begin{equation}
	n_2 = r\gamma, \hspace{7mm}-m_1 = r\delta \;.
\end{equation}
Note that $\gamma > 0$. Now define $\alpha\in\mathbb{Z}/\gamma\mathbb{Z}$ and $\beta$ to satisfy
$\alpha\delta-\beta\gamma=1$. Then, 
inspection of \eqref{polesig} shows that 
the $\mathrm{SL}(2,\mathbb{Z})_{\sigma}$ transformation 
\begin{equation}
	\begin{pmatrix}
		\delta & -\beta \\
		-\gamma & \alpha
	\end{pmatrix}
\end{equation}
takes the generic pole \eqref{n2p} to the linear pole 
\begin{equation}
	\rho r +jv+\sigma(n_1\delta -m_2\gamma)-n_1\beta + m_2 \alpha = 0.
	\label{linp}
\end{equation}
This is equivalent to starting from \eqref{n2p} and mapping 
\bea
\begin{pmatrix} n_2 & n_1 \\ - m_1 & m_2 \end{pmatrix} 
\mapsto
\begin{pmatrix} n'_2 & n'_1 \\ - m'_1 & m'_2 \end{pmatrix} = 
\begin{pmatrix} 0 & n_1\delta -m_2\gamma \\ r & \;  - n_1 \beta + m_2 \alpha \end{pmatrix} = 
\begin{pmatrix} \delta & -\gamma \\ -\beta & \alpha \end{pmatrix} \begin{pmatrix}  r \gamma & n_1 \\  r \delta & m_2 \end{pmatrix} \;.
\eea
Since the determinant of the matrix on the left hand side equals $(1-j^2)/4$ by virtue of \eqref{conmnj}, we infer
\bea
\frac{1-j^2}{4} = \frac{(1-j)}{2} \ \frac{(1+j)}{2} = r \left(  m_2 \gamma - n_1 \delta \right) \;.
\label{detcond}
\eea
Now we define $-c = \gcd(r, (j-1)/2)$ and $d= \gcd(m_2 \gamma - n_1 \delta, (j+1)/2)$. Then $\gcd(-c, d) = 1$. Consequently, we write 
\bea
r &=& -ac \nonumber\\
\tfrac12 (j-1) &=& B c \;\;,\;\; \tfrac12 (j+1) = A d \; \Longrightarrow \;
j = Ad+Bc \text{ with } Ad-Bc=1 \nonumber\\
m_2 \gamma - n_1 \delta &=& b d \;.
\eea
Note that $c<0$ and $a> 0$, as well as 
$\gcd(a,  (j-1)/2) = 1 = \gcd(b,  (j+1)/2)$. It is straightforward to see\footnote{A common prime factor $p$ of $a$ and $B$ must divide $(j-1)/2$. Hence, $p | (\gcd (a, (j-1)/2) = 1 )$   implying $(a, B) =1$. Similarly $(b, A)=1$.} that 
\bea
(a, B) &=& 1,\nonumber\\
(b, A) &=& 1 \;.
\label{gcdaabb}
\eea
The determinant equation \eqref{detcond} gives
\bea
a c b d = A c B d \,,
\eea
which, using \eqref{gcdaabb},
implies $a = \pm A$ and $b = \pm B$, where the signs are correlated. We pick the $+$ sign.
Then, the pole \eqref{linp} can be written as
\bea
-a c  \rho + (a d + b c ) v - b d  \sigma -n _1 \beta + m_2 \alpha = 0 \;.
\label{linpp}
\eea
This allows us to deploy the $\mathrm{SL}(2, \mathbb{Z})$ matrix $\begin{pmatrix} d & - b \\ - c & a \end{pmatrix}$ to implement an S-duality transformation on $\Omega$,
as in \eqref{S-dual}, to yield 
\bea
v -n _1 \beta + m_2 \alpha = 0 \;.
\eea

We finally execute a simple translation operation $v\rightarrow v + \Sigma$, where $\Sigma = n_1 \beta - m_2 \alpha \in \mathbb{Z}$, to generate the pole $v=0$. 
We now solve 
\bea 
m_2 \gamma - n_1 \delta &=& b d \nonumber\\
n_1 \beta - m_2 \alpha &=&  \Sigma \;.
\eea
If $\alpha =0$, we infer using $\beta \gamma = -1$,
\bea
n_1 \beta &=&  \Sigma \longrightarrow n_1 = - \gamma \Sigma \nonumber\\
m_2 &=& - b d \beta - \delta  \Sigma  \;.
\eea
On the other hand, if $\alpha \neq 0$,
then we may multiply the first equation with $\alpha$ and the second equation with $\gamma$
to get 
\bea
n_1 = - b d \alpha - \gamma \Sigma \;.
\eea
Inserting this result back into the first equation determines $m_2$ to equal
\bea
m_2 = - b d \beta - \delta  \Sigma  \;.
\eea
These expressions reproduce the values of $n_1$ and $m_2$ when $\alpha =0$.

Thus, we have shown that the five integers $m_1, m_2, j, n_1, n_2$ can be parametrized as in \eqref{set1set2enl},
and any pole \eqref{n2p} with $n_2 \geq 1$ can be mapped to the pole $v=0$ by the sequence of $\mathrm{Sp}(2, \mathbb{Z})$ transformations
given above.

\end{proof}

Using the parametrization \eqref{set1set2enl}, 
each of the poles \eqref{n2p} with $n_2 \geq 1$ corresponds to an element of the set $P \cup \{ \Sigma \in \mathbb{Z} \}$, where $P$ is
defined by
\bea
	{P} &=& \left\{ \begin{pmatrix}
		a & b \\
		c&d
	\end{pmatrix}, \begin{pmatrix}
		\alpha & \beta \\
		\gamma & \delta
	\end{pmatrix} \in \mathrm{SL}(2,\mathbb{Z})
		 \; \vert  
	 a,\gamma>0, c<0, \alpha\in\mathbb{Z}/\gamma\mathbb{Z} \right\}.
	 \label{setP}
\eea
Defining the sets of matrices, 
\begin{equation}
\label{sgsg}
	S_\Gamma = \left\{ \begin{pmatrix}
		\alpha & \beta \\
		\gamma & \delta 
	\end{pmatrix}\in {\rm SL}(2,\mathbb{Z}): \gamma>0 \right\}, \hspace{2mm} 
	S_G = \left\{ \begin{pmatrix}
		a & b \\
		c & d
	\end{pmatrix} \in {\rm SL}(2,\mathbb{Z}): a>0,c<0\right\},
\end{equation}
we will be summing over elements in $\Gamma_\infty \backslash S_\Gamma$, $S_G$, and in $ \{ \Sigma \in \mathbb{Z} \}$,
where
\begin{equation}
	\Gamma_\infty = \left\{ \begin{pmatrix}
 1 & k \\
 0 & 1	
 \end{pmatrix}\in {\rm SL}(2,\mathbb{Z}) : k\in\mathbb{Z}
 \right\}.
\end{equation}
We can therefore parametrize the sum over poles in the set $P$ as
\begin{equation}
	\sum_{P} = \sum_{\substack{
		\Gamma_\infty \backslash S_\Gamma\\ S_G
				 }}. 
\end{equation}

\subsection{Residues at $v'=0$}
As shown above, any pole \eqref{n2p} with $n_2 \geq 1$ can be mapped to the pole $v=0$ under the $\mathrm{Sp}(2, \mathbb{Z})$ transformation
\begin{equation}
	\frac{1}{\Phi_{10}(\rho,\sigma,v)} = (\gamma\sigma + \delta)^{10}\frac{1}{\Phi_{10}(\rho',\sigma',v')} \;,
\end{equation}
with $\rho', \sigma',v' $ given by \eqref{rsvp}.

Defining 
\begin{equation}
	\Lambda (\sigma, v) = \frac{\gamma v^2}{\gamma \sigma + \delta} 
-\frac{b d}{ac}   \left( \frac{\alpha \sigma + \beta}{\gamma \sigma + \delta} \right)  + \frac{(a d + b c )}{ac}\; \frac{v}{\gamma \sigma + \delta} - \frac{1}{ac} \Sigma,
\label{lamb}
\end{equation}
we infer
\begin{equation}
	v' = -ac \left(\rho-\Lambda  (\sigma, v) \right).
	\label{vprl}
\end{equation}
From the above we see that for a $\Lambda$ satisfying the constraint $\rho = \Lambda$, a translation of $\Sigma$ in integral units of $ac$ will modify $\Lambda$ such that it will no longer satisfy the constraint in the given range of  $\rho_1$.
Hence we restrict $\Sigma$ to take values
\begin{equation}
	\Sigma \in \mathbb{Z}/|ac|\mathbb{Z}.
\end{equation}

Next we evaluate the $\rho$-integral in \eqref{degg}. We only consider the contributions from the residues associated with the 
poles of  $1/\Phi_{10}$ with $n_2 \geq 1$. Since any such pole can be mapped to the pole $v' =0$, we compute the residue associated with 
$\rho = \Lambda  (\sigma, v) $, for fixed $\sigma$ and $v$ by noting that 
in the neighbourhood of $v=0$, 
$1/\Phi_{10}$ behaves as
\begin{equation}
	\frac{1}{\Phi_{10}(\rho,\sigma,v)}  \xrightarrow[v \to 0]{}  -\frac{1}{4\pi^2}\frac{1}{v^2}\frac{1}{\eta^{24}(\rho)}\frac{1}{\eta^{24}(\sigma)} \;.
	\label{phi10v0}
\end{equation}
Therefore, for fixed $\sigma$ and $v$, evaluating the residue at $\rho = \Lambda  (\sigma, v) $ using \eqref{phi10v0}, we obtain
\begin{equation}
	-2\pi i \lim_{\rho \to \Lambda(\sigma,v)} (-1)^{\ell+1}\frac{\partial}{\partial\rho}\left( (\rho-\Lambda(\sigma, v))^2 \, (\gamma\sigma+\delta)^{10} \, \frac{(-1)}{4\pi^2}\frac{1}{v'^2}\frac{1}{\eta^{24}(\rho')}\frac{1}{\eta^{24}(\sigma')} \, e^{-2\pi i (m\rho+n\sigma+\ell v)}
 \right). 
\end{equation}
Using
\be 
\frac{d}{d\tau} \log \eta^{24}(\tau) = 2\pi i\, E_2(\tau) \;,
\ee 
we obtain
\begin{equation}
	(-1)^{\ell+1}\frac{(\gamma\sigma+\delta)^{10}}{ac}\left( \frac{m}{ac} + \frac{a}{c}E_2(\rho'_*)+\frac{c}{a}E_2(\sigma'_*)\right)\frac{1}{\eta^{24}(\rho'_*)}\frac{1}{\eta^{24}(\sigma'_*)}e^{-2\pi i (m\Lambda+n\sigma+\ell v)} \;,
	\label{resrv0}
\end{equation}
where 
\bea
\rho'_* &=& - \frac{b}{c} \left( \frac{\alpha \sigma + \beta}{\gamma \sigma + \delta}
\right) + \frac{a}{c} \left( \frac{v}{\gamma \sigma + \delta} \right)
  - \frac{a}{c} \Sigma \nonumber\\
\sigma'_* &=&  \frac{d}{a} \left( \frac{\alpha \sigma + \beta}{\gamma \sigma + \delta}
\right) - \frac{c}{a} \left( \frac{v}{\gamma \sigma + \delta} \right)
  - \frac{c}{a} \Sigma \;.
  \nonumber\\
v'_* &=& 0  \;.
\label{rsvst}
\eea

\section{Integrating over $v$
 \label{sec:vsint}}

Next, we 
perform the integration over $v$ in \eqref{2}. We will first define the contour $\Gamma_v (D)$ for a given $n_2 \geq 1$ pole. For notational
simplicity we will henceforth refer to this contour by $\Gamma_v$.

Recall that we are considering poles with $n_2 \geq 1$.
The pole $v' _*=0$ specified in \eqref{rsvst} will be in the Siegel upper half plane provided
\begin{equation}
	\text{Im}\rho'_*>0, \; \; \; \text{Im}\sigma'_*>0 \;.
	\label{stcond}
\end{equation}
Evaluating
\bea
 {\rm Im} \rho'_* &=& \frac{1}{n_2 | \gamma \sigma + \delta|^2} \left( \gamma a b \sigma_2 -  \gamma \delta a^2 v_2 + \gamma^2 a^2 \left( v_1 \sigma_2 - v_2 \sigma_1 \right) \right) , \nonumber\\
 {\rm Im} \sigma'_* &=& \frac{1}{n_2 | \gamma \sigma + \delta|^2} \left( - \gamma c d \sigma_2 + \gamma \delta c^2 v_2 - \gamma^2 c^2 \left( v_1 \sigma_2 - v_2 \sigma_1 \right) \right) ,
 \label{imrims}
 \eea
 and imposing \eqref{stcond} results in 
\begin{equation}
	\frac{v_2}{\sigma_2} \, \sigma_1 + \frac{\delta}{\gamma} \frac{v_2}{\sigma_2} - \frac{b}{\gamma a } < v_1 < \frac{v_2}{\sigma_2} \, \sigma_1 + \frac{\delta}{\gamma} \frac{v_2}{\sigma_2} - \frac{d}{\gamma c }.
\label{v1b}
\end{equation}
Using the value for $v_2/\sigma_2$ given in \eqref{rch1}, we obtain
\begin{equation}
	{ \cal V} < v_1 < {\cal V} + \frac{1}{n_2}
\label{intV}
\end{equation}
with 
\begin{equation}
{\cal V} = 	-\frac{\ell}{2m\gamma}(\gamma\sigma_1+\delta) - \frac{b}{\gamma a} \;.
\label{calV}
\end{equation}
The above defines a contour $\Gamma_v$ of integration for the $v$-integral that goes from 
\begin{equation}
	-\frac{\ell}{2m\gamma}(\gamma\sigma+\delta) - \frac{b}{\gamma a} \hspace{4mm} \text{to }\hspace{4mm} -\frac{\ell}{2m\gamma}(\gamma\sigma+\delta) - \frac{b}{\gamma a} - \frac{1}{ac\gamma}.
	\label{attrcontour}
\end{equation}
Since $n_2 \geq 1$, the range of integration specified by \eqref{intV} lies in the original unit interval length integration contour for $v_1$, as required.

\subsection{The $v$ integral }

The residue associated with the pole $v_*'=0$ was given in \eqref{resrv0}. We now turn to the $v$-integral over the sum of the residues
associated with the poles that belong to the set $P \cup \{ \Sigma \in \mathbb{Z} \}$
given in \eqref{setP}. 
We follow the prescription given in \eqref{2},
\begin{equation}
\sum_{\substack{P\\ \Sigma \in \mathbb{Z} }} \, (-1)^{\ell+1}\frac{(\gamma\sigma+\delta)^{10}}{ac} \,
\int_{\Gamma_v} dv
\left( \frac{m}{ac} + \frac{a}{c}E_2(\rho'_*)+\frac{c}{a}E_2(\sigma'_*)\right)\frac{1}{\eta^{24}(\rho'_*)}\frac{1}{\eta^{24}(\sigma'_*)}e^{-2\pi i (m\Lambda+n\sigma+\ell v)}.
	\label{vintres}
\end{equation}
Since $\rho'_*$ and $\sigma'_*$ satisfy \eqref{stcond}, we may Fourier expand $1/\eta^{24}$ and $E_2$ using
\begin{equation}
	\frac{1}{\eta^{24}(\tau)} = \sum_{n=-1}^\infty d(n) e^{2\pi i \tau n}
\end{equation}
and
\begin{equation}
	\frac{E_2(\tau)}{\eta^{24}(\tau)} = -\sum_{n=-1}^\infty nd(n) e^{2\pi i \tau n} \;.
\end{equation}
Then, \eqref{vintres} becomes
\bea
	&& \sum_{\substack{P\\ \Sigma \in \mathbb{Z} }}  (-1)^{\ell} \frac{(\gamma\sigma+\delta)^{10}}{ac}
	\int_{\Gamma_v} dv \, \sum_{M,N\geq -1}
	\left( -\frac{m}{ac} + \frac{a}{c}M+\frac{c}{a}N\right) \, d(M) \, d(N) \nonumber\\ 
	&& \qquad \qquad e^{2\pi i \rho'_* M}e^{2\pi i \sigma'_* N}e^{-2\pi i (m\Lambda+n\sigma+\ell v)} .
\eea
Substituting the values for $\rho'_*$ and $\sigma'_*$ given in \eqref{rsvst} results in
\begin{equation}
	(-1)^\ell \sum_{\substack{P\\ \Sigma \in \mathbb{Z} }}   \frac{(\gamma\sigma+\delta)^{10}}{ac} \int_{\Gamma_v} dv \, \sum_{M,N\geq -1}
	\left( -\frac{m}{ac} + \frac{a}{c}M+\frac{c}{a}N\right)d(M)d(N)\exp\left(-2\pi i A \right) \;,
	\label{residue-expanded-1}
\end{equation}
where
\begin{align}
	&A =\left(\frac{b}{c}M-\frac{d}{a}N - \frac{bd}{ac}m\right) \left(\frac{\alpha\sigma+\beta}{\gamma\sigma+\delta}\right)+\frac{v}{\gamma\sigma+\delta}\left(-\frac{a}{c}M+\frac{c}{a}N+\frac{ad+bc}{ac} m \right) 
	 \nonumber\\
	  & \quad + m \frac{\gamma v^2}{\gamma\sigma+\delta}+\ell v +n\sigma + \Sigma\left(  -\frac{m}{ac} + \frac{a}{c}M+\frac{c}{a}N\right).
\end{align}
This leads us to define the combinations
\begin{align}
	L & =  -\frac{m}{ac}+\frac{a}{c}M+\frac{c}{a}N 
		\nonumber\\
	\tilde{\ell} & =  -\frac{ad+bc}{ac} m+\frac{a}{c}M-\frac{c}{a}N 
		 \nonumber\\
	\tilde{n} & =  \frac{bd}{ac}m-\frac{b}{c}M+\frac{d}{a}N . 
	\label{L-lt-n-tilde-definition}
\end{align}
Notice that \eqref{residue-expanded-1} contains the following sum over $\Sigma$,
\begin{equation}
	\sum_{\Sigma \in \mathbb{Z}/|ac|\mathbb{Z}} e^{-2\pi i \Sigma L} \;,
\end{equation}
which is only non-vanishing provided that the combination $L$ is an integer, in which case this sum equals 
$-ac$. In other words, the only contributions to \eqref{residue-expanded-1} will come from poles that satisfy the divisibility
condition
\begin{equation}
	ac \mid -m + a^2 M +c^2 N .
\end{equation}
Thus, restricting to $L\in\mathbb{Z}$,  \eqref{residue-expanded-1} becomes 
\bea
	&&(-1)^{\ell+1}\sum_{P} {(\gamma\sigma+\delta)^{10}} \,\int_{\Gamma_v} dv \, 
		\sum_{\substack{M,N\geq -1\\  L\in\mathbb{Z} }}  L \, d(M) \, d(N) \nonumber\\
	&&\qquad \exp\left(-2\pi i \left[ - \tilde{n}\left(\frac{\alpha\sigma+\beta}{\gamma\sigma+\delta}\right)-\tilde{\ell} \frac{v}{\gamma\sigma+\delta}+m \frac{\gamma v^2}{\gamma\sigma+\delta}+\ell v +n\sigma \right] \right) \;,
	\label{ppse}
\eea

Using \eqref{L-lt-n-tilde-definition}, we obtain
\bea
\label{setgam}
N &=& a^2 {\tilde n} + b^2 m + ab {\tilde \ell} \;, \nonumber\\
L &=& (ad + bc ) {\tilde \ell} + 2  ac {\tilde n} + 2 bd m 
 \;, \nonumber\\
M &=& c^2 {\tilde n} + d^2 m + cd {\tilde \ell} \;. 
\eea
We note that the triplets $(m,\tilde{n},\tilde{\ell})$ and $(M,N,L)$ are related by the following $\mathrm{SL}(2,\mathbb{Z})$ transformation,
\begin{equation}
	\begin{pmatrix}
	N \\
	L \\
	M	
	\end{pmatrix}=
\begin{pmatrix}
		a^2 & ab & b^2 \\
		2ac & ad+bc & 2bd \\
		c^2 & cd & d^2
	\end{pmatrix}
	\begin{pmatrix}
		\tilde{n} \\
		\tilde{\ell}\\
		m
	\end{pmatrix},
\end{equation}
or equivalently,
\begin{equation}
	\begin{pmatrix}
		\tilde{n} \\
		\tilde{\ell}\\
		m
	\end{pmatrix}=
\begin{pmatrix}
		d^2 & -bd & b^2 \\
		-2cd & ad+bc & -2ab \\
		c^2 & -ac & a^2
	\end{pmatrix} 
	\begin{pmatrix}
	N \\
	L \\
	M	
	\end{pmatrix} \;.
\end{equation}
Therefore, since the triplet $(M,N,L)$ consists of integers, also 
$\tilde{n}$ and $\tilde{\ell}$ have to be integers. Thus, we have two triplets of integers that are related by the
$\mathrm{SL}(2,\mathbb{Z})$ matrices $G$ given in \eqref{ggm}.

Next, we define 
\bea
\tilde{\Delta} = 4 m {\tilde n} - {\tilde \ell}^2 \;.
\label{deltdel}
\eea
Note that $\tilde{\Delta}$ is invariant under the $\mathrm{SL}(2,\mathbb{Z})$ transformation $G$ given in \eqref{ggm},
\bea
\tilde{\Delta} = 4 m {\tilde n} - {\tilde \ell}^2  = 4 MN - L^2 \;.
\eea
Now we recall that we are considering BPS dyons with $\Delta = 4 m n - \ell^2> 0$, and hence $m>0$, in which case we may
perform the following rewriting of the exponent in \eqref{ppse},
\begin{align}
	&-\tilde{\ell} \frac{v}{\gamma\sigma+\delta}+m \frac{\gamma v^2}{\gamma\sigma+\delta}+\ell v \\
	& = \frac{m\gamma}{\gamma\sigma+\delta} \left( v+\frac{1}{2m\gamma}\left( (\gamma\sigma+\delta)\ell-\tilde{\ell}\right)\right)^2 -\frac{1}{4m\gamma}(\gamma\sigma+\delta)\ell^2 - \frac{1}{4m\gamma}\frac{\tilde{\ell}^2}{\gamma\sigma+\delta}+\frac{1}{2m\gamma}\ell\tilde{\ell}. \nonumber
\end{align}
Using this, \eqref{ppse} can be written as 
\begin{equation} \label{whole-expr-2}
\begin{gathered}
	(-1)^{\ell+1}\sum_{{P}}{(\gamma\sigma+\delta)^{10}} \,\int_{\Gamma_v} dv \, 
		\sum_{\substack{M,N\geq -1\\  L\in\mathbb{Z} }}
	L d(M)d(N)
	e^{2\pi i \left(\frac{\alpha}{\gamma}\frac{\tilde{\Delta}}{4m}+ \frac{\delta}{\gamma}\frac{\Delta}{4m}  \right)}
	e^{2\pi i \left(\frac{\alpha}{\gamma}\frac{\tilde{\ell}^2}{4m}-\frac{\ell\tilde{\ell}}{2m\gamma} +\frac{\delta}{\gamma}\frac{\ell^2}{4m}   \right)}
\\
	e^{-2\pi i \left[  
	\frac{\tilde{\Delta}}{4m}\frac{1}{\gamma}\frac{1}{\gamma\sigma+\delta}
	+\frac{\Delta}{4m}\frac{\gamma\sigma+\delta}{\gamma}
	+\frac{m\gamma}{\gamma\sigma+\delta}\left(v+ \frac{1}{2m\gamma}\left((\gamma\sigma+\delta)\ell-\tilde{\ell}\right)\right)^2 \right]}.
	\end{gathered}
\end{equation}

\subsubsection{$T$-shifts}

In the following, we rewrite the expression \eqref{whole-expr-2} by
recasting the sum over  $b\in \mathbb{Z}$
as a sum over
a new integer $T$, as follows. Let $T \in \mathbb{Z}$ and consider the matrix 
$\begin{pmatrix}
		1 & T \\
		0 & 1
	\end{pmatrix}$.
First we note 
that the operation
\begin{equation}
	\begin{pmatrix}
		a & b \\
		c & d
	\end{pmatrix}\begin{pmatrix}
		1 & T \\
		0 & 1
	\end{pmatrix} = \begin{pmatrix}
		a & aT+b \\
		c & cT +d
	\end{pmatrix}
\end{equation}
induces the change
\begin{equation}
	b/a \mapsto b/a + T, \hspace{5mm}d/c \mapsto d/c + T,
	\label{bdT}
\end{equation}
while leaving $a$ and $c$ invariant. Therefore, expressions such as $L$ that only depend on $a$ and $c$ are invariant under this change.
Hence, also $\tilde \Delta$ is invariant under this change. On the other hand, $\tilde{\ell}$ and $\tilde{n}$ transform as follows,
\bea
	\tilde{\ell} &\mapsto& \tilde{\ell} -2mT \nonumber\\
		\tilde{n} &\mapsto& \tilde{n}-\tilde{\ell}T +mT^2 \;.
\eea

Next, note that $\tilde{\ell} $ can be written as 
\begin{equation}
	\tilde{\ell} = -\frac{1}{ac} m+\frac{a}{c}M-\frac{c}{a}N -2m\frac{b}{a} \;,
	\label{lba}
\end{equation} 
and hence the condition $b\in \mathbb{Z}/a\mathbb{Z}$ translates to $\tilde{\ell} \in \mathbb{Z}/2m\mathbb{Z}$.
Thus we can express \eqref{whole-expr-2} as 
\bea
\label{whole-expr-3} 
&&	 (-1)^{\ell+1} 
	\sum_{\substack{P\\ b\in \mathbb{Z}/a\mathbb{Z}, \, T \in \mathbb{Z}  }}
	{(\gamma\sigma+\delta)^{10}} \,  \int_{\Gamma_v} dv \, 
		\sum_{\substack{M,N\geq -1\\  L\in\mathbb{Z} }} \, L \, d(M) \, d(N) \\
	&& \qquad \qquad 
	e^{2\pi i \left(\frac{\alpha}{\gamma}\frac{\tilde{\Delta}}{4m}+ \frac{\delta}{\gamma}\frac{\Delta}{4m}  \right)}
	e^{2\pi i \left(\frac{\alpha}{\gamma}\frac{(\tilde{\ell}-2mT)^2}{4m}-\frac{\ell(\tilde{\ell}-2mT)}{2m\gamma} +\frac{\delta}{\gamma}\frac{\ell^2}{4m}   \right)}
\nonumber\\
&&
	 \exp\left(-2\pi i \left[  
	\frac{\tilde{\Delta}}{4m}\frac{1}{\gamma}\frac{1}{\gamma\sigma+\delta}
	+\frac{{\Delta}}{4m}\frac{\gamma\sigma+\delta}{\gamma}
	+\frac{m\gamma}{\gamma\sigma+\delta}\left(v+ \frac{1}{2m\gamma}\left((\gamma\sigma+\delta)\ell-\tilde{\ell}\right)+\frac{T}{\gamma}\right)^2 \right] \right). \nonumber
	\eea
Using \eqref{bdT},  the integration contour \eqref{attrcontour} goes from 
\begin{equation}
	-\frac{\ell}{2m\gamma}(\gamma\sigma+\delta) - \frac{b}{\gamma a} - \frac{T}{\gamma} \hspace{4mm} \text{to }\hspace{4mm} -\frac{\ell}{2m\gamma}(\gamma\sigma+\delta) - \frac{b}{\gamma a} - \frac{1}{ac\gamma}- \frac{T}{\gamma} \;,
	\label{vrangewithT2}
\end{equation}
and the range of integration specified by \eqref{intV} requires restricting the values of $T$ to
\begin{equation}
	T \in \mathbb{Z}/\gamma\mathbb{Z}.
\end{equation}

\subsubsection{Performing the $v$-integral}

We now perform the $v$-integral in \eqref{whole-expr-3} along the contour $\Gamma_v$ specified by \eqref{vrangewithT2}.
To do so, we assume the legitimacy of interchanging the integration with the summation over $M,N$ in  \eqref{whole-expr-3}.
Note that the dependence on $v$ is contained in the last line of \eqref{whole-expr-3}, only,
\begin{equation}
	\int_{\Gamma_v} dv \, \exp \left(-2\pi  i\frac{m\gamma}{\gamma\sigma+\delta} \left( v+\frac{1}{2m\gamma}\left( (\gamma\sigma+\delta)\ell-\tilde{\ell}\right)+\frac{T}{\gamma}\right)^2\right) \;.
	\label{intvT}
\end{equation}
Using the expression for the error function 
\begin{equation}
	\int^x dt \,e^{-a(t+b)^2} = \frac{1}{2}\sqrt{\frac{\pi}{a}}\text{Erf}\left[\sqrt{a}(x+b) \right]+C,
\end{equation}
we obtain for \eqref{intvT}, 
\begin{align}
	& \frac{1}{2}\frac{\sqrt{\gamma\sigma+\delta}}{\sqrt{2m\gamma i}}\left(\text{Erf}\left[\sqrt{\frac{2\pi i m\gamma}{\gamma\sigma+\delta}}\left(-\frac{b}{a\gamma}-\frac{1}{ac\gamma}-\frac{\tilde{\ell}}{2m\gamma} \right)\right] -\text{Erf}\left[\sqrt{\frac{2\pi i m\gamma}{\gamma\sigma+\delta}}\left(-\frac{b}{a\gamma}-\frac{\tilde{\ell}}{2m\gamma} \right)\right]\right),
\end{align}
where we take the principal branch of the square roots.

Hence, integrating  \eqref{whole-expr-3} over $v$ results in 
\begin{equation}
\begin{gathered}
	(-1)^{\ell+1}
	\sum_{\substack{P\\  b\in \mathbb{Z}/a\mathbb{Z}, \, T \in \mathbb{Z}/\gamma\mathbb{Z}}} \;
	{(\gamma\sigma+\delta)^{10}} \,
		\sum_{\substack{M,N\geq -1\\  L\in\mathbb{Z} }}  \, L \, d(M) \, d(N) \,
	e^{2\pi i \left(\frac{\alpha}{\gamma}\frac{\tilde{\Delta}}{4m}+ \frac{\delta}{\gamma}\frac{\Delta}{4m}  \right)}
\\
	e^{2\pi i \left(\frac{\alpha}{\gamma}\frac{(\tilde{\ell}-2mT)^2}{4m}-\frac{\ell(\tilde{\ell}-2mT)}{2m\gamma} +\frac{\delta}{\gamma}\frac{\ell^2}{4m}   \right)}   
	e^{\left(-2\pi i \left[  
	\frac{\tilde{\Delta}}{4m}\frac{1}{\gamma}\frac{1}{\gamma\sigma+\delta}
	+\frac{{\Delta}}{4m}\frac{\gamma\sigma+\delta}{\gamma}
	\right] \right)}
	\\
	\frac{1}{2}\frac{\sqrt{\gamma\sigma+\delta}}{\sqrt{2m\gamma i}}\left(\text{Erf}\left[\sqrt{\frac{2\pi i m\gamma}{\gamma\sigma+\delta}}\left(-\frac{b}{a\gamma}-\frac{1}{ac\gamma}-\frac{\tilde{\ell}}{2m\gamma} \right)\right] -\text{Erf}\left[\sqrt{\frac{2\pi i m\gamma}{\gamma\sigma+\delta}}\left(-\frac{b}{a\gamma}-\frac{\tilde{\ell}}{2m\gamma} \right)\right]\right).
	\end{gathered} \label{after-v-integration}
\end{equation}

\subsubsection{$\mathrm{SL}(2,\mathbb{Z})$ multiplier system}

Inspection of \eqref{after-v-integration} reveals the presence of the multiplier system 
associated with the $\mathrm{SL}(2,\mathbb{Z})$ matrix $\Gamma = \begin{pmatrix}
	\alpha & \beta \\
	\gamma & \delta
\end{pmatrix}$
\cite{Zagier1989,Gomes:2017bpi},
\begin{equation}
	\psi(\Gamma)_{\ell j} = \frac{1}{\sqrt{2m\gamma i}}\sum_{T\in\mathbb{Z}/\gamma\mathbb{Z}} e^{2\pi i \left(\frac{\alpha}{\gamma}\frac{(\ell-2mT)^2}{4m}-\frac{j(\ell-2mT)}{2m\gamma} +\frac{\delta}{\gamma}\frac{j^2}{4m}   \right)} \;.
\end{equation}
 It has the property
$\psi_{(\ell+ 2 m k) j}(\Gamma) = \psi_{ {\ell} j}(\Gamma)$ (with $k \in \mathbb{Z}$) \cite{kloosterman}.
This multiplier system arises when considering  the transformation law of the standard Jacobi theta function $\vartheta_{m,\ell}(\sigma,v)$ of
 weight $1/2$ and index $m$
under modular transformations \cite{Dabholkar:2012nd},
\begin{equation}
	\vartheta_{m,\ell}\left( \frac{\alpha\sigma+\beta}{\gamma\sigma+\delta}, \frac{v}{\gamma\sigma+\delta} \right) = (\gamma\sigma+\delta)^{1/2}e^{2\pi i m \frac{\gamma v^2}{\gamma\sigma+\delta}} \sum_{j\in\mathbb{Z}/2m\mathbb{Z}} \psi(\Gamma)_{\ell j}\;\vartheta_{m,j}(\sigma,v) .
\end{equation}
The multiplier system that enters in \eqref{after-v-integration} is $ \psi(\Gamma)_{\tilde{\ell}\ell}$.

\subsubsection{Splitting the error functions}	

The error functions in \eqref{after-v-integration} can be split into three different terms. We refer to Appendix \ref{appendix-error-functions} for the details on this.
Here we briefly summarise its salient features.

Using the property $\text{Erf}(-x) = -\text{Erf}(x)$, we write the difference of the error functions in \eqref{after-v-integration} as
\begin{equation} \label{2-erfs}
\text{Erf}\left[\sqrt{\frac{2\pi i m}{\gamma(\gamma\sigma+\delta)}}\left(\frac{b}{a}+\frac{\tilde{\ell}}{2m} \right)\right]+\text{Erf}\left[\sqrt{\frac{2\pi i m}{\gamma(\gamma\sigma+\delta)}}\left(-\frac{b}{a}-\frac{1}{ac}-\frac{\tilde{\ell}}{2m} \right)\right].
\end{equation}
Next we define the following functions,
\begin{equation}
	E_{m,\gamma(\gamma\sigma+\delta)}(x) = - \frac{1}{\sqrt{\pi}}\frac{1}{\sqrt{\frac{2\pi i m}{\gamma(\gamma\sigma+\delta)}}x}e^{-\frac{2\pi i m}{\gamma(\gamma\sigma+\delta)}x^2}
	\label{e-function}
\end{equation}
and
\begin{equation}
	I_{m,\gamma(\gamma\sigma+\delta)}(x) =-\frac{1}{2\sqrt{\pi}} \frac{1}{\sqrt{2\pi i m}x}e^{-\frac{2\pi i m}{\gamma(\gamma\sigma+\delta)}x^2}\int_0^{i\infty}\left(\frac{1}{\gamma(\gamma\sigma+\delta)}-z \right)^{-3/2}e^{2\pi i m x^2 z}dz \;.
	\label{i-function}
\end{equation}

We introduce 
\begin{equation} \label{def-capital-x-y-1}
	X = \frac{b}{a}+\frac{\tilde{\ell}}{2m}, \hspace{6mm}Y = -\frac{b}{a}-\frac{1}{ac}-\frac{\tilde{\ell}}{2m},
\end{equation}
and we note that since $-1/ac >0$, $X$ and $Y$ cannot be both negative. We now consider the product
\begin{equation} \label{product-cont-fract}
		XY = \left(\frac{b}{a}+\frac{\tilde{\ell}}{2m} \right)\left(-\frac{b}{a}-\frac{1}{ac}-\frac{\tilde{\ell}}{2m} \right) \;.
\end{equation}
Depending on the sign of $X Y$, the expression for the sum \eqref{2-erfs} will take a different form, as follows.
If $X Y > 0$, \eqref{2-erfs} becomes
\begin{equation}  \label{both-positive-1}
 2 + E_{m,\gamma(\gamma\sigma+\delta)} \left(X \right) + I_{m,\gamma(\gamma\sigma+\delta)} \left(X\right)+ E_{m,\gamma(\gamma\sigma+\delta)} \left(Y \right) + I_{m,\gamma(\gamma\sigma+\delta)} \left(Y \right).
\end{equation}
If $X Y = 0$, \eqref{2-erfs} has the form
\begin{equation} \label{one-zero-1}
1 +  E_{m,\gamma(\gamma\sigma+\delta)} \left(-\frac{1}{ac}\right) + I_{m,\gamma(\gamma\sigma+\delta)} \left(-\frac{1}{ac}\right),
\end{equation}
where we used that when  $X=0$, $Y$ is given by $Y=- 1/ac$, and when $Y=0$, $X$ is given by $X=-1/ac$.
Since $- 1/ac > 0$, we can re-express the condition $X Y \geq 0$ as 
\begin{equation} \label{cond-positive-cont-frac}
	0 \leq \frac{b}{a}+\frac{\tilde{\ell}}{2m} \leq -\frac{1}{ac} \;.
\end{equation}
This condition \cite{LopesCardoso:2020pmp} is satisfied by all the convergents of the continued fraction of
\begin{equation}
	\frac{\tilde{\ell}}{2m}.
\end{equation}

Finally, when  $X Y < 0$ , \eqref{2-erfs} takes the form
\begin{equation} \label{both-negative-1}
E_{m,\gamma(\gamma\sigma+\delta)}\left(X \right) + I_{m,\gamma(\gamma\sigma+\delta)} \left(X \right) + E_{m,\gamma(\gamma\sigma+\delta)} \left(Y\right) + I_{m,\gamma(\gamma\sigma+\delta)}\left(Y \right).
\end{equation}

The above shows that the error functions in \eqref{after-v-integration} give rise to three distinct types of terms, namely constants, terms involving
$E_{m,\gamma(\gamma\sigma+\delta)}$ and terms involving $I_{m,\gamma(\gamma\sigma+\delta)}$. As we will see below, 
upon performing the $\sigma$-integral, 
these three distinct contributions
will give rise to terms involving the Bessel function $I_{23/2}$, the Bessel function $I_{12}$ and 
the integral of 
the Bessel function $I_{25/2}$, respectively.

\subsection{A symmetry}

The expression \eqref{after-v-integration} possesses a symmetry that can be used to simplify the terms that involve 
$E_{m,\gamma(\gamma\sigma+\delta)}(x)$ and $I_{m,\gamma(\gamma\sigma+\delta)}(x)$, as follows.
Let us first consider the case when $X Y >0$ or $XY<0$. Then, \eqref{after-v-integration} will contain terms of the form (c.f. \eqref{both-positive-1} and \eqref{both-negative-1}) 
\begin{equation}
\begin{gathered}
	(-1)^{\ell+1} \sum_{\substack{P\\ b\in \mathbb{Z}/a\mathbb{Z} }}
		{(\gamma\sigma+\delta)^{10}} 
	\sum_{\substack{M,N\geq -1\\  L\in\mathbb{Z} }} 
		L \, d(M) \, d(N) \\
	e^{2\pi i \left(\frac{\alpha}{\gamma}\frac{\tilde{\Delta}}{4m}+ \frac{\delta}{\gamma}\frac{\Delta}{4m}  \right)}
	\, \psi(\Gamma)_{\tilde{\ell}\ell} \, e^{\left(-2\pi i \left[  
	\frac{\tilde{\Delta}}{4m}\frac{1}{\gamma}\frac{1}{\gamma\sigma+\delta}
	+\frac{{\Delta}}{4m}\frac{\gamma\sigma+\delta}{\gamma}
	\right] \right)}
	\\
	\frac{1}{2}\sqrt{\gamma\sigma+\delta}\, 
	\left( E_{m,\gamma(\gamma\sigma+\delta)} \left(Y\right) + I_{m,\gamma(\gamma\sigma+\delta)}\left(Y \right) \right) \;.
	\end{gathered} \label{symmetry-1}
\end{equation}
We observe that under the mapping, 
\bea
	\begin{pmatrix}
		a & b \\
		c & d
	\end{pmatrix} & \mapsto & \begin{pmatrix}
		a' & b' \\
		c' & d'
	\end{pmatrix}
=
		\begin{pmatrix}
		-c & d \\
		-a & b
	\end{pmatrix} \nonumber\\
	\nonumber\\
		(M,N)  
				&\mapsto & (M',N') = (N,M) \;,
		\label{relmn}
\eea
the property $a',-c'>0$ as well as the unit determinant property are preserved.
Further, the following quantities remain invariant:
\begin{equation}
	L' = L, \hspace{4mm} \tilde{\Delta}' = \tilde{\Delta} \;,
\end{equation}
while $\tilde{\ell}, X$ and $Y$ are mapped\footnote{Note that this mapping might need to be supplemented by a $\Gamma_{\infty}$ operation on the right to ensure that we are still in the chosen gauge for $b \in \mathbb{Z}/a \mathbb{Z}$. } 
\begin{equation}
	\tilde{\ell}' = -\tilde{\ell}, \hspace{4mm} X' = Y, \hspace{4mm} Y' = X.
\end{equation}
Thus, we can write \eqref{symmetry-1} as 
\begin{equation}
\begin{gathered}
	(-1)^{\ell+1} \sum_{\substack{P\\ b\in \mathbb{Z}/a\mathbb{Z}}}
		{(\gamma\sigma+\delta)^{10}} 
	\sum_{\substack{M,N\geq -1\\  L\in\mathbb{Z} }} 
		L \,  d(M) \, d(N) \\
	e^{2\pi i \left(\frac{\alpha}{\gamma}\frac{\tilde{\Delta}}{4m}+ \frac{\delta}{\gamma}\frac{\Delta}{4m}  \right)} \, \psi(\Gamma)_{-\tilde{\ell}\ell} \,
	e^{\left(-2\pi i \left[  
	\frac{\tilde{\Delta}}{4m}\frac{1}{\gamma}\frac{1}{\gamma\sigma+\delta}
	+\frac{{\Delta}}{4m}\frac{\gamma\sigma+\delta}{\gamma}
	\right] \right)}
	\\
	\frac{1}{2}\sqrt{\gamma\sigma+\delta}
	\left( E_{m,\gamma(\gamma\sigma+\delta)} \left(X\right) + I_{m,\gamma(\gamma\sigma+\delta)}\left(X \right) \right),
	\end{gathered} \label{symmetry-2}
\end{equation}
and hence we can express the combination 
\begin{equation}
	\psi(\Gamma)_{\tilde{\ell}\ell}\left( E_{m,\gamma(\gamma\sigma+\delta)} \left(X\right) + I_{m,\gamma(\gamma\sigma+\delta)}\left(X \right) + E_{m,\gamma(\gamma\sigma+\delta)} \left(Y\right) + I_{m,\gamma(\gamma\sigma+\delta)}\left(Y \right) \right) \;,
\end{equation}
which occurs in \eqref{after-v-integration} when $X Y >0$  or $XY < 0$, as
\begin{equation}
	\left(\psi(\Gamma)_{\tilde{\ell}\ell} +\psi(\Gamma)_{-\tilde{\ell}\ell} \right)\left( E_{m,\gamma(\gamma\sigma+\delta)} \left(X\right) + I_{m,\gamma(\gamma\sigma+\delta)}\left(X \right)  \right).
	\label{pspssu}
\end{equation}
On the other hand, when $XY=0$, 
 \eqref{after-v-integration} will contain terms of the form  \eqref{one-zero-1}, which, using the mapping \eqref{relmn},
 can be brought to the form 
 \eqref{symmetry-2}. Hence, the combined contributions from $X=0$ (in which case $Y=-1/ac$) and from $Y= 0$ (in which case 
 $X=-1/ac$) take again the form \eqref{pspssu}. Thus, irrespective of whether $XY >0, XY=0$ or $XY<0$, 
 the contributions to  \eqref{after-v-integration}
  from the terms involving $E_{m,\gamma(\gamma\sigma+\delta)}(x)$ and $I_{m,\gamma(\gamma\sigma+\delta)}(x)$
  takes the form
    \begin{equation}
\begin{gathered}
	(-1)^{\ell+1} \sum_{\substack{P\\ b\in \mathbb{Z}/a\mathbb{Z} }} \;
		{(\gamma\sigma+\delta)^{10}} 
	\sum_{\substack{M,N\geq -1\\  L\in\mathbb{Z}, \; X \neq 0}} 
		L \, d(M) \, d(N) \\
	e^{2\pi i \left(\frac{\alpha}{\gamma}\frac{\tilde{\Delta}}{4m}+ \frac{\delta}{\gamma}\frac{\Delta}{4m}  \right)}\left(\psi(\Gamma)_{\tilde{\ell}\ell}+\psi(\Gamma)_{-\tilde{\ell}\ell}\right)
	e^{\left(-2\pi i \left[  
	\frac{\tilde{\Delta}}{4m}\frac{1}{\gamma}\frac{1}{\gamma\sigma+\delta}
	+\frac{{\Delta}}{4m}\frac{\gamma\sigma+\delta}{\gamma}
	\right] \right)}
	\\
	\frac{1}{2}\sqrt{\gamma\sigma+\delta}
	\left( E_{m,\gamma(\gamma\sigma+\delta)} \left(X\right) + I_{m,\gamma(\gamma\sigma+\delta)}\left(X \right) \right).
	\end{gathered} \label{symmetry-3}
\end{equation}

\section{Integrating over $\sigma$ \label{sec:sigint}}

Next, we turn to the $\sigma$-integral of \eqref{after-v-integration}. We first explicate our
choice of integration contour $\Gamma_\sigma (D)$ in \eqref{2} for a given $n_2 \geq 1$ pole $D$ in $P \cup \{\Sigma \in \mathbb{Z} \}$.

\subsection{Contour of integration}

The contour $\Gamma_\sigma(D)$ defined in the $\sigma$ upper half plane restricted to 
$\sigma_1 \in [0,1)$ crosses the locus \eqref{n2p} associated with quadratic poles.
Writing out the real and imaginary parts of \eqref{n2p} gives
\bea
&&	-ac\gamma(\rho_1\sigma_1-v_1^2-\rho_2\sigma_2+v_2^2)+(ad+bc)v_1+(-bd\alpha-\gamma\Sigma)\sigma_1-ac\delta\rho_1-bd\beta-\delta\Sigma =0 ,
	\nonumber\\
&&		-ac\gamma(\rho_1\sigma_2+\rho_2\sigma_1-2v_1v_2)+(ad+bc)v_2+(-bd\alpha-\gamma\Sigma)\sigma_2-ac\delta\rho_2= 0. 
\label{riloc}
\eea
Solving the second equation for $\rho_1$ yields
\begin{equation}
\rho_1=	\frac{-a c\rho_2(\gamma\sigma_1+ \delta )+2 a c \gamma v_1 v_2+(a d+bc) v_2+(-bd\alpha-\gamma\Sigma)\sigma_2}{a c \gamma \sigma_2}.
\label{exp-r1}
\end{equation}
Now we recall that $v_1$ lies in the range \eqref{v1b}, which we write as
\begin{equation}
	v_1 = \frac{v_2}{\gamma\sigma_2}(\gamma\sigma_1+\delta)-\frac{b}{\gamma a}+x, 
	\label{v1x}
\end{equation}
where $x\in (0,-1/ac\gamma)$. Inserting \eqref{v1x} as well as \eqref{exp-r1} into the first equation of \eqref{riloc} gives
\bea
\left( \rho_2 \sigma_2 - v_2^2 \right) \left( \sigma_1 + \frac{\delta}{\gamma} \right)^2 + \sigma_2^2 \left( x + \frac{1}{2 a c \gamma} \right)^2 = 
\sigma_2^2 \left[ \frac{1}{4 (a c \gamma)^2 } - \left( \rho_2 \sigma_2 - v_2^2 \right) \right].
\label{ellip}
\eea
This describes an ellipse in the $(\sigma_1, x)$-plane provided that the right hand side of this equation is non-vanishing and positive,
\begin{equation}
	\rho_2\sigma_2-v_2^2 = \rho_2\sigma_2 - \left(\frac{\ell}{2m } \right)^2 \sigma_2^2 	
	<\frac{1}{4a^2c^2\gamma^2},
	\label{rangs2v2}
\end{equation}
where we set
\bea
\frac{v_2}{\sigma_2} = - \frac{\ell}{2m} .
\eea
Combining the $\sigma_2$-direction with \eqref{ellip} results in an ellipsoid in three dimensions, with
$\sigma_2$ taking values in the range specified by \eqref{rangs2v2}.

On the other hand, we note that the point $(\sigma_1 = - \delta/\gamma, x, \sigma_2 =0)$ lies in the locus \eqref{ellip}. This point,
which lies on the boundary of the Siegel upper-half plane, is the anchoring point of a curve in the complex $\sigma$-plane at fixed 
$x\in (0,-1/ac\gamma)$, as follows. Since $\sigma_2 >0$, we may divide \eqref{ellip} to arrive at
\bea
 \left( \sigma_1 +  \, 
\frac{\delta}{\gamma} \right)^2 + \left( \sigma_2  - \frac{X(x)}{2 \lambda} \right)^2   =  \frac{(X(x))^2}{4 \lambda^2 },
\label{circlam}
\eea
where we defined
\bea
\lambda \equiv  \rho_2  - \left(\frac{\ell}{2m } \right)^2 \sigma_2 > 0 \;\;\;,\;\;\; X(x) \equiv \frac{1}{(2 a c \gamma)^2} - \left( x + \frac{1}{2 a c \gamma} \right)^2 .
\eea
Note that $X(x)> 0$ for $x\in (0,-1/ac\gamma)$, and that the positivity of $\lambda $ can be enforced by taking $\rho_2$ to be sufficiently large.
At fixed $x$, the above equation \eqref{circlam} describes a circle in the complex $\sigma$-plane anchored at $(\sigma_1 = - \delta/\gamma, \sigma_2 =0)$,
provided that $\lambda $ is kept fixed. The latter is compatible with the condition  \eqref{rangs2v2}, as follows.
We write \eqref{rangs2v2}
as
\bea
0 < \sigma_2 < \frac{1}{4a^2c^2\gamma^2 \, \lambda}.
\label{condsigl}
\eea
The bound on the right hand side is precisely saturated when $\sigma_2 = X_{\max} /\lambda$, where  $X_{\max} $ is the maximal value of $X$,
which is attained for $x = - 1/(2 a c \gamma)$. The associated point on the circle is $(\sigma_1 = - \delta/\gamma, \sigma_2 = X_{\max} /\lambda)$,
which is the point on the circle that intersects the line $\sigma_1 = - \delta/\gamma$ in the complex $\sigma$-plane. Thus,
we see that keeping $\lambda$ fixed is compatible with restricting the range of $\sigma_2$ to
\bea
0 < \sigma_2 \leq  \frac{1}{4a^2c^2\gamma^2 \, \lambda}.
\eea

At fixed $\lambda$, the circle described by \eqref{circlam} 
is homotopic to a Ford circle $\mathcal{C}( -\delta, \gamma)$ in the complex $\sigma$-plane anchored on the real axis at $\sigma_1 = - \delta/\gamma$ (see Appendix \ref{sec:forad} for details). Note that since the homotopy is between circles, the leading behaviour of the integrand in \eqref{after-v-integration2} when approaching the point 
$(\sigma_1 = - \delta/\gamma, \sigma_2 =0)$ along any of these two circles in the same.
The chosen range $\sigma_1 \in [0,1)$  constrains the poles contributing to \eqref{after-v-integration} to those associated with $0 \leq -\frac{\delta}{\gamma}< 1$. Since this holds for any  $x\in (0,-1/ac\gamma)$, our integration contour over $\sigma$ for a given pole is $\Gamma_\sigma (D)
= \mathcal{C}( -\delta, \gamma) $, which for notational simplicity we will denote by $\Gamma_{\sigma}$.

The interpretation of this construction is the one given in \cite{Sen:2007qy}. When $\sigma_2$ is large, the integration contour does not
intersect the ellipsoid described above. When lowering the value of $\sigma_2$, the integration contour will cross some of the poles in
the Siegel upper half plane 
described by \eqref{circlam}. This will cease to be the case when $\sigma_2$ reaches the boundary $\sigma_2 = 0$ of the Siegel upper half plane.
Fixing the value of $\lambda$ to be large enough, 
we note that as we decrease $\sigma_2$ we continue to remain in the ${\cal R}$-chamber, ensuring that the integration contour does not cross
any $n_2=0$ pole.

We will now perform the  $\sigma$-integral of \eqref{after-v-integration} over the Ford circle $\Gamma_{\sigma}$ described above, 
following the prescription given 
in \eqref{2}, which results in 
\begin{equation}
\begin{gathered}
	(-1)^{\ell+1}
	\sum_{\substack{P\\ b \in\mathbb{Z}/a\mathbb{Z}}} 
		\int_{\Gamma_{\sigma}} d \sigma
	\; {(\gamma\sigma+\delta)^{10}} 
	\sum_{\substack{M,N\geq -1\\  L\in\mathbb{Z} }}  \, L \, d(M) \, d(N) \\
	e^{2\pi i \left(\frac{\alpha}{\gamma}\frac{\tilde{\Delta}}{4m}+ \frac{\delta}{\gamma}\frac{\Delta}{4m}  \right)} \, \psi(\Gamma)_{\tilde{\ell}\ell} \;
		e^{\left(-2\pi i \left[  
	\frac{\tilde{\Delta}}{4m}\frac{1}{\gamma}\frac{1}{\gamma\sigma+\delta}
	+\frac{{\Delta}}{4m}\frac{\gamma\sigma+\delta}{\gamma}
	\right] \right)} \\
	\frac{1}{2}
		\sqrt{\gamma\sigma+\delta}
		\left(\text{Erf}\left[\sqrt{\frac{2\pi i m\gamma}{\gamma\sigma+\delta}}\left(-\frac{b}{a\gamma}-\frac{1}{ac\gamma}-\frac{\tilde{\ell}}{2m\gamma} \right)\right] -\text{Erf}\left[\sqrt{\frac{2\pi i m\gamma}{\gamma\sigma+\delta}}\left(-\frac{b}{a\gamma}-\frac{\tilde{\ell}}{2m\gamma} \right)\right]\right).
	\end{gathered} \label{after-v-integration2}
\end{equation}
To perform this integral, we will use the 
decomposition of the error functions given above.

\subsection{Bessel function $I_{23/2}$}

We first focus 
on the constant terms  in the decompositions
\eqref{both-positive-1} and \eqref{one-zero-1}. 

Firstly we show that the two cases corresponding to 
$X=0$ and  $Y=0$ give rise to the same contribution, as follows. The condition $X=0$ yields $m -a^2 M = - c^2 N$, and hence
$L = 2 c N / a$. We will show later that only terms with ${\tilde \Delta} <0$ contribute. Therefore combining $L = 2 c N / a$ with ${\tilde \Delta} <0$ results in $L > 0$, which in turn implies $N=-1$. The latter implies $a | 2$.
Consequently ${\tilde \ell} = - b \, 2 m /a = k m, k \in \mathbb{Z}$. Therefore ${\tilde \ell} = - {\tilde \ell} \, \text{mod} \, 2m $ and hence
$ \psi(\Gamma)_{-\tilde{\ell}\ell} =  \psi(\Gamma)_{\tilde{\ell}\ell} $. Using the mapping \eqref{relmn}, 
this shows that the contribution from the sector $X=0$ equals the one
from the sector $Y=0$. Therefore, the combined contribution from the sectors $X Y \geq  0$ can be expressed as follows, 
\begin{equation}
\begin{gathered}
	(-1)^{\ell+1} \sum_{\substack{P\\ b \in\mathbb{Z}/a\mathbb{Z} }} 
 \int_{\Gamma_{\sigma}} d \sigma \; 
		{(\gamma\sigma+\delta)^{21/2}} 
	\sum_{\substack{M,N\geq -1\\  L\in\mathbb{Z} \\ 0 \leq \frac{b}{a}+\frac{\tilde{\ell}}{2m} < -\frac{1}{ac}  }} 
		L \, d(M) \, d(N) \\
	e^{2\pi i \left(\frac{\alpha}{\gamma}\frac{\tilde{\Delta}}{4m}+ \frac{\delta}{\gamma}\frac{\Delta}{4m}  \right)} \,  \psi(\Gamma)_{\tilde{\ell}\ell} \;
		\, e^{\left(-2\pi i \left[  
	\frac{\tilde{\Delta}}{4m}\frac{1}{\gamma}\frac{1}{\gamma\sigma+\delta}
	+\frac{{\Delta}}{4m}\frac{\gamma\sigma+\delta}{\gamma}
	\right] \right)} \;,
	\end{gathered} \label{symmetry-5}
\end{equation}
where the continued fraction condition \eqref{cond-positive-cont-frac} now takes the form given in \eqref{symmetry-5}.
To proceed, we interchange the integration with the summation over $M,N$. This is allowed by the following arguments.

First we note that the condition  $0 \leq \frac{b}{a}+\frac{\tilde{\ell}}{2m} < -\frac{1}{ac}$ in the summation above
can be written as 
\begin{equation} \label{cont-fract-transformed}
	-m < a^2 M -c^2N \leq m.
\end{equation}
Then, using the expression for $L$ given in \eqref{L-lt-n-tilde-definition}, we write out $\tilde{\Delta} = 4MN-L^2$ and obtain
\begin{equation}
	\tilde{\Delta} =
	\frac{1}{a^2c^2} \left[-(a^2M-c^2N)^2-\left( m^2-2m(a^2M+c^2N) \right) \right] \;.
	\label{tdelMN}
	\end{equation}
Now let us consider terms that satisfy $\tilde{\Delta}\leq 0$, in which case we obtain from \eqref{tdelMN},
\begin{equation}
	2m(a^2M+c^2N) \leq (a^2M-c^2N)^2+m^2 \;.
\end{equation}
Combining this with \eqref{cont-fract-transformed} we infer
\begin{equation}
	a^2M+c^2N \leq m.
\end{equation}
Then, by combining this last inequality with \eqref{cont-fract-transformed} we obtain the bounds 
\begin{equation}
	a^2 M  \leq m , \hspace{7mm} c^2 N < m \;.
\end{equation}
Therefore, for a given $a,c$ there is only a finite set of values $M,N$  which satisfy $\tilde{\Delta} \leq 0 $ as well as 
the continued fraction condition \eqref{cond-positive-cont-frac} and the condition $L \in \mathbb{Z}$. Thus, in this case, we can interchange 
 the integration with the summation over $M,N$. 
 
 Next, let us consider the terms with $\tilde{\Delta} >0 $. For large values of $M$, the Fourier coefficients of $1/\eta^{24}$
grow exponentially as 
\begin{equation}
	d(M) \sim e^{4 \pi \sqrt{M}}.
\end{equation}
{From} \eqref{tdelMN}, and using  \eqref{cont-fract-transformed},
 we infer that for large $M,N$, $\tilde{\Delta} $ behaves, schematically, as $\tilde{\Delta} \sim M + N$, and hence becomes large.
 Parametrizing the Ford circle $\Gamma_{\sigma}$ in \eqref{symmetry-5} by
 \begin{equation}
	\sigma(\theta) = -\frac{\delta}{\gamma}+\frac{i}{\gamma^2}\left(\frac{1+e^{i\theta}}{2} \right)
\end{equation}
or, equivalently, by
\begin{equation}
	\sigma_1(\theta) = -\frac{\delta}{\gamma}-\frac{1}{\gamma^2}\frac{\sin\theta}{2}, \hspace{4mm} 
	\sigma_2(\theta) = \frac{1}{\gamma^2}\frac{1+\cos\theta}{2},
\end{equation}
where $\theta\in[0,\pi)  \cup (\pi, 2 \pi)$, we infer that on the contour $\Gamma_{\sigma}$, 
\bea
\Big \vert 
 e^{-2\pi i \left[  
	\frac{\tilde{\Delta}}{4m}\frac{1}{\gamma}\frac{1}{\gamma\sigma+\delta}
	+\frac{{\Delta}}{4m}\frac{\gamma\sigma+\delta}{\gamma}
	\right] } \Big \vert \leq e^{- \frac{2 \pi}{4m} {\tilde \Delta} + \frac{2 \pi}{4m \gamma^2 } \Delta} .
	\eea	
Since $d(M) \, d(N) \,  e^{- \frac{2 \pi}{4m} {\tilde \Delta}} $ is exponentially suppressed for large $M,N$, 
the sum over $M,N$ in \eqref{symmetry-5}  is uniformly convergent on $\Gamma$ by the Weierstrass M test, and since each summand is integrable,
we conclude that interchanging the integration with the summation over $M,N$ is justified also when $\tilde{\Delta} >0 $.

Thus, interchanging the integration  with the summation over $M,N$  
results in 
\begin{equation}
\begin{gathered}
	(-1)^{\ell+1} \sum_{\substack{P\\ b \in\mathbb{Z}/a\mathbb{Z} }}  \;\;
	\sum_{\substack{M,N\geq -1 \\  -m < a^2M - c^2N \leq m \\  L\in\mathbb{Z}  }} 
		L \, d(M) \, d(N) \,
	e^{2\pi i \left(\frac{\alpha}{\gamma}\frac{\tilde{\Delta}}{4m}+ \frac{\delta}{\gamma}\frac{\Delta}{4m}  \right)} \,\psi(\Gamma)_{\tilde{\ell}\ell} \\
		 \int_{\Gamma_{\sigma}} d \sigma \; 
		{(\gamma\sigma+\delta)^{21/2}} \, e^{-2\pi i \left(
	\frac{\tilde{\Delta}}{4m}\frac{1}{\gamma}\frac{1}{\gamma\sigma+\delta}
	+\frac{{\Delta}}{4m}\frac{\gamma\sigma+\delta}{\gamma}
	\right)} \;.
	\end{gathered} \label{symmetry-6}
\end{equation}

\subsubsection{Bessel integral \label{sec:bess23/2}} 

We perform the $\sigma$-integration over
the Ford circle $\Gamma_{\sigma}$ that skirts the point  $- \delta/\gamma$. This Ford circle has radius $1/(2 \gamma^2)$, is
centered
at $\sigma =  - \frac{\delta}{\gamma} + i \frac{1}{2 \gamma^2}$,
\bea
\label{cont-sig}
\Gamma_{\sigma}: \vert \sigma + \frac{\delta}{\gamma} - i \frac{1}{2 \gamma^2} \vert = \frac{1}{4\gamma^4} \;,
\eea
and is oriented counter clockwise. Then, 
\eqref{symmetry-6} becomes replaced by  
\begin{equation}
\begin{gathered}
	(-1)^{\ell + 1} \sum_{\substack{P' \\ b \in\mathbb{Z}/a \mathbb{Z} }}  \;\; 
	\sum_{\substack{M,N\geq -1 \\ -m < a^2M - c^2N \leq m  \\  L\in\mathbb{Z} }} 
		L \, d(M) \, d(N) \,
	e^{2\pi i \left(\frac{\alpha}{\gamma}\frac{\tilde{\Delta}}{4m}+ \frac{\delta}{\gamma}\frac{\Delta}{4m}  \right)} \, \psi(\Gamma)_{\tilde{\ell}\ell}\\
		\int_{\Gamma_{\sigma}} d \sigma
	{(\gamma\sigma+\delta)^{21/2}} \, e^{-2\pi i \left(
	\frac{\tilde{\Delta}}{4m}\frac{1}{\gamma}\frac{1}{\gamma\sigma+\delta}
	+\frac{{\Delta}}{4m}\frac{\gamma\sigma+\delta}{\gamma}
	\right)} \;,
		\end{gathered} \label{symmetry-ford}
\end{equation}
where $P'$ denotes the set $P$, but with $\delta$ restricted to lie in the range $0 \leq -\delta < \gamma$.
We change the integration variable to
\bea
{\tilde \sigma} = \gamma \left( \gamma \sigma + \delta \right) \;.
\label{tilsig}
\eea
The essential singularity is now located at the origin ${\tilde \sigma} = 0$.
We choose the branch cut, which originates at ${\tilde \sigma} = 0$, to lie along the negative imaginary axis of the ${\hat \sigma}$-plane.
 Next, we change the integration variable once more,
\bea
w = \frac{\gamma}{\Delta} \frac{i}{\tilde \sigma} \;.
\label{wts}
\eea
Now the branch cut originates at $w = 0$ and lies along the negative real axis of the $w$-plane. The integration contour now runs
along along a line parallel to the imaginary axis,
\bea
 - \frac{i^{23/2}}{\gamma \, \Delta^{23/2}}
  \int_{{\tilde \epsilon}-i\infty}^{{\tilde \epsilon}+i \infty} 
    \frac{d w}{w^{25/2}} \; 
	e^{2\pi  \left( -
	\frac{\tilde{\Delta} \Delta }{4m \gamma} w 
	+\frac{{1}}{4m \gamma \; w}
	\right)} 
	\label{intfo}
	\eea
with ${\tilde \epsilon} >0$. Now recall that $\Delta >0$. When ${\tilde \Delta } \geq 0$,
the coefficient ${\tilde{\Delta} \Delta } / (4m \gamma)$ in the exponent is $\geq 0$, and hence
the integration contour can be closed in the half plane ${\rm Re} \, w > 0$, where the integrand is analytic
and hence the integral vanishes. Thus, we now take ${\tilde \Delta} < 0$.

Then, performing the redefinition
\bea
t = \frac{\pi |{\tilde \Delta}| \Delta}{2 m \gamma} \, w \;
\eea
and defining $z = \frac{\pi}{m\gamma}\sqrt{|\tilde\Delta|\Delta}$,
we obtain for the integral \eqref{intfo}, with $\epsilon > 0$,
\bea
 - \frac{i^{23/2}}{\gamma } \left( \frac{ \pi {|\tilde \Delta}|  }{ 2 m \gamma } \right)^{23/2}
  \ \int_{ \epsilon-i\infty}^{\epsilon+i \infty}   \frac{d t }{t^{25/2}} \;
	e^{ t
	+  \frac{z^2}{4 t}
	} = - 2 \pi \,  \frac{i^{1/2}}{\gamma} \left( \frac{|{\tilde \Delta} |}{\Delta} \right)^{23/4} \, I_{23/2}  \left(\frac{\pi }{m \gamma} \, \sqrt{ {|\tilde \Delta}| \Delta } 
	\right) \;, \nonumber\\
\eea
where $I_\nu(z)$ denotes the modified Bessel function of first kind of index $\nu$,
\begin{equation}
	I_\nu(z) = \frac{(\frac{1}{2}z)^\nu}{2\pi i }\int_{\epsilon-i\infty}^{\epsilon+i \infty} d t \,\,t^{-\nu-1} e^{t + (z^2/4 t)} ,
	\label{bess-int}
\end{equation}
where $\epsilon > 0$. Then, \eqref{symmetry-6} becomes
\begin{equation}
\begin{gathered}
	(-1)^{\ell} \,  i^{1/2}  \, 2 \pi \sum_{\substack{P' \\ b \in\mathbb{Z}/a \mathbb{Z} }}  \;\;\; 
	\sum_{\substack{M,N\geq -1 \\ -m < a^2M - c^2N \leq m  \\  L\in\mathbb{Z}, \;  {\tilde \Delta} < 0 }} 
		\psi(\Gamma)_{\tilde{\ell}\ell} \;
	L \, d(M) \, d(N) \,
	 \frac{
	e^{2\pi i \left(\frac{\alpha}{\gamma}\frac{\tilde{\Delta}}{4m}+ \frac{\delta}{\gamma}\frac{\Delta}{4m}  \right)}}{\gamma}  
	\left( \frac{|{\tilde \Delta} |}{\Delta} \right)^{23/4} \,\\ I_{23/2}  \left(\frac{\pi }{m \gamma} \, \sqrt{ {|\tilde \Delta}| \Delta } \right) . \\
		\end{gathered} \label{symmetry-ext2}
\end{equation}

Note that in \eqref{symmetry-ext2} the dependence on 
$\alpha$ and $\delta$ is entirely encoded in 
multiplier system $\psi(\Gamma)_{\tilde{\ell}\ell}$ and in the phase 
\begin{equation}
		e^{2\pi i \left(\frac{\alpha}{\gamma}\frac{\tilde{\Delta}}{4m}+ \frac{\delta}{\gamma}\frac{\Delta}{4m}  \right)}.
\end{equation}
Since $0\leq - \delta < \gamma$ and $\alpha\in\mathbb{Z}/\gamma\mathbb{Z}$, and since $\alpha$ is the modular inverse of $\delta$, 
i.e. $\alpha\delta = 1 \text{ mod } \gamma$,
each $\delta$ uniquely specifies one $\alpha$. Thus, the sum over $\delta$ yields the generalized Kloosterman sum
${\rm Kl}( \frac{\Delta}{4m}, \frac{\tilde{\Delta}}{4m} ;\gamma,\psi)_{\ell\tilde{\ell}}$,
\begin{equation}
		{\rm Kl} ( \frac{\Delta}{4m}, \frac{\tilde{\Delta}}{4m} ;\gamma,\psi)_{\ell\tilde{\ell}} = \sum_{\substack{0\leq -\delta <\gamma\\ (\delta,\gamma)=1, \alpha\delta =  1 \text{ mod } \gamma}}e^{2\pi i \left( \frac{\alpha}{\gamma}\frac{\tilde{\Delta}}{4m} +\frac{\delta}{\gamma}\frac{\Delta}{4m}\right)}\psi(\Gamma)_{\tilde{\ell}\ell} .
	\label{kloo}
\end{equation}
Thus, \eqref{symmetry-ext2} can be written as 
\begin{equation}
\begin{gathered}
		(-1)^{\ell} i^{1/2} \, 2 \pi
		\sum_{\gamma =1}^{\infty} \; 
		\sum_{\substack{S_G \\ b \in \mathbb{Z}/a \mathbb{Z}  }} \;\;
		\sum_{\substack{M, N \geq -1 \\ -m < a^2M - c^2N \leq  m \\L \in \mathbb{Z} , \, \tilde{\Delta}<0 }} 
		L \, d(M)d(N) \; \frac{{\rm Kl}(\frac{\Delta}{4m}
	\frac{\tilde{\Delta}}{4m},\gamma,\psi)_{\ell\tilde{\ell}}}{\gamma} \\
	  \left(\frac{ \vert \tilde{\Delta} \vert}{\Delta} \right)^{23/4} I_{23/2}\left(\frac{\pi}{\gamma m}\sqrt{\Delta\vert\tilde{\Delta}\vert} \right),
	  \label{lsum1}
\end{gathered} 
\end{equation}
where $\tilde{\Delta} = 4 M N - L^2$, with $L$ given in \eqref{L-lt-n-tilde-definition}. Note that the sum over the allowed $M,N$ is finite.

Next, using \eqref{setgam}, we express the triplet $(N, L, M)$ in terms of the triplet $({\tilde n}, m, {\tilde \ell})$. We then trade the 
sum over $M,N$ for a sum over ${\tilde n}, {\tilde \ell}$. In section \ref{sec:tilden} we will show that ${\tilde n}$ is bounded by ${\tilde n} \geq -1$.
Writing ${\tilde \Delta}$ as ${\tilde \Delta} = 4 m {\tilde n} - \ell^2$, we rewrite \eqref{lsum1} as 
\begin{equation}
\begin{gathered}
		(-1)^{\ell} i^{1/2} \, 2 \pi
		\sum_{\gamma =1}^{\infty} \; \sum_{\substack{ {\tilde n} \geq -1 \\ {\tilde \ell}  \in \mathbb{Z}/2m \mathbb{Z}    \\ {\tilde \Delta} < 0  }}\;\;
		\sum_{\substack{a>0, c<0    \\ b \in \mathbb{Z}/ a \mathbb{Z}, \; 
		a d - b c = 1 \\ 0 \leq \frac{b}{a} + \frac{\tilde \ell}{2m} < - \frac{1}{ac  }  }} \\
		\left( (ad + bc ) {\tilde \ell} + 2  ac {\tilde n} + 2 bd m  \right) d( c^2 {\tilde n} + d^2 m + cd {\tilde \ell}) \, d(a^2 {\tilde n} + b^2 m + ab {\tilde \ell}) \\
\frac{{\rm Kl}(\frac{\Delta}{4m}
	\frac{\tilde{\Delta}}{4m},\gamma,\psi)_{\ell\tilde{\ell}}}{\gamma} 
	  \left(\frac{ \vert \tilde{\Delta} \vert}{\Delta} \right)^{23/4} I_{23/2}\left(\frac{\pi}{\gamma m}\sqrt{\Delta\vert\tilde{\Delta}\vert} \right),
	  \label{lsum}
\end{gathered} 
\end{equation}
which we write as
\begin{equation}
\label{fin-23/2}
\begin{gathered}
		(-1)^{\ell} i^{1/2} \, 2 \pi
		\sum_{\gamma =1}^{\infty} \; 
		\sum_{\substack{  {\tilde n} \geq -1 \\ {\tilde \ell}  \in \mathbb{Z}/2m \mathbb{Z} \ \\ {\tilde \Delta} < 0  }} \;\;
		c_m^F(\tilde{n},\tilde{\ell})  \, \frac{{\rm Kl}(\frac{\Delta}{4m} \, 
	\frac{\tilde{\Delta}}{4m},\gamma,\psi)_{\ell\tilde{\ell}}}{\gamma} 
	  \left(\frac{ \vert \tilde{\Delta} \vert}{\Delta} \right)^{23/4} I_{23/2}\left(\frac{\pi}{\gamma m}\sqrt{\Delta\vert\tilde{\Delta}\vert} \right),
\end{gathered} 
\end{equation}
where $c_m^F(\tilde{n},\tilde{\ell}) $
is defined by 
\bea
\label{ccf}
	c_m^F(\tilde{n},\tilde{\ell}) =
			\sum_{\substack{a>0, c<0    \\ b \in \mathbb{Z}/ a \mathbb{Z}, \; 
		a d - b c = 1 \\  0 \leq \frac{b}{a} + \frac{\tilde \ell}{2m} < - \frac{1}{ac }  }} 
	\left( (ad + bc ) {\tilde \ell} + 2  ac {\tilde n} + 2 bd m  \right) 
d( c^2 {\tilde n} + d^2 m + cd {\tilde \ell}) \, d(a^2 {\tilde n} + b^2 m + ab {\tilde \ell}) .\nonumber\\
\eea
Note that the above sum includes two subsets of matrices in $S_G$. The first subset contains matrices satisfying ${\tilde \ell}/2m=-b/a$, while the second subset contains matrices  that correspond to the continued fraction expansion of ${\tilde \ell}/2m$. The latter subset is finite by definition, while  the bounds $M,N \geq -1$ and \eqref{cont-fract-transformed}  can be used to show the finiteness of the first subset.
This is consistent with the proofs of finiteness of \cite{Chowdhury:2019mnb,LopesCardoso:2020pmp}.

\subsection{Lower bound on $\tilde{n}$ \label{sec:tilden}}

We next show that $\tilde{n} \geq -1$ whenever the condition 
\bea
0 \leq \frac{b}{a} + \frac{\tilde \ell}{2m} < - \frac{1}{ac}
\label{ccond}
\eea is true.
To this end, we first recall that \eqref{ccond} can be written as
\bea
-m <  a^2 M - c^2 N \leq m .
\label{inmm}
\eea
and further $M, N \geq -1$. 
Then beginning with 
\bea
{\tilde n} = \frac{bd}{ac} m - \frac{b}{c} M + \frac{d}{a} N,
\label{exptn}
\eea
we study three cases:
\begin{enumerate}

\item $d=0$:  Then, \eqref{exptn} becomes
\bea
{\tilde n} = - \frac{b}{c} M = \frac{M}{c^2},
\eea
which satisfies ${\tilde n} \geq -1$, since $M \geq -1$ and $c^2 \geq 1$.

\item  $bc<0$:
In this case, we obtain, 
\bea
(ac)^2 {\tilde n} &=& a b c d m - b c a^2 M + d a c^2 N \nonumber\\
&=&  a b c d m + c^2 N - b c (a^2 M - c^2 N ) \nonumber\\
&=& a b c d m + c^2 N + |  b c |  (a^2 M - c^2 N ).
\eea
Using the lower bound in \eqref{inmm}, we get,
\bea
(ac)^2 {\tilde n} > a b c d m + c^2 N - |  b c |  m &=& c^2 N + m b c \left( a d + 1 \right)  \nonumber\\
&=& c^2 N + m  \left( (a d)^2 - 1 \right) \geq c^2  N,
\eea
Since $(ad)^2 \geq 1$,
\bea 
\tilde{n} > \frac{N}{a^2} \geq -1 .
\eea
\item $bc \geq 0$: 
Here we have, using the upper bound in \eqref{inmm}, 
\bea
(ac)^2 {\tilde n} &=& a b c d m - b c a^2 M + d a c^2 N \nonumber\\
&=&  a b c d m + c^2 N - b c (a^2 M - c^2 N ) \nonumber\\
&\geq & a b c d m + c^2 N - b c m \nonumber\\
&=& c^2 N + b c m (a d -1) = c^2 N + (b c)^2  m \geq c^2 N .
\eea
Hence, 
\bea
\tilde{n} \geq \frac{N}{a^2} \geq -1 .
\eea 
\end{enumerate}

\subsection{Bessel function $I_{12}$ \label{sec:be12}}

Now we focus on the terms $E_{m,\gamma (\gamma \sigma + \delta)}$
in the decompositions
\eqref{both-positive-1}, \eqref{one-zero-1} and  \eqref{both-negative-1}. In section \ref{sec:canc},
we will show that these terms only give a non-vanishing contribution when $X Y \neq 0$, whereas when $X Y =0$ they cancel out.
Thus, we will assume $X Y \neq 0$ in the following. 

We proceed as in section \ref{sec:bess23/2}, namely, 
we restrict the sum over $\delta$ to the range $0 \leq -\delta < \gamma$ and perform the integration over
the Ford circle $\Gamma_{\sigma}$ in \eqref{cont-sig}.
Collecting the terms proportional $E_{m,\gamma (\gamma \sigma + \delta)}$ in \eqref{after-v-integration2}, and changing 
the integration variable to ${\tilde \sigma}$ given 
in \eqref{tilsig}, we obtain
\begin{equation} \label{after-kloosterman-12}
\begin{gathered}
	(-1)^{\ell+1}\sum_{P_{0}} \, {\rm Kl} ( \frac{\Delta}{4m}, \frac{\tilde{\Delta}}{4m} ;\gamma,\psi)_{\ell\tilde{\ell}}
	\; L  \, d(M) \, d(N) \; 
	\int_{\tilde \Gamma} d\tilde{\sigma} \;\frac{{\tilde{\sigma}^{21/2}}}{\gamma^{25/2}}\,
	e^{\left(-2\pi i \left[  
	\frac{\tilde{\Delta}}{4m}\frac{1}{\tilde{\sigma}}
	+\frac{{\Delta}}{4m}\frac{\tilde{\sigma}}{\gamma^2}
	\right] \right)}
	\\
	\frac{1}{2}\left(E_{m,\tilde{\sigma}}\left( \frac{b}{a}+\frac{\tilde{\ell}}{2m}\right)+E_{m,\tilde{\sigma}}\left(-\frac{b}{a}-\frac{1}{ac}-\frac{\tilde{\ell}}{2m}\right) \right),
	\end{gathered}
\end{equation}
where $L$ and ${\tilde \ell}$ are given in \eqref{L-lt-n-tilde-definition}, and where 
the generalized Kloosterman sum ${\rm Kl} ( \frac{\Delta}{4m}, \frac{\tilde{\Delta}}{4m} ;\gamma,\psi)_{\ell\tilde{\ell}}$ is given in \eqref{kloo}.
The set $P_0$ is given by 
\bea
\label{setP0} 
	{P}_0 = \left\{ \begin{pmatrix}
		a & b \\
		c&d
	\end{pmatrix} \in \mathrm{SL}(2,\mathbb{Z}), M, N, \gamma \in \mathbb{Z} \; \vert  
	 a,\gamma>0, c<0,  M  \geq -1, N \geq -1, L \in \mathbb{Z}, X Y \neq 0 \right\}.
	 \nonumber\\
	 \eea
The integration contour ${\tilde \Gamma}$ denotes a Ford circle centered at ${\tilde \sigma} = i/2$ that skirts the origin ${\tilde \sigma} =0$.

Using the symmetry property \eqref{symmetry-3}, we write \eqref{after-kloosterman-12} as 
\begin{equation} \label{after-kloosterman-123}
\begin{gathered}
	(-1)^{\ell+1}\sum_{P_{0}} \, \left[ {\rm Kl} ( \frac{\Delta}{4m}, \frac{\tilde{\Delta}}{4m} ;\gamma,\psi)_{\ell\tilde{\ell}}
	+ {\rm Kl} ( \frac{\Delta}{4m}, \frac{\tilde{\Delta}}{4m} ;\gamma,\psi)_{\ell ( - \tilde{\ell}) } \right]
	\; L  \, d(M) \, d(N) \; \\
	\int_{\tilde \Gamma} d\tilde{\sigma} \;\frac{{\tilde{\sigma}^{21/2}}}{\gamma^{25/2}}\,
	e^{\left(-2\pi i \left[  
	\frac{\tilde{\Delta}}{4m}\frac{1}{\tilde{\sigma}}
	+\frac{{\Delta}}{4m}\frac{\tilde{\sigma}}{\gamma^2}
	\right] \right)} \;
	\frac{1}{2}
		E_{m,\tilde{\sigma}}\left( \frac{b}{a}+\frac{\tilde{\ell}}{2m}\right) \;.
		\end{gathered}
\end{equation}
Using the expression for $E_{m,\tilde{\sigma}}$ given in \eqref{e-function} and the relation
\begin{equation}
\label{relDN}
	\frac{\tilde{\Delta}}{4m}+m\left(\frac{b}{a}+\frac{\tilde{\ell}}{2m}\right)^2 = \frac{N}{a^2} \,,
\end{equation}
we obtain
\begin{equation}
\begin{gathered}
	(-1)^{\ell}\sum_{P_{0}} \, \left[ {\rm Kl} ( \frac{\Delta}{4m}, \frac{\tilde{\Delta}}{4m} ;\gamma,\psi)_{\ell\tilde{\ell}}
	+ {\rm Kl} ( \frac{\Delta}{4m}, \frac{\tilde{\Delta}}{4m} ;\gamma,\psi)_{\ell ( - \tilde{\ell}) } \right]
	 \,  \frac{L}{\left( \frac{b}{a}+\frac{\tilde{\ell}}{2m} \right)} \, d(M) \, d(N) \\
		\int_{\tilde \Gamma} d\tilde{\sigma} \;\frac{{\tilde{\sigma}^{11}}}{\gamma^{25/2}}\frac{1}{\sqrt{8\pi^2i m}}\,
	e^{-2\pi i   \left( \frac{{\Delta}}{4m} \frac{\tilde{\sigma}}{\gamma^2} + \frac{N}{a^2} \frac{1}{\tilde{\sigma}} \right) } .
		\end{gathered}
	\label{i12nm}
\end{equation}
Note that the integrand does not exhibit a branch cut. Performing the variable change given in \eqref{wts}, 
the integral over the Ford circle $\tilde \Gamma$ takes a form similar to \eqref{intfo}, with ${\tilde \Delta}/4m$ replaced by $N/a^2$.
The integral will be non-vanishing provided  $N/a^2 <0$ .
This in turn implies 
$N=-1$.
As shown in section \ref{sec:canc}, the only contributions to the sum come from the terms in set the $P_0$
satisfying $a=1$ and $M=m$. Using the expression for $L$ and ${\tilde \ell}$ given in \eqref{L-lt-n-tilde-definition}, we infer that in this case, 
\bea
\frac{b}{a}+\frac{\tilde{\ell}}{2m} = \frac{c}{2m} \;\;\;,\;\;\; L = -c>0, 
\label{btlLc}
\eea
in which case
\bea
	\frac{L}{\frac{b}{a}+\frac{\tilde{\ell}}{2m}}  = -2m ,
\eea
so that \eqref{i12nm} yields (using $d(-1) = 1$)
\begin{equation}
\begin{gathered}
	(-1)^{\ell+1}\sum_{P_{0}'}  \left[ {\rm Kl} ( \frac{\Delta}{4m}, \frac{\tilde{\Delta}}{4m} ;\gamma,\psi)_{\ell\tilde{\ell}}
	+ {\rm Kl} ( \frac{\Delta}{4m}, \frac{\tilde{\Delta}}{4m} ;\gamma,\psi)_{\ell ( - \tilde{\ell}) } \right]
	 \; d(m) \\
	\int_{\tilde \Gamma}  d\tilde{\sigma} \;\frac{{\tilde{\sigma}^{11}}}{\gamma^{25/2}}\sqrt{\frac{m}{2\pi^2i}}\,
	e^{-2\pi i \left( \frac{{\Delta}}{4m}\frac{\tilde{\sigma}}{\gamma^2} - \frac{1}{\tilde{\sigma}} \right) } ,
	\end{gathered}
	\label{klml}
\end{equation}
where the set $P_{0}'$ is the set $P_0$ subject to the restrictions
$a=1,N=-1,M=m$. 

{From} \eqref{btlLc} we infer
\begin{equation}
	\tilde{\Delta} = 
		 4MN-L^2 = -4m-c^2 \;,
\end{equation}
in which case
\begin{equation}
	\frac{\tilde{\Delta}}{4m} = -1-\frac{c^2}{4m}.
	\label{tdcm}
\end{equation}
Using \eqref{btlLc}, we can write the Kloosterman sums in \eqref{klml} as (recall that $b \in \mathbb{Z}/a\mathbb{Z} $ with $a=1$ fixes $b$
to a single value) 
\begin{equation}
	(-1)^{\ell+1} 
		\sum_{\substack{c<0}} \;\; 
		\sum_{\substack{0\leq -\delta <\gamma\\ (\delta,\gamma)=1, \alpha\delta =  1 \text{ mod } \gamma}}e^{2\pi i \left( \frac{\alpha}{\gamma}\frac{\tilde{\Delta}}{4m} +\frac{\delta}{\gamma}\frac{\Delta}{4m}\right)} \left( \psi(\Gamma)_{(c - 2m b) {\ell}} + \psi(\Gamma)_{(- c + 2mb)  {\ell}} \right) \;.
	\label{klppdiv}
\end{equation}
Using 
\begin{equation}
\label{cj}
	\sum_{\substack{c<0}} f(c ) = \sum_{j \in \mathbb{Z}/2m\mathbb{Z}} \;\;  \sum_{\substack{ c < 0 \\ c =  j \;  \text{mod} \; 2m }}  \, f(c) \;.
	\end{equation}
and using the multiplier system property $\psi(\Gamma)_{(j + 2 m k) {\ell}} = \psi(\Gamma)_{j {\ell}}$ (with $k \in \mathbb{Z}$) \cite{kloosterman}, we
write \eqref{klppdiv} as 
\begin{equation}
	(-1)^{\ell+1} \sum_{j \in\mathbb{Z}/2m\mathbb{Z}}\sum_{\substack{c<0\\c =  j \text{ mod } 2m}}\sum_{\substack{0\leq -\delta <\gamma\\ (\delta,\gamma)=1, \alpha\delta =  1 \text{ mod } \gamma}}e^{2\pi i \left( \frac{\alpha}{\gamma}\frac{\tilde{\Delta}}{4m} +\frac{\delta}{\gamma}\frac{\Delta}{4m}\right)} \left( \psi(\Gamma)_{j {\ell}}+ \psi(\Gamma)_{(-j) {\ell}} \right) \;.
	\label{klppdiv2}
\end{equation}
Using \eqref{tdcm}, this becomes
\begin{equation}
	(-1)^{\ell+1} \sum_{j \in\mathbb{Z}/2m\mathbb{Z}}\sum_{\substack{c<0\\c =  j \text{ mod } 2m}}\sum_{\substack{0\leq -\delta <\gamma\\ (\delta,\gamma)=1, \alpha\delta =  1 \text{ mod } \gamma}} e^{-2\pi i  \frac{\alpha}{\gamma}\frac{c^2}{4m}} \, e^{2\pi i \left( - \frac{\alpha}{\gamma} 
	+\frac{\delta}{\gamma}\frac{\Delta}{4m}\right)} \left( \psi(\Gamma)_{j {\ell}}+ \psi(\Gamma)_{(-j) {\ell}} \right) \;.
	\label{klppdiv3}
\end{equation}

Next we focus on the sum over $c$ in this expression, 
\begin{equation}
		\sum_{j \in\mathbb{Z}/2m\mathbb{Z}}\sum_{\substack{c<0\\c =  j \text{ mod } 2m}} e^{-2\pi i  \frac{\alpha}{\gamma}\frac{c^2}{4m}}
		\left( \psi(\Gamma)_{j {\ell}} + \psi(\Gamma)_{(-j) {\ell}}\right) \;,
	\end{equation}
which we write as 
\begin{equation}
\frac12	
\sum_{j \in\mathbb{Z}/2m\mathbb{Z}}\sum_{\substack{c \in \mathbb{Z} \backslash\{0\} \\c =  j \text{ mod } 2m}} e^{-2\pi i  \frac{\alpha}{\gamma}\frac{c^2}{4m}}
		\left( \psi(\Gamma)_{j {\ell}} + \psi(\Gamma)_{(-j) {\ell}} \right) \;,
		\label{sumcdiv}
	\end{equation}
where we made use of the symmetry $j \to - j$ to obtain an expression with the symmetry $c \to -c$.

The expression \eqref{sumcdiv} is divergent. Below we will discuss a regularization procedure to extract a finite part of this expression.

\subsubsection{Regularization}

We now regard the sum over $c \in \mathbb{Z} \backslash\{0\} $ in \eqref{sumcdiv},
\bea
\sum_{\substack{c \in \mathbb{Z} \backslash\{0\} \\c =  j \text{ mod } 2m}} e^{-2\pi i  \frac{\alpha}{\gamma}\frac{c^2}{4m}},
\label{formexpsumc}
\eea
as
\begin{equation}
		 \lim_{\tau \to -\alpha/\gamma} \left( \vartheta_{m,j}(\tau)-\delta_{0,j} \right),
		 \label{thetmone}
\end{equation}
where $  \vartheta_{m,j}(\tau) $ denotes the standard weight $1/2$ index $m$ Jacobi theta function. Using this, we then regard \eqref{sumcdiv} as
\begin{equation}
\frac12	
\lim_{\tau \to -\alpha/\gamma }
 \sum_{j \in\mathbb{Z}/2m\mathbb{Z}}
  \left( \vartheta_{m,j}(\tau)-\delta_{0,j} \right)
		\left( \psi(\Gamma)_{j {\ell}} + \psi(\Gamma)_{(-j) {\ell}} \right) \;.
		\label{sumcdivv2}
	\end{equation}

We focus on the combination 
\begin{equation}
	 \sum_{j \in\mathbb{Z}/2m\mathbb{Z}}
	\left( \vartheta_{m,j}(\tau)-\delta_{0,j} \right) \psi(\Gamma)_{j {\ell}},
	\label{combthepsi}
	\end{equation}
which is one of the combinations contained in \eqref{sumcdivv2}. Using that  $\psi(\Gamma)_{j {\ell}}$ are the components of a unitary matrix, i.e.
\bea
\psi (\Gamma) = \overline{\left( \psi (\Gamma^{-1}) \right)^T}, 
\eea
we infer the property
\bea
\psi(\Gamma)_{j {\ell}}  = \psi({\tilde \Gamma})_{{\ell} j} , 
\label{GtG}
\eea
where
\bea
\Gamma = \begin{pmatrix}
	\alpha & \beta \\
	\gamma & \delta
\end{pmatrix} \quad , \quad 
{\tilde \Gamma} = \begin{pmatrix}
		\delta& \beta\\
		\gamma & \alpha
	\end{pmatrix} \, .
\eea
Now we recall 
the transformation
property under ${\rm SL}(2,\mathbb{Z})$ transformations, 
\begin{equation}
	\vartheta_{m, \ell }\left( \frac{\alpha\tau+\beta}{\gamma\tau+\delta}, \frac{v}{\gamma\tau+\delta} \right) = (\gamma\tau+\delta)^{1/2}e^{2\pi i m \frac{\gamma v^2}{\gamma\tau+\delta}} \sum_{j\in\mathbb{Z}/2m\mathbb{Z}} \psi(\Gamma)_{\ell j}\;\vartheta_{m,j}(\tau,v).
\end{equation}
Setting $v=0$ and choosing the  ${\rm SL}(2,\mathbb{Z})$ transformation ${\tilde \Gamma}$, this becomes
 \begin{equation}
	\vartheta_{m,\ell}\left( \frac{\delta \tau+\beta}{\gamma\tau+\alpha} \right) = (\gamma\tau+\alpha)^{1/2} \sum_{j\in\mathbb{Z}/2m\mathbb{Z}} \psi 
	({\tilde \Gamma})_{\ell j}\;\vartheta_{m,j}(\tau).
	\label{trans-tGam}
\end{equation}
Then, by combining \eqref{combthepsi} with \eqref{GtG} and \eqref{trans-tGam}, we obtain
\bea
	 \sum_{j \in\mathbb{Z}/2m\mathbb{Z}}
	\left( \vartheta_{m,j}(\tau)-\delta_{0,j} \right) \psi(\Gamma)_{j {\ell}} = 
	(\gamma\tau+\alpha)^{-1/2} \, \vartheta_{m,\ell}\left( \frac{\delta \tau+\beta}{\gamma\tau+\alpha} \right) - \psi(\Gamma)_{0 {\ell}} .
	\label{relthetheGtG}
\eea
Now we study this equation in the limit $\tau \to -\alpha/\gamma$. Using 
	\begin{equation}
	\lim_{\tau \to +i\infty}
	\vartheta_{m,\ell}(\tau) = \begin{cases}
		0 \;\; \text{for }\ell \neq 0 \mod 2m, \\
		1  \;\; \text{for }\ell = 0 \mod 2m, \\
	\end{cases}
\end{equation}
we infer that in the limit $\tau \to -\alpha/\gamma$, 
the right hand side of \eqref{relthetheGtG} tends to $- \psi(\Gamma)_{0 {\ell}}$ when $\ell \neq 0$, whereas when $\ell = 0$ it diverges and behaves as
$(\gamma\tau+\alpha)^{-1/2} - \psi(\Gamma)_{0 0}$. 
We note that the divergent term $(\gamma\tau+\alpha)^{-1/2}$ that arises when $\ell =0$, is due to the presence of a constant term in $\vartheta_{m,0}(\tau)$.
Then, by subtracting this constant term
we obtain the following regularized expression,
\bea
\left( \lim_{\tau \to -\alpha/\gamma} \sum_{j \in\mathbb{Z}/2m\mathbb{Z}}
	\left( \vartheta_{m,j}(\tau)-\delta_{0,j} \right) \psi(\Gamma)_{j {\ell}}  \right) \vert_{\text{regularized}} = - \psi(\Gamma)_{0 {\ell}} .
\eea
A similar reasoning applies to the other combination, proportional to $\psi(\Gamma)_{(-j) {\ell}}$, contained in \eqref{sumcdivv2}.

Thus, we are led to the following regularized expression for \eqref{sumcdivv2},
\begin{equation}
\frac12	
  \lim_{\tau \to -\alpha/\gamma }
 \sum_{j \in\mathbb{Z}/2m\mathbb{Z}}
  \left( \vartheta_{m,j}(\tau)-\delta_{0,j} \right)
		\left( \psi(\Gamma)_{j {\ell}} + \psi(\Gamma)_{(-j) {\ell}} \right) = 
			- \psi(\Gamma)_{0 {\ell}} .
			\end{equation}
Using this in \eqref{klppdiv3}, we arrive at our proposal for the regularized expression for \eqref{klppdiv3}, 
\begin{equation}
	(-1)^{\ell} \sum_{\substack{0\leq -\delta <\gamma\\ (\delta,\gamma)=1, \alpha\delta =  1 \text{ mod } \gamma}}e^{2\pi i \left( - \frac{\alpha}{\gamma}
	 +\frac{\delta}{\gamma}\frac{\Delta}{4m}\right)} \psi(\Gamma)_{0 {\ell}} = (-1)^{\ell} \, {\rm Kl} ( \frac{\Delta}{4m}, -1;\gamma,\psi)_{\ell 0 } \;.
	\label{bessel-12-delta-2}
		\end{equation}

To summarize, 
the regularization procedure described above removes one divergent contribution that arises when $\ell =0$. 
At present	
we do not have neither a physics nor a mathematics justification for using precisely this regulator.

\subsubsection{Bessel integral}
Now we return to \eqref{klml}, which we regularize using the expression \eqref{bessel-12-delta-2},
\begin{equation}
\begin{gathered}
	(-1)^{\ell}  \; d(m) \, \sqrt{\frac{m}{2\pi^2i}} \,
	\sum_{\gamma=1}^{\infty} {\rm Kl} ( \frac{\Delta}{4m}, -1;\gamma,\psi)_{\ell 0 } \,
		\int_{\tilde \Gamma}  d\tilde{\sigma} \;\frac{{\tilde{\sigma}^{11}}}{\gamma^{25/2}} \,
	e^{-2\pi i \left( \frac{{\Delta}}{4m}\frac{\tilde{\sigma}}{\gamma^2} - \frac{1}{\tilde{\sigma}} \right) } .
	\end{gathered}
	\end{equation}
Changing the integration variable to 
\begin{equation}
  t = 2\pi \frac{i}{\tilde{\sigma}}, 
\end{equation}
we get
\begin{equation}
\begin{gathered}
	(-1)^{\ell + 1 }  \; d(m) \, \sqrt{\frac{m}{2\pi^2i}} \,  \frac{(2\pi)^{12}}{\gamma^{25/2}} \, 
	\sum_{\gamma=1}^{\infty} {\rm Kl} ( \frac{\Delta}{4m}, -1;\gamma,\psi)_{\ell 0 } \,
		\int_{\epsilon-i\infty}^{\epsilon+i \infty}
	d t  \frac{1}{t^{13}} \,  e^{\frac{1}{4t}\left( \frac{2\pi \sqrt{\Delta}}{\gamma\sqrt{m}} \right)^2+t} \;.
	 	\end{gathered}
	\end{equation}
Using \eqref{bess-int} we obtain,
\begin{equation} 
\label{fin-12}
	(-1)^{\ell+1}\sqrt{2m}\,i^{1/2}\, d(m) 
	\sum_{\gamma>0} \frac{{\rm Kl} ( \frac{\Delta}{4m}, -1 ;\gamma,\psi)_{\ell 0}}{\sqrt{\gamma}}\left(\frac{4m}{\Delta} \right)^6 I_{12}\left(\frac{2\pi}{\gamma}\sqrt{\frac{\Delta}{m}} \right) \;.
\end{equation}

\subsection{Isolating non-vanishing contributions \label{sec:canc}}
We return to \eqref{symmetry-3}, show that the sum exhibits cancellations between various terms, thereby identifying non-vanishing contributions.
Taking into account the form of  
$ E_{m,\gamma(\gamma\sigma+\delta)} \left(X\right)$ and $I_{m,\gamma(\gamma\sigma+\delta)}\left(X \right)$ given in \eqref{e-function} and \eqref{i-function},
we write \eqref{symmetry-3}  as
\begin{equation}
\begin{gathered}
	(-1)^{\ell+1} \sum_{\substack{P\\ b \in\mathbb{Z}/a \mathbb{Z} }} \;
		{(\gamma\sigma+\delta)^{10}} 
	\sum_{\substack{M,N\geq -1\\  L\in\mathbb{Z}, \; X \neq 0}} 
		\frac{L}{X} \, d(M) \, d(N) \\
	e^{2\pi i \left(\frac{\alpha}{\gamma}\frac{\tilde{\Delta}}{4m}+ \frac{\delta}{\gamma}\frac{\Delta}{4m}  \right)}\left(\psi(\Gamma)_{\tilde{\ell}\ell}+\psi(\Gamma)_{-\tilde{\ell}\ell}\right)
	e^{\left(-2\pi i \left[  
	\frac{\tilde{\Delta}}{4m}\frac{1}{\gamma}\frac{1}{\gamma\sigma+\delta}
	+\frac{{\Delta}}{4m}\frac{\gamma\sigma+\delta}{\gamma}
	\right] \right)}
	F_{{m,\gamma(\gamma\sigma+\delta)}}(X^2) \;,
	\end{gathered} \label{cancellations-2}
\end{equation}
where $F_{m,\gamma(\gamma\sigma+\delta)}$ is a function that only depends on $X^2$. 
Consider changing the variable $\sigma$ to ${\tilde \sigma} = \gamma (\gamma \sigma + \delta)$, and integrating along
a Ford circle ${\tilde \Gamma}$, centered at ${\tilde \sigma} = i/2$,  that skirts the origin ${\tilde \sigma} =0$. Then, as described
in section \ref{sec:be12}, only when $N=-1$ do we get a non-vanishing contribution. This was shown in section \ref{sec:be12} for
the case of the contribution proportional to $ E_{m,\gamma(\gamma\sigma+\delta)} \left(X\right)$, but the same argument also applies
 to $I_{m,\gamma(\gamma\sigma+\delta)}\left(X \right)$.
Thus, in the following, we set $N=-1$.

We evaluate
\begin{equation}
	\frac{L}{X} =\frac{L}{\left( \frac{b}{a}+\frac{\tilde{\ell}}{2m}\right)}= 2m \frac{ \frac{m -a^2M}{c} +c }{\frac{m-a^2M}{c}-c}.
	\label{LXc}
\end{equation}
We recall that the condition $L\in\mathbb{Z}$ in \eqref{cancellations-2} implies the divisibilty condition
$ac \mid   m -a^2M +c^2$. Hence, if we fix $M$ and $a$, the sum over $c$ runs only over those $c<0$ that satisfy
\begin{equation}
	c \mid m-a^2M.
\end{equation}

We now fix $M$ and $a$.
There are three cases to be considered. First, consider the case when 
\begin{equation}
	m-a^2M <0.
\end{equation}
Let  $c<0$ be an integer that divides $m-a^2M$ and contributes the amount \eqref{LXc}. For any such integer there
exists another integer
$c' = -\frac{m-a^2M}{c}<0$, which also divides $m-a^2M$ and contributes the amount
\begin{equation}
	 \frac{L'}{\left( \frac{b}{a}+\frac{\tilde{\ell}}{2m}\right)'} =2m \frac{ \frac{m -a^2M}{c'} +c' }{\frac{m-a^2M}{c'}-c'} =2m \frac{ -c -\frac{m -a^2M}{c} }{-c +\frac{m-a^2M}{c}}  =  -\frac{L}{\left( \frac{b}{a}+\frac{\tilde{\ell}}{2m}\right)}.
\end{equation}
Thus, in \eqref{cancellations-2}, for a fixed $M$ and $a$, any summand $c<0$ with an associated value $L$ comes accompanied
by a summand $c'<0$ with an associated value $L' = -L$. Both these summands give rise to the same value $\tilde{\Delta} = 4MN-L^2 = - 4 M -L^2$.
Since both $c$ and $c'$ have the same value  $\frac{b}{a}+\frac{\tilde{\ell}}{2m}$, we infer that the value $\tilde{\ell}' $ associated to $c'$ is
\begin{equation}
	\tilde{\ell}'  = \tilde{\ell}+2m\frac{b-b'}{a} \;.
\end{equation}

We now show that it's possible to pick $b=b'$ in the above as follows. Observing that $a|(m+c^2)$, we write
\bea 
m = - c^2 + a k,\,\,\,\, k \in \mathbb{Z}
\label{relmc}
\eea 
and hence 
\bea 
c' = - \frac{m-a^2 M}{c} = c - \frac{a}{c} (k- a M).
\eea
Hence the operation 
\bea
\begin{pmatrix} 1 & 0 \\ -\frac{(k -a M)}{c} & 1\end{pmatrix} \begin{pmatrix} a & b \\ c & d\end{pmatrix} = \begin{pmatrix} a' & b' \\ c' & d'\end{pmatrix}
\eea
yields a new matrix with the same $a$ and $b'=b$ and with the required $c'$. 
Hence we can choose $b'=b$, in which case $\tilde{\ell}'  = \tilde{\ell}$. Since the multiplier systems are the same, we 
get a cancellation between the contributions from $c$ and $c'$.

Next, let us consider the case when
\begin{equation}
	m-a^2M >0.
\end{equation}
Let  $c<0$ be an integer that divides $m-a^2M$ and contributes the amount \eqref{LXc}. For any such integer there
exists another integer
 $c' = \frac{m-a^2M}{c}<0$, which also divides $m-a^2M$ and contributes the amount
\begin{equation}
	 \frac{L'}{\left( \frac{b}{a}+\frac{\tilde{\ell}}{2m}\right)'} =2m \frac{ \frac{m -a^2M}{c'} +c' }{\frac{m-a^2M}{c'}-c'} =2m \frac{ c +\frac{m -a^2M}{c} }{c -\frac{m-a^2M}{c}}  =  -\frac{L}{\left( \frac{b}{a}+\frac{\tilde{\ell}}{2m}\right)}.
	 \label{cancellations2}
\end{equation}
Note that now $L' = L$, while
\begin{equation}
	\left( \frac{\tilde{\ell}}{2m}+\frac{b}{a} \right)' = -\left( \frac{\tilde{\ell}}{2m}+\frac{b}{a} \right),
	\label{llp}
\end{equation}
Using \eqref{relmc}, we obtain 
\bea 
c' = \frac{m-a^2 M}{c} = - c + \frac{a}{c} (k- a M).
\eea
We then choose 
\bea
a' &=& a \nonumber\\
b' &=& - b + j a \nonumber\\
d' &=& d +  \frac{(j a-b)}{c}  (k- a M) - j c, \;\;\; j \in \mathbb{Z},
\eea
which satisfies $a' d' - b' c' = 1$. The shift by $j$ returns $b'$ to $b' \in \mathbb{Z}/a \mathbb{Z}$. Then, from \eqref{llp}, we infer
\bea
{\tilde \ell}' = - {\tilde \ell} - 2 m \frac{ b' + b}{a}  = - {\tilde \ell} - 2 m j.
\eea
Using the property $\psi(\Gamma)_{(-{\tilde \ell} - 2 m j) \ell} = \psi (\Gamma)_{- {\tilde \ell} \ell}$, and recalling that \eqref{cancellations-2} uses
 the two multiplier systems $\psi(\Gamma)_{ \tilde{\ell}\ell}$ and $\psi(\Gamma)_{-\tilde{\ell}\ell }$, also in this case
 we get a cancellation between the contributions from $c$ and $c'$.

Finally, let us consider the case when $m-a^2M = 0$. Then, the divisibility condition $ac \mid   m -a^2M +c^2$
becomes
\begin{equation}
	ac \mid   c^2,
\end{equation}
and since $\gcd(a,c)=1$, this imposes
\begin{equation}
	a = 1
\end{equation}
and therefore
\begin{equation}
	M = m.
\end{equation}
Thus, when $N=-1$ the only terms that contribute satisfy $M=m$ and $a=1$. 

Finally, we note that when $N=-1, M=m$ and $a=1$, then
\begin{equation}
	\left(\frac{\tilde{\ell}}{2m}+\frac{b}{a} \right) = \frac{1}{2m}\left(-\frac{1}{ac}\left(m-a^2M-c^2\right)\right) = \frac{c}{2m} .
\end{equation}
Since $c<0$, we infer $\frac{\tilde{\ell}}{2m}+\frac{b}{a} \notin [0, - \frac{1}{ac}]$, and hence $X \neq 0, - \frac{1}{ac}$.

Thus, we conclude that only when $X Y \neq 0$, do we get non-vanishing contributions in  \eqref{cancellations-2}.

\subsection{Bessel function $I_{25/2}$}

Finally, we focus on the terms $I_{m,\gamma (\gamma \sigma + \delta)}$
in the decompositions
\eqref{both-positive-1}, \eqref{one-zero-1} and  \eqref{both-negative-1}. In the previous subsection 
we showed that these terms only give a non-vanishing contribution when $X Y \neq 0$.
Thus, we will assume $X Y \neq 0$ in the following. 

We proceed as in section \ref{sec:bess23/2}, namely, 
we restrict the sum over $\delta$ to the range $0 \leq -\delta < \gamma$ and perform the integration over
the Ford circle $\Gamma_{\sigma}$ in \eqref{cont-sig}.
Collecting the terms proportional $I_{m,\gamma (\gamma \sigma + \delta)}$ in \eqref{after-v-integration2}, and changing 
the integration variable to ${\tilde \sigma}$ given 
in \eqref{tilsig}, we obtain
\begin{equation} \label{after-kloosterman-25/2}
\begin{gathered}
	(-1)^{\ell+1}\sum_{P_{0}} \, {\rm Kl} ( \frac{\Delta}{4m}, \frac{\tilde{\Delta}}{4m} ;\gamma,\psi)_{\ell\tilde{\ell}}
	\; L  \, d(M) \, d(N) \; 
	\int_{\tilde \Gamma} d\tilde{\sigma} \;\frac{{\tilde{\sigma}^{21/2}}}{\gamma^{25/2}}\,
	e^{\left(-2\pi i \left[  
	\frac{\tilde{\Delta}}{4m}\frac{1}{\tilde{\sigma}}
	+\frac{{\Delta}}{4m}\frac{\tilde{\sigma}}{\gamma^2}
	\right] \right)}
	\\
	\frac{1}{2}\left(I_{m,\tilde{\sigma}}\left( \frac{b}{a}+\frac{\tilde{\ell}}{2m}\right)+I_{m,\tilde{\sigma}}\left(-\frac{b}{a}-\frac{1}{ac}-\frac{\tilde{\ell}}{2m}\right) \right),
	\end{gathered}
\end{equation}
where $L$ and ${\tilde \ell}$ are given in \eqref{L-lt-n-tilde-definition}, and where 
the generalized Kloosterman sum ${\rm Kl} ( \frac{\Delta}{4m}, \frac{\tilde{\Delta}}{4m} ;\gamma,\psi)_{\ell\tilde{\ell}}$ is given in \eqref{kloo}.
The set $P_0$ is given in \eqref{setP0}. 
The integration contour ${\tilde \Gamma}$ denotes a Ford circle centered at ${\tilde \sigma} = i/2$ that skirts the origin ${\tilde \sigma} =0$.

Using the symmetry property \eqref{symmetry-3}, we write \eqref{after-kloosterman-25/2} as 
\begin{equation} \label{after-kloosterman-25/2-2}
\begin{gathered}
	(-1)^{\ell+1}\sum_{P_{0}} \, \left( {\rm Kl} ( \frac{\Delta}{4m}, \frac{\tilde{\Delta}}{4m} ;\gamma,\psi)_{\ell\tilde{\ell}}
	+  {\rm Kl} ( \frac{\Delta}{4m}, \frac{\tilde{\Delta}}{4m} ;\gamma,\psi)_{ \ell (-\tilde{\ell})} \right)
	\; L  \, d(M) \, d(N) \; \\
	\int_{\tilde \Gamma} d\tilde{\sigma} \;\frac{{\tilde{\sigma}^{21/2}}}{\gamma^{25/2}}\,
	e^{\left(-2\pi i \left[  
	\frac{\tilde{\Delta}}{4m}\frac{1}{\tilde{\sigma}}
	+\frac{{\Delta}}{4m}\frac{\tilde{\sigma}}{\gamma^2}
	\right] \right)}
	\frac{1}{2} I_{m,\tilde{\sigma}}\left( \frac{b}{a}+\frac{\tilde{\ell}}{2m}\right) \;.
		\end{gathered}
\end{equation}
Using the expression for $I_{m,\tilde{\sigma}}$ given in \eqref{i-function} and the relation \eqref{relDN}, we obtain 
\begin{equation} \label{after-kloosterman-25/2-subst}
\begin{gathered}
	(-1)^{\ell} \frac{1}{4\sqrt{2\pi^2 i m}} \sum_{P_{0}} 
	 \left[ {\rm Kl} ( \frac{\Delta}{4m}, \frac{\tilde{\Delta}}{4m} ;\gamma,\psi)_{\ell\tilde{\ell}}
	+ {\rm Kl} ( \frac{\Delta}{4m}, \frac{\tilde{\Delta}}{4m} ;\gamma,\psi)_{\ell ( - \tilde{\ell}) } \right]
	 \,  \frac{L}{\left( \frac{b}{a}+\frac{\tilde{\ell}}{2m} \right)} \, d(M) \, d(N) \\
	 \int_{\tilde \Gamma} d\tilde{\sigma} \;\frac{{\tilde{\sigma}^{21/2}}}{\gamma^{25/2}}\,
	e^{\left(-2\pi i \left[  
	\frac{N}{a^2}\frac{1}{\tilde{\sigma}}
	+\frac{{\Delta}}{4m}\frac{\tilde{\sigma}}{\gamma^2}
	\right] \right)}
	\int_0^{i\infty}\left(\frac{1}{\tilde{\sigma}}-z \right)^{-3/2}e^{2\pi i m \left( \frac{b}{a}+\frac{\tilde{\ell}}{2m}\right)^2 z}dz \;.
	\end{gathered}
\end{equation}
Now consider the integral
\begin{equation}
	\int_0^{i\infty}\left(\frac{1}{\tilde{\sigma}}-z \right)^{-3/2}e^{2\pi i m \left( \frac{b}{a}+\frac{\tilde{\ell}}{2m}\right)^2 z}dz,
	\label{intzlj}
\end{equation}
which we rewrite using
 \cite{2007arXiv0708.0692B} 
 \begin{equation} \label{Bringmann-Lovejoy-1}
	\int_\mathbb{R} \frac{e^{-2\pi t x^2}}{(x-is)^2} \, dx = -  \sqrt{2}\pi t \int_0^\infty \frac{e^{-2\pi u s^2}}{(u+t)^{3/2}}du,
\end{equation}
where $s \in \mathbb{R} \backslash \{0\}$, and where $t  \in  \mathbb{C}$ with ${\rm Re} \, t  > 0$. 
Setting $u = - i m z$ and $t = im /{\tilde \sigma}$ (note that ${\rm Re} \, t = m \sigma_2/|{\tilde \sigma}|^2 > 0$), 
the integral \eqref{intzlj} can be expressed as 
\begin{equation}
	\frac{1}{\sqrt{2 i m} \, \pi} \, \tilde{\sigma}\int_\mathbb{R} \frac{e^{-2\pi i m\frac{1}{\tilde{\sigma}}x^2}}{\left( x-i  \left( \frac{b}{a}+\frac{\tilde{\ell}}{2m}\right) \right)^2}dx \;.
\end{equation}
Then, collecting the exponential terms in  \eqref{after-kloosterman-25/2-subst}
with $1/\tilde{\sigma}$ in the exponent gives
\begin{equation}
	e^{-2\pi i \left(\frac{N}{a^2}+mx^2 \right)/{\tilde \sigma}  } \;.
\end{equation}
Interchanging the two integrations in  \eqref{after-kloosterman-25/2-subst}, and performing the integration over ${\tilde \sigma}$ along the
Ford circle ${\tilde \Gamma}$ as described in section \ref{sec:be12}, the latter will only be non-vanishing provided that
$\left(\frac{N}{a^2}+mx^2 \right) < 0$, which in turn implies $N=-1$. Then, as shown in the previous subsection, the only non-vanishing contribution
to \eqref{after-kloosterman-25/2-subst}
stems from $N=-1, a=1, M=m$.

Then, setting $N=-1, a=1, M=m$, we have
\bea
\tilde{\ell} 	= -2bm+c \;\;,\;\;
\frac{b}{a}+\frac{\tilde{\ell}}{2m} = \frac{c}{2m} \;\;,\;\; L  = -c>0 , \nonumber\\
\frac{L}{\frac{b}{a}+\frac{\tilde{\ell}}{2m}} = -2m \;\;,\;\; \tilde{\Delta} = 4MN-L^2 = -4m-c^2 \;.
\eea

Using \eqref{cj} and the multiplier system property $\psi_{(j + 2 m k) {\ell}}(\Gamma) = \psi_{j {\ell}}(\Gamma)$ (with $k \in \mathbb{Z}$) \cite{kloosterman}, we
write  \eqref{after-kloosterman-25/2-subst} as 
\begin{equation} 
\label{exp-interm1}
\begin{gathered}
	(-1)^{\ell+1} \frac{  \sqrt{2m } \, d(m) }{4\sqrt{\pi^2 i }}\,  \sum_{\gamma = 1 }^{\infty}
	 \sum_{j \in\mathbb{Z}/2m\mathbb{Z}}\sum_{\substack{c<0\\c =  j \text{ mod } 2m}} \\
	 \left[ {\rm Kl} ( \frac{\Delta}{4m}, -1 - \frac{c^2}{4m}  ;\gamma,\psi)_{\ell j }
	+ {\rm Kl} ( \frac{\Delta}{4m}, -1 - \frac{c^2}{4m} ;\gamma,\psi)_{\ell ( - j ) } \right]\\
	 \int_{\tilde \Gamma} d\tilde{\sigma} \;\frac{{\tilde{\sigma}^{21/2}}}{\gamma^{25/2}}\,
	e^{\left(2\pi i \left[  
	\frac{1}{\tilde{\sigma}}
	-\frac{{\Delta}}{4m}\frac{\tilde{\sigma}}{\gamma^2}
	\right] \right)}
	\int_0^{i\infty}\left(\frac{1}{\tilde{\sigma}}-z \right)^{-3/2}e^{2\pi i  \frac{c^2}{4m}  z}dz \;.
	\end{gathered}
\end{equation}
Noticing that the dependence on $c$ is quadratic, we write 
\begin{equation}
 \sum_{j \in\mathbb{Z}/2m\mathbb{Z}} \sum_{\substack{c<0\\ c =  j  \text{ mod }2m}}  f(c) = 
  \sum_{j \in\mathbb{Z}/2m\mathbb{Z}} \frac{1}{2}\sum_{\substack{c\in\mathbb{Z}\setminus\{0\}\\ c = j  \text{ mod }2m}}  f(c),
\end{equation}
where we made use of the symmetry $j \to - j$ to obtain an expression with the symmetry $c \to -c$. We now re-express the sum over $c$ as
\cite{Ferrari:2017msn} 
\begin{equation}
	\frac{1}{2}\sum_{\substack{c\in\mathbb{Z}\setminus\{0\}\\ c = j  \text{ mod }2m}}  f(c) = \frac{1}{2}\sum_{\substack{p\in\mathbb{Z}\\ g \in \mathbb{Z}/2m\gamma \mathbb{Z} \\2m\gamma p+ g\neq 0 \\ g = j \text{ mod }2m}}  f(2m\gamma p + g).
\end{equation}
Then, the $c$ depending part of the Kloosterman sum \eqref{kloo} gets re-expressed as
\begin{equation}
	e^{2\pi i \frac{\alpha}{\gamma} \left(\frac{-c^2}{4m} \right) } = e^{2\pi i \frac{\alpha}{\gamma} \left( \frac{-(2m\gamma p + g)^2}{4m} \right) } = e^{2\pi i \frac{\alpha}{\gamma} \left( \frac{-g^2}{4m}\right) },
\end{equation}
while the exponential term in the $dz$-integral becomes
\begin{equation}
	e^{2\pi i \frac{c^2}{4m} z} = e^{2\pi i \frac{(2m\gamma p + g)^2}{4m} z} = e^{2\pi i m \left(\gamma p +  \frac{g}{2m} \right)^2 z} \;.
\end{equation}
Thus we rewrite \eqref{exp-interm1} as 
\begin{equation} 
\label{exp-interm}
\begin{gathered}
	(-1)^{\ell+1} \frac{  \sqrt{2m } \, d(m) }{8 \sqrt{\pi^2 i }}\,  \sum_{\gamma = 1 }^{\infty}
	 \sum_{j \in\mathbb{Z}/2m\mathbb{Z}}
	 \sum_{\substack{p\in\mathbb{Z}\\ g \in \mathbb{Z}/2m\gamma \mathbb{Z} \\2m\gamma p+ g\neq 0 \\ g = j  \text{ mod }2m}}
\\
	 \left[ {\rm Kl} ( \frac{\Delta}{4m}, -1 - \frac{g^2}{4m}  ;\gamma,\psi)_{\ell j }
	+ {\rm Kl} ( \frac{\Delta}{4m}, -1 - \frac{g^2}{4m} ;\gamma,\psi)_{\ell ( - j ) } \right]\\
	 \int_{\tilde \Gamma} d\tilde{\sigma} \;\frac{{\tilde{\sigma}^{21/2}}}{\gamma^{25/2}}\,
	e^{\left(2\pi i \left[  
	\frac{1}{\tilde{\sigma}}
	-\frac{{\Delta}}{4m}\frac{\tilde{\sigma}}{\gamma^2}
	\right] \right)}
	\int_0^{i\infty}\left(\frac{1}{\tilde{\sigma}}-z \right)^{-3/2}e^{2\pi i  m \left(\gamma p +  \frac{g}{2m} \right)^2   z}dz \;.
	\end{gathered}
\end{equation}
Note that the sum over $p$ and $g$ builds up the standard weight $1/2$ index $m$ Jacobi theta function, with the exception of the term 
$2m\gamma p+ g =0$,
\begin{equation} \label{theta-function}
	\vartheta_{m,  j }\left(i\tau - \frac{\alpha}{\gamma};0\right) = \sum_{\substack{g \in\mathbb{Z}/2m \gamma \mathbb{Z}  \\ g =   j
	 \text{mod }2m}} 
	e^{-2\pi i \frac{\alpha}{\gamma \, 4m}g^2}\sum_{p\in\mathbb{Z}}e^{-2\pi\tau m\left( \frac{g}{2m}+\gamma p \right)^2}.
\end{equation}

Next, we follow \cite{Bringmann:2010sd,Ferrari:2017msn}. Using the results reviewed in Appendix \ref{sec:sinh},
we rewrite \eqref{exp-interm} as 
\begin{equation} 
\label{int-rv-f}
\begin{gathered}
	(-1)^{\ell} \, i  \frac{d(m)}{8 \pi^2} \, \sum_{\gamma>0}\sum_{j \in\mathbb{Z}/2m\mathbb{Z}} \sum_{\substack{ g \in \mathbb{Z}/2m\gamma \mathbb{Z} \\ g = j \text{ mod }2m}}
	 \left[ {\rm Kl} ( \frac{\Delta}{4m}, -1 - \frac{g^2}{4m}  ;\gamma,\psi)_{\ell j }
	+ {\rm Kl} ( \frac{\Delta}{4m}, -1 - \frac{g^2}{4m} ;\gamma,\psi)_{\ell ( - j ) } \right]\\
		\int_{\tilde \Gamma} d\tilde{\sigma} \;\frac{{\tilde{\sigma}^{23/2}}}{\gamma^{25/2}} \, 
e^{-2\pi i\frac{{\Delta}}{4m}\frac{\tilde{\sigma}}{\gamma^2}
	}e^{2\pi i\frac{1}{\tilde{\sigma}}}
	\int_\mathbb{R} e^{-2\pi i m\frac{1}{\tilde{\sigma}}x'^2}f_{\gamma,g,m}(x') \frac{1}{\gamma^2}dx',
	\end{gathered}
\end{equation}
where
\begin{equation} 
\label{fsinh}
	 f_{\gamma,g,m}(x') = \sum_{\substack{p\in\mathbb{Z}\\2m\gamma p+ g\neq 0}}	 \frac{\gamma^2}{\left( x'-i  \gamma p -i\frac{g}{2m} \right)^2} = \left\{\begin{aligned}
	 	\frac{\pi^2}{\sinh^2\left( \frac{\pi x'}{\gamma} - \frac{\pi i g}{2m \gamma} \right)} &\;\; \text{ if}\;\;\; g \neq 0 \text{ mod } 2m\gamma \\
	 	\frac{\pi^2}{\sinh^2\left( \frac{\pi x'}{\gamma} \right)} - \frac{\gamma^2}{x'^2} & \;\;\text{ if}\;\;\; g = 0 \text{ mod } 2m\gamma 
	 \end{aligned}\right.
\end{equation}

We now turn our attention to the $\tilde{\sigma}$ and $x'$ integrals 
\begin{equation} 
	\int_{\tilde \Gamma}  d\tilde{\sigma} \;\frac{{\tilde{\sigma}^{23/2}}}{\gamma^{29/2}}\,
	e^{-2\pi i\frac{{\Delta}}{4m}\frac{\tilde{\sigma}}{\gamma^2}
	}
	\int_\mathbb{R} e^{2\pi i\frac{1}{\tilde{\sigma}}(1- mx'^2)}f_{\gamma,g,m}(x') dx'.
\end{equation}
We perform the change of variables
\begin{equation}
	w = \frac{i}{\tilde{\sigma}}, 
	\end{equation}
and obtain (${\tilde \epsilon} > 0$) 
\begin{equation} \label{t-integral-25}
	 (-1) \int_{{\tilde \epsilon}-i\infty}^{{\tilde \epsilon}+i \infty}  dw \;\frac{{i^{25/2}}}{\gamma^{29/2}}\frac{1}{w^{27/2}}\,
	e^{2\pi \frac{{\Delta}}{4m\gamma^2}\frac{1}{w}
	}
	\int_\mathbb{R} e^{2\pi(1- mx'^2)w} f_{\gamma,g,m}(x') dx'.
\end{equation}
Interchanging the two integrations and performing the integration over $w$ first, we only get a non-vanishing 
Bessel integral provided that we restrict the range of integration over $x'$ to $1- mx'^2 > 0$, that is to
\begin{equation}
	\int_\mathbb{R} dx' \to \int_{-1/\sqrt{m}}^{1/\sqrt{m}} dx' \;.
\end{equation}
Performing another change of variables,
\begin{equation}
	t = 2\pi (1-mx'^2) w, 
	\end{equation}
the integral \eqref{t-integral-25} becomes ($\epsilon >0$) 
\begin{equation} 
	 (-1) (2\pi)^{25/2} \;\frac{{i^{1/2}}}{\gamma^{29/2}}
	 \int_{-1/\sqrt{m}}^{1/\sqrt{m}}dx'f_{\gamma,g,m}(x') \, (1-mx'^2)^{25/2}
	 \int_{{ \epsilon}-i\infty}^{{\epsilon}+i \infty}  d t 
	 \frac{1}{t^{27/2}}\,
	e^{4\pi^2 \frac{{\Delta}}{4m\gamma^2}(1-mx'^2)\frac{1}{t}
	}
	e^{t} ,
\end{equation}
which, using \eqref{bess-int}, equals
\begin{equation} 
	 (-1)i^{3/2}2\pi\int_{-1/\sqrt{m}}^{1/\sqrt{m}}dx'f_{\gamma,g,m}(x')  (1-mx'^2)^{25/4}\left(\frac{4m}{\Delta} \right)^{25/4} \;\frac{{1}}{\gamma^{2}}\,
	I_{25/2}\left(\frac{2\pi}{\gamma\sqrt{m}}\sqrt{\Delta(1-mx'^2)}\right).
\end{equation}
Thus, \eqref{int-rv-f} becomes 
\begin{equation} 
\begin{gathered}
	(-1)^{\ell} \, i^{1/2} \frac{d(m)}{4 \pi} \, \sum_{\gamma>0}\sum_{j \in\mathbb{Z}/2m\mathbb{Z}} \sum_{\substack{ g \in \mathbb{Z}/2m\gamma \mathbb{Z} \\ g = j  \text{ mod }2m}} \frac{
	 \left[ {\rm Kl} ( \frac{\Delta}{4m}, -1 - \frac{g^2}{4m}  ;\gamma,\psi)_{\ell j }
	+ {\rm Kl} ( \frac{\Delta}{4m}, -1 - \frac{g^2}{4m} ;\gamma,\psi)_{\ell ( - j ) } \right] }{\gamma^2} \\
	\left(\frac{4m}{\Delta} \right)^{25/4} \, 
\int_{-1/\sqrt{m}}^{1/\sqrt{m}}dx' f_{\gamma,g,m}(x')  (1-mx'^2)^{25/4} \;
	I_{25/2}\left(\frac{2\pi}{\gamma\sqrt{m}}\sqrt{\Delta(1-mx'^2)}\right).
	\end{gathered}
\end{equation}
The above expression involves two Kloosterman sums. 
Using the property  $f_{\gamma, - g,m}(- x') =  f_{\gamma,g,m}(x')$, we can show that both Kloosterman sums give rise to the same contribution, 
because the sign change of $j$ in the multiplier system can be compensated by a sign change in $g$ and $x'$. 
Thus, we arrive at the expression
\begin{equation} 
\label{fin-25/2}
\begin{gathered}
	(-1)^{\ell} \, i^{1/2} \frac{d(m)}{2 \pi} \, \sum_{\gamma>0}\sum_{j \in\mathbb{Z}/2m\mathbb{Z}} \sum_{\substack{ g \in \mathbb{Z}/2m\gamma \mathbb{Z} \\ g = j  \text{ mod }2m}} \frac{
	  {\rm Kl} ( \frac{\Delta}{4m}, -1 - \frac{g^2}{4m}  ;\gamma,\psi)_{\ell j }   }{\gamma^2} \\
	\left(\frac{4m}{\Delta} \right)^{25/4} \, 
\int_{-1/\sqrt{m}}^{1/\sqrt{m}}dx'f_{\gamma,g,m}(x')  (1-mx'^2)^{25/4} \;
	I_{25/2}\left(\frac{2\pi}{\gamma\sqrt{m}}\sqrt{\Delta(1-mx'^2)}\right).
	\end{gathered}
\end{equation}


\subsection{Complete result for $\Delta > 0$}

Combining the contributions \eqref{fin-23/2}, \eqref{fin-12} and \eqref{fin-25/2}, and relabelling $j$ by $\tilde \ell$, 
we obtain 
the following expression for the degeneracy $d(m,n,\ell)$ defined in \eqref{2}:

{\theorem

\bea
\label{crbs}
&& d(m,n,\ell) =  \nonumber\\
		&& (-1)^{\ell} \, i^{1/2} \,  \sum_{\gamma=1}^{+\infty}\sum_{\tilde{\ell}\in\mathbb{Z}/2m\mathbb{Z}} \left( 2\pi \sum_{\substack{\tilde{n}\geq -1,\nonumber\\
		\tilde{\Delta}<0}} c^F_m(\tilde{n},\tilde{\ell})\frac{{\rm Kl}(\frac{\Delta}{4m},\frac{\tilde{\Delta}}{4m},\gamma,\psi)_{\ell\tilde{\ell}}}{\gamma}
	  \left(\frac{ \vert\tilde{\Delta} \vert}{\Delta} \right)^{23/4} I_{23/2}\left(\frac{\pi}{\gamma m}\sqrt{\Delta\vert\tilde{\Delta}\vert} \right) \right.
	  \nonumber\\
	&&\qquad \qquad   -\delta_{\tilde{\ell},0}\sqrt{2m}\, d(m) 
	\frac{{\rm Kl} ( \frac{\Delta}{4m}, -1 ;\gamma,\psi)_{\ell 0}}{\sqrt{\gamma}}\left(\frac{4m}{\Delta} \right)^6 I_{12}\left(\frac{2\pi}{\gamma}\sqrt{\frac{\Delta}{m}} \right) \nonumber\\
	&&\qquad \qquad +\frac{1}{2\pi} d(m) \sum_{\substack{ g \in \mathbb{Z}/2m\gamma \mathbb{Z} \\ g = \tilde{\ell} \text{ mod }2m}}
	\frac{{\rm Kl }( \frac{\Delta}{4m}, -1-\frac{g^2}{4m} ;\gamma,\psi)_{\ell\tilde{\ell}}}{\gamma^2} \\
&& \qquad \qquad  \left.  \left(\frac{4m}{\Delta} \right)^{25/4}\int_{-1/\sqrt{m}}^{1/\sqrt{m}}dx'  f_{\gamma,g,m}(x')  (1-mx'^2)^{25/4}
	I_{25/2}\left(\frac{2\pi}{\gamma\sqrt{m}}\sqrt{\Delta(1-mx'^2)}\right) \right), \nonumber
\eea

\noindent
with 

\bea
&&c^F_m(\tilde{n},\tilde{\ell} )=
			\sum_{\substack{a>0, c<0    \\ b \in \mathbb{Z}/ a \mathbb{Z}, \; 
		a d - b c = 1 \\  0 \leq \frac{b}{a} + \frac{\tilde \ell}{2m} < - \frac{1}{ac }  }} 
	\left( (ad + bc ) {\tilde \ell} + 2  ac {\tilde n} + 2 bd m  \right)
	d( c^2 {\tilde n} + d^2 m + cd {\tilde \ell}) \, d(a^2 {\tilde n} + b^2 m + ab {\tilde \ell}) 
	\nonumber\\
	&& \frac{1}{\eta^{24}(\tau)} = \sum_{n=-1}^\infty d(n) \, e^{2\pi i \tau n} \; \;\;,\;\; \;
{\rm Kl}(\frac{\Delta}{4m},\frac{\tilde{\Delta}}{4m},\gamma,\psi)_{\ell\tilde{\ell}} = 
 \sum_{\substack{0\leq -\delta <\gamma\\ (\delta,\gamma)=1, \alpha\delta = 1 \text{ mod } \gamma}}e^{2\pi i \left( \frac{\alpha}{\gamma}\frac{\tilde{\Delta}}{4m} +\frac{\delta}{\gamma}\frac{\Delta}{4m}\right)} \, {\psi} (\Gamma)_{\tilde{\ell}\ell} 
\nonumber\\
	&& {\psi} (\Gamma)_{{\tilde \ell} \ell} = \frac{1}{\sqrt{2m\gamma i }}\sum_{T\in\mathbb{Z}/\gamma\mathbb{Z}} e^{2\pi i \left(\frac{\alpha}{\gamma}\frac{({\tilde \ell }-2mT)^2}{4m}-\frac{\ell ({\tilde \ell} -2mT)}{2m\gamma} +\frac{\delta}{\gamma}\frac{\ell^2}{4m}   \right)} 
\\
&& f_{\gamma,g,m}(x') = \sum_{\substack{p\in\mathbb{Z}\\2m\gamma p+ g\neq 0}}	 \frac{\gamma^2}{\left( x'-i  \gamma p -i\frac{g}{2m} \right)^2} = \left\{\begin{aligned}
	 	\frac{\pi^2}{\sinh^2\left( \frac{\pi x'}{\gamma} - \frac{\pi i g}{2m \gamma} \right)} &\;\; \text{ if}\;\;\; g \neq 0 \text{ mod } 2m\gamma \\
	 	\frac{\pi^2}{\sinh^2\left( \frac{\pi x'}{\gamma} \right)} - \frac{\gamma^2}{x'^2} & \;\;\text{ if}\;\;\; g = 0 \text{ mod } 2m\gamma 
	 \end{aligned}\right. \nonumber
\eea

}

Note that the continued fraction structure is encoded in a subset of $S_G$, while the 
Kloosterman sums ${\rm Kl}$ and Bessel functions are built up from $\Gamma_{\infty} \backslash S_{\Gamma}$ and $S_G$.

The expression \eqref{crbs}, without the continued fraction structure of $c^F_m(\tilde{n},\tilde{\ell})$, was first derived in \cite{Ferrari:2017msn} by
viewing $d(m, n, \ell)$ as Fourier coefficients of a mixed Mock Jacobi form $\psi_m^F$ \cite{Dabholkar:2012nd}. 
The above result is an exact expression for $d(m,n,\ell)$ which can be viewed as the non-perturbative completion of previous results in 
\cite{David:2006yn,Banerjee:2008ky,Murthy:2009dq}.

\section{The case $\Delta = m= \ell = 0$}

Now we consider the case $m= \ell = 0$, which implies $\Delta =0$.

We return to \eqref{ppse} and set $m=\ell =0$,
\bea
	&&(-1) \sum_{P} {(\gamma\sigma+\delta)^{10}} \,\int_{\Gamma_v} dv \, 
		\sum_{\substack{M,N\geq -1\\  L\in\mathbb{Z} }}  L \, d(M) \, d(N) \nonumber\\
	&&\qquad \exp\left(-2\pi i \left[ - \tilde{n}\left(\frac{\alpha\sigma+\beta}{\gamma\sigma+\delta}\right)-\tilde{\ell} \frac{v}{\gamma\sigma+\delta}+n\sigma \right] \right) \;,
	\label{ppseml0}
\eea
where the integers $L, {\tilde \ell}$ and ${\tilde n}$ are given by
\begin{align}
	L & = \frac{a}{c}M+\frac{c}{a}N \nonumber\\
	\tilde{\ell} & =  \frac{a}{c}M-\frac{c}{a}N \nonumber\\
	\tilde{n} & =  -\frac{b}{c}M+\frac{d}{a}N . 
	\label{Lln}
\end{align}
We apply the $T$-shift transformation \eqref{bdT} to $b$ and to $d$, 
which leaves $L, \tilde{\ell}$ invariant and changes $\tilde n$ into
\begin{equation}
	\tilde{n} \to \tilde{n}-\tilde{\ell}T.
\end{equation}
Then, as before, 
we split the sum over $b$ into a sum over $b\in\mathbb{Z}/a\mathbb{Z}$ and a sum over $T \in \mathbb{Z}$.

We pick the following integration contour $\Gamma_v$. We  set $v_2 = 0$. 
This choice is motivated by noting that setting $\ell =0$ in \eqref{rch1}
one obtains $v_2 =0$.
 Then, setting $v_2 = 0$ in the expressions \eqref{imrims}, we infer that the range
of $v_1$ is restricted to 
\begin{equation}
	 - \frac{b}{\gamma a} - \frac{T}{\gamma} < v_1 < 
	 - \frac{b}{\gamma a} - \frac{1}{ac\gamma}- \frac{T}{\gamma} =  - \frac{b}{\gamma a} + \frac{1}{n_2} - \frac{T}{\gamma},
	\label{vrangewithT}
\end{equation}
where we have made use of the $T$-shift. Demanding that the range of integration over $v_1$ is contained in an interval of 
length $1$  constrains $T$ to take values in $T\in\mathbb{Z}/\gamma\mathbb{Z}$.
Then, \eqref{ppseml0} takes the form
\bea
\label{expr-ml0} 
&&	 (-1)
	\sum_{\substack{P\\ b\in\mathbb{Z}/a\mathbb{Z},  \; T \in \mathbb{Z}/\gamma\mathbb{Z} }}
	{(\gamma\sigma+\delta)^{10}} \,  \int_{\Gamma_v} dv \, 
	\sum_{\substack{M,N\geq -1\\  L\in\mathbb{Z} }} \, L \, d(M) \, d(N) \nonumber\\
	&& \exp\left(-2\pi i \left[ - \tilde{n}\left(\frac{\alpha\sigma+\beta}{\gamma\sigma+\delta}\right)+T\tilde{\ell}\left(\frac{\alpha\sigma+\beta}{\gamma\sigma+\delta}\right)-\tilde{\ell} \frac{v}{\gamma\sigma+\delta} +n\sigma \right] \right) \;.
		\eea

We interchange the integration over $v_1$ with the summation over $M,N$.
The integration over $v_1$ will produce a different result 
depending on whether ${\tilde \ell}$ vanishes or not. Therefore, we discuss both cases
separately.

\subsection{$\tilde{\ell} = 0$}

Setting ${\tilde \ell} = 0$ implies
\bea
 \frac{a}{c}M = \frac{c}{a}N \;\;\;,\;\;\;
 \tilde{n} = \frac{N}{a^2}.
 \label{cond-tl0}
 \eea
The integral over $v_1$ is trivial and independent of $T$. The sum over $T$ yields a factor $\gamma$. Thus, 
 we obtain for \eqref{expr-ml0},
\bea
\label{expr-ml0-1}
	 \frac{1}{ac} 
	\sum_{\substack{P\\ b\in\mathbb{Z}/a\mathbb{Z} }}
	{(\gamma\sigma+\delta)^{10}} \,  
	\sum_{\substack{M,N\geq -1\\  L\in\mathbb{Z}, \;   {\tilde \ell} = 0 }} \, L \, d(M) \, d(N) \,
	 \exp\left(-2\pi i \left[ - \tilde{n}\left(\frac{\alpha\sigma+\beta}{\gamma\sigma+\delta}\right) +n\sigma \right] \right) \;.
		\eea
Using 
\begin{equation}
	- \tilde{n}\left(\frac{\alpha\sigma+\beta}{\gamma\sigma+\delta}\right) = - \tilde{n}\frac{\alpha}{\gamma} + \tilde{n}\frac{1}{\gamma(\gamma\sigma+\delta)}, \hspace{6mm}n\sigma = \frac{1}{\gamma^2}n\gamma(\gamma\sigma+\delta)-n\frac{\delta}{\gamma}
\end{equation}
as well as \eqref{cond-tl0}, we get

\bea
\label{expr-ml0-2}
	 \frac{1}{ac} 
	\sum_{\substack{P\\ b\in\mathbb{Z}/a\mathbb{Z} }}
	{(\gamma\sigma+\delta)^{10}} \,  
	\sum_{\substack{M,N\geq -1\\  L\in\mathbb{Z}, \;   {\tilde \ell} = 0 }} \, L \, d(M) \, d(N) \,
	e^{2\pi i\left(\frac{\alpha}{\gamma}\frac{N}{a^2} +n\frac{\delta}{\gamma}\right) }e^{-2\pi i \left( \frac{N}{a^2}\frac{1}{\gamma(\gamma\sigma+\delta)}
	 +\frac{n}{\gamma^2}\gamma(\gamma\sigma+\delta) \right)}.
		\eea
		
Next, we integrate over $\sigma_1$.
We proceed as in section \ref{sec:bess23/2}, namely, 
we restrict the sum over $\delta$ to the range $0 \leq -\delta < \gamma$ and perform the integration over
the Ford circle $\Gamma_{\sigma}$ in \eqref{cont-sig}.
 We interchange the integration over $\sigma$ with the sum over $M$ and $N$.
 The resulting integral is similar to the one in \eqref{symmetry-ford}, but without a branch cut, and will
 only be non-vanishing provided that $N <0$, i.e. when $N=-1$. Inserting $N=-1$ into the expression for $L - {\tilde \ell}$ given in \eqref{Lln}, and 
 recalling that $\gcd (a, c)=1$, fixes $a=1$. Then, from \eqref{cond-tl0}, we infer  $M = -c^2$, which in turn implies $M=-1$ and $c=-1$, and hence
 $L=2$. Moreover, since $b\in\mathbb{Z}/a\mathbb{Z} $ with $a=1$, this fixes $b$ to a single value. Thus, integrating \eqref{expr-ml0-2} over $\sigma$, we obtain,
 using $d(-1)=1$, 
\begin{equation}
	-2 
 \sum_{\substack{\gamma \geq 1\\ 0 \leq \alpha, -\delta < \gamma\\
	\alpha\delta  = 1 \text{ mod } \gamma
	}}  \int_{\Gamma_{\sigma}} d\sigma \,  {(\gamma\sigma+\delta)^{10}}
e^{2\pi i\left(-\frac{\alpha}{\gamma} +n\frac{\delta}{\gamma}\right) }
e^{-2\pi i \left( -\frac{1}{\gamma(\gamma\sigma+\delta)}
	 +\frac{n}{\gamma^2}\gamma(\gamma\sigma+\delta) \right)}.
\end{equation}
Performing the change of variables
\begin{equation}
	\tilde{\sigma} = \gamma(\gamma\sigma+\delta) , 
	\end{equation}
the Ford circle $\Gamma_{\sigma}$ 
becomes the Ford circle ${\tilde \Gamma}$ of radius $1/2$ and of center $\tilde{\sigma} = i/2$,
\begin{equation}
	-2\,\sum_{\substack{\gamma \geq 1\\ 0 \leq \alpha, -\delta < \gamma\\
	\alpha\delta  = 1 \text{ mod } \gamma
	}}
\int_{\tilde \Gamma}  d\tilde{\sigma}  \,  \frac{\tilde{\sigma}^{10}}{\gamma^{12}}\,
e^{2\pi i\left(-\frac{\alpha}{\gamma} +n\frac{\delta}{\gamma}\right) }
e^{-2\pi i \left( -\frac{1}{\tilde{\sigma}}
	 +\frac{n}{\gamma^2}\tilde{\sigma} \right)}.
\end{equation}
Performing a further change of variables,
\begin{equation}
	t = \frac{2 \pi i}{\tilde{\sigma}},
\end{equation}
we obtain, with $\epsilon > 0$,
\begin{equation}
	\,2\, i^{11} \, (2\pi)^{11} \sum_{\substack{\gamma \geq 1\\ 0 \leq \alpha, -\delta < \gamma\\
	\alpha\delta  = 1 \text{ mod } \gamma
	}} \,  \frac{1}{\gamma^{12}}\,  e^{2\pi i\left(-\frac{\alpha}{\gamma} +n\frac{\delta}{\gamma}\right) }
\int_{\epsilon-i\infty}^{\epsilon+i \infty}  \frac{dt}{t^{12} } 
\,   
e^{t +4\pi^2 \frac{n}{\gamma^2}\frac{1}{t}}.
\end{equation}
Using \eqref{bess-int} we obtain,
\begin{equation}
	4\pi\sum_{\substack{\gamma \geq 1\\ 0 \leq \alpha, -\delta < \gamma\\
	\alpha\delta  = 1 \text{ mod } \gamma
	}}e^{2\pi i\left(-\frac{\alpha}{\gamma} +n\frac{\delta}{\gamma}\right) }\frac{1}{\gamma \,n^{11/2}} I_{11}\left(\frac{4\pi\sqrt{n}}{\gamma} \right) \;,
\end{equation}
which equals
\begin{equation}
	4\pi\sum_{\gamma=1}^{\infty} \frac{ {\rm Kl} (n,-1,\gamma) }{\gamma} \, \frac{1}{n^{11/2}}\,  I_{11}\left(\frac{4\pi\sqrt{n}}{\gamma} \right)
	\label{bessel11}
\end{equation}
when  written in terms of the classical Kloosterman sum
\bea
\label{claKL}
{\rm Kl} (n,-1,\gamma) = 
\sum_{\substack{0 \leq \alpha, -\delta < \gamma\\
	\alpha\delta  = 1 \text{ mod } \gamma
	}}e^{2\pi i\left(-\frac{\alpha}{\gamma} +n\frac{\delta}{\gamma}\right) } \;.
	\eea

\subsection{$\tilde{\ell} \neq 0$}

Now we take $\tilde{\ell}\neq 0$ in \eqref{expr-ml0}, and obtain 
\begin{equation}
\begin{aligned}
	&\frac{(-1)}{2\pi i}
	\sum_{\substack{P\\ b\in\mathbb{Z}/a\mathbb{Z},  \; T \in \mathbb{Z}/\gamma\mathbb{Z} }}
	{(\gamma\sigma+\delta)^{11}} \,  \\
		&\sum_{M,N\geq -1} \frac{L}{\tilde{\ell}} \, d(M) \, d(N) \, \left( e^{-2\pi i \frac{\tilde{\ell}}{\gamma\sigma+\delta}\left(\frac{b}{\gamma a}+\frac{1}{ac\gamma}+\frac{T}{\gamma}\right)}- e^{-2\pi i \frac{\tilde{\ell}}{\gamma\sigma+\delta}\left(\frac{b}{\gamma a}+\frac{T}{\gamma}\right)} \right) \\ 
	& \qquad \qquad  \exp\left(-2\pi i \left[ - \tilde{n}\left(\frac{\alpha\sigma+\beta}{\gamma\sigma+\delta}\right)+T\tilde{\ell}\left(\frac{\alpha\sigma+\beta}{\gamma\sigma+\delta}\right) +n\sigma \right] \right) \;.
	\end{aligned}
\end{equation}
Using
\begin{equation}
	\frac{\alpha\sigma+\beta}{\gamma\sigma+\delta} = \frac{\alpha}{\gamma}-\frac{1}{\gamma(\gamma\sigma+\delta)}
\end{equation}
and the expressions for $\tilde{n},\tilde{\ell}$ given in \eqref{Lln}, we get 
\begin{equation}
\label{exp-MN-bs}
\begin{aligned}
	&\frac{(-1)}{2\pi i}
	\sum_{\substack{P\\ b\in\mathbb{Z}/a\mathbb{Z},  \; T \in \mathbb{Z}/\gamma\mathbb{Z} }} {(\gamma\sigma+\delta)^{11}}
		\sum_{M,N\geq -1} \frac{L}{\tilde{\ell}} \, d(M) \, d(N) \, e^{-2\pi i \left( - \tilde{n}\frac{\alpha}{\gamma}+T\tilde{\ell}\frac{\alpha}{\gamma} \right)} \\
	&\left[ e^{-2\pi i\left( n\sigma+\frac{1}{\gamma(\gamma\sigma+\delta)}\left(\frac{1}{c^2}M\right)\right)}- e^{-2\pi i\left( n\sigma+\frac{1}{\gamma(\gamma\sigma+\delta)}\left(\frac{1}{a^2}N\right)\right)} \right].
\end{aligned}
\end{equation}
We will integrate over $\sigma_1$ following the procedure described below \eqref{expr-ml0-2}. The integration over the Ford circle $\Gamma$ will select the values
$M=-1$ and $N=-1$ in \eqref{exp-MN-bs}, respectively. Let us then focus on these terms in \eqref{exp-MN-bs}. We
first consider the contribution from $M=-1$. 
 Inserting $M=-1$ into the expression for $L$ and  ${\tilde \ell} \in \mathbb{Z}$ given in \eqref{Lln}, we see immediately that the integrality of these two variables, combined with $\gcd (a, c)=1$, fixes $c=-1$. Then, we obtain the contribution
\begin{equation}
\label{exp-MN-bs2}
	\frac{(-1)}{2\pi i}
	\sum_{\substack{P\\ b\in\mathbb{Z}/a\mathbb{Z},  \; T \in \mathbb{Z}/\gamma\mathbb{Z} }} {(\gamma\sigma+\delta)^{11}}
		\sum_{\substack{N> -1\\M=-1}} \frac{L}{\tilde{\ell}} \, d(N) \, e^{-2\pi i \left( - \tilde{n}\frac{\alpha}{\gamma}+T\tilde{\ell}\frac{\alpha}{\gamma}  \right)} 
	 e^{-2\pi i\left( n\sigma-\frac{1}{\gamma(\gamma\sigma+\delta)}\right)}.
\end{equation}
Using \eqref{Lln} we infer
\bea
{\tilde n} = - b + \frac{(1-b)}{a^2} N = -1 + \frac{(1-b)}{a} \, {\tilde \ell} ,
\eea
where $a | 1-b$, since $a d - b c = a d + b = 1$. We perform the decomposition
\bea
\frac{(1-b)}{a} = A + k \,  \gamma, 
\eea
where $A \in \mathbb{Z}/ \gamma \mathbb{Z}$ and $k \in \mathbb{Z}$. Then, the exponent in \eqref{exp-MN-bs2} becomes
\bea
e^{- 2 \pi i  \frac{\alpha}{\gamma} \left( 1 + \left( - A + T \right) {\tilde \ell} \right) } .
\eea
Since both $A$ and $T$ lie in $\mathbb{Z}/ \gamma \mathbb{Z}$, we can absorb $A$ into $T$, thereby arriving at
\bea
e^{- 2 \pi i  \frac{\alpha}{\gamma} \left( 1 + T  {\tilde \ell} \right) } \;\;\;,\;\;\; T \in \mathbb{Z}/ \gamma \mathbb{Z}.
\label{expitn}
\eea
Thus, we may replace the exponent in \eqref{exp-MN-bs2} with the exponent \eqref{expitn}, which is independent of $\tilde n$.
This we use in the cancellation argument that we now present.

Let us consider
\begin{equation}
	\frac{L}{\tilde{\ell}} = \frac{a-\frac{N}{a}}{a+\frac{N}{a}},
	\label{LlNa}
\end{equation}
where $a | N$, since $L, {\tilde \ell} \in \mathbb{Z}$. If $N>0$, then 
for each divisor $a$ of $N$, there will be another divisor $a' = \frac{N}{a}>0$ which contributes
\begin{equation}
	\frac{L'}{\tilde{\ell}'} = \frac{a'-\frac{N}{a'}}{a'+\frac{N}{a'}} = \frac{\frac{N}{a}-a}{\frac{N}{a}+a} = -\frac{L}{\tilde{\ell}}.
\end{equation}
Since $\tilde{\ell}' = \tilde{\ell}$, this contribution will cancel against the contribution \eqref{LlNa} in \eqref{exp-MN-bs2}.
Thus, \eqref{exp-MN-bs2} does not receive contributions from $N >0$. 

When
$N=0$, we have $L/\tilde{\ell} = 1$, ${\tilde \ell} = a$, ${\tilde n} = - b$. Now we recall that the $T$-transformation
\eqref{bdT} can be viewed as either imposing the restriction $b\in\mathbb{Z}/a\mathbb{Z}$ or $d\in \mathbb{Z}/(-c\mathbb{Z})$.
We choose the latter viewpoint, in which case we may set $d=0$, since $c=-1$. Then $a d - bc =1$ implies $b=1$, and 
we obtain from \eqref{exp-MN-bs2}, 
\begin{equation}
\label{exp-MN-bs3}
	\frac{(-1)}{2\pi i} d(0) \sum_{\gamma>0} \;
	\sum_{\substack{0 \leq \alpha, -\delta < \gamma\\
	\alpha\delta  = 1 \text{ mod } \gamma
	}}\\
		\; \sum_{T\in\mathbb{Z}/\gamma\mathbb{Z}}  {(\gamma\sigma+\delta)^{11}} \, e^{-2\pi i \left( 
		\frac{\alpha}{\gamma}+T a\frac{\alpha}{\gamma}  \right)} 
	 e^{-2\pi i\left( n\sigma-\frac{1}{\gamma(\gamma\sigma+\delta)}\right)}.
\end{equation}

Summing over $T$ restricts $a$ to 
\begin{equation}
	\sum_{T\in\mathbb{Z}/\gamma\mathbb{Z}}e^{-2\pi i T a\frac{\alpha}{\gamma}} = 
	\left\{\begin{aligned}
 	& 0 && \gamma \nmid a \\
 	& \gamma && \gamma\mid a
 \end{aligned}\right.
\end{equation}
Therefore, only when $a=k \,\gamma$, with $k \in \mathbb{N}$, do we get a contribution 
to \eqref{exp-MN-bs3},
\begin{equation}
	\frac{(-1)}{2\pi i}  d(0)   \, \sum_{\gamma>0}  
	\sum_{\substack{0 \leq \alpha, -\delta < \gamma\\
	\alpha\delta  = 1 \text{ mod } \gamma
	}}\\
	\sum_{\substack{k>0}}  \gamma \, {(\gamma\sigma+\delta)^{11}} \, e^{-2\pi i \frac{\alpha}{\gamma}} 
	 e^{-2\pi i\left( n\sigma-\frac{1}{\gamma(\gamma\sigma+\delta)}\right)}.
\end{equation}
The sum over $k>0$ is divergent. We regularize this sum by replacing it with
\begin{equation}
 \zeta(0) = -\frac{1}{2},
 \label{zet}
\end{equation}
where $\zeta(s)$ denotes the Riemann zeta function.

Next, performing the $\sigma$-integral over the Ford circle $\Gamma_{\sigma}$ following the procedure described below \eqref{expr-ml0-2},
we obtain
\begin{equation}
	 -\frac{1}{2}d(0)\frac{1}{2\pi i} \sum_{\gamma>0} \;
	  \sum_{\substack{0 \leq \alpha, -\delta < \gamma\\
	\alpha\delta  = 1 \text{ mod } \gamma
	}}\\
		e^{2\pi i \left(-\frac{\alpha}{\gamma} + n\frac{\delta}{\gamma}\right)}
	\, \frac{(2\pi)^{12}}{\gamma^{12}}
	 \int_{\epsilon-i\infty}^{\epsilon+i \infty}  \frac{dt}{t^{13} } 
\,   
e^{t +4\pi^2 \frac{n}{\gamma^2}\frac{1}{t}},
\end{equation}
which, when written in terms of the classical Kloosterman sum \eqref{claKL} and the modified Bessel function \eqref{bess-int},
equals
\begin{equation}
	-12\sum_{\gamma>0} {\rm Kl} (n,-1,\gamma)\frac{1}{n^6} I_{12}\left(\frac{4\pi \sqrt{n}}{\gamma} \right),
\end{equation}
where we have used $d(0) = 24$. 

This is the contribution stemming from the $M=-1, N>-1$ terms.
 The contribution from the $M>-1, N=-1$ terms is exactly the same, so that in total we obtain
\begin{equation}
	-24\sum_{\gamma>0} {\rm Kl} (n,-1,\gamma)\frac{1}{n^6} I_{12}\left(\frac{4\pi \sqrt{n}}{\gamma} \right).
	\label{bessel12-total}
\end{equation}

\subsection{Complete result for $\Delta = m = \ell =0$}

Adding up the contributions \eqref{bessel11} and \eqref{bessel12-total} 
 from the two sectors
$\tilde{\ell}=0$ and $\tilde{\ell}\neq 0$, we get for the degeneracy $d(0,n,0)$ of BPS dyons with charge bilinears $(m=0, n, \ell =0)$,
\begin{equation}
\label{imm-dd-ml0}
	d(0,n,0)= \sum_{\gamma \geq 1} {\rm Kl}(n,-1,\gamma)\left(\frac{4\pi}{\gamma \,n^{11/2}} I_{11}\left(\frac{4\pi\sqrt{n}}{\gamma} \right)-\frac{24}{n^6} I_{12}\left(\frac{4\pi \sqrt{n}}{\gamma} \right) \right).
\end{equation}
This is in agreement with the expression for the degeneracy of immortal BPS dyons generated by the quasi-modular form 
\cite{Dabholkar:2012nd,Bossard:2018rlt,LopesCardoso:2020pmp}
\begin{equation}
	2\frac{E_{2}(\sigma)}{\eta^{24}(\sigma)} = \sum_{n\geq -1} c(n)q^n \;.
\end{equation}
Its Rademacher expansion, which we review in Appendix \ref{sec:rade},  can be obtained
by means of the quasi-modular transformation property
\begin{equation}
	2\frac{E_{2}(\sigma)}{\eta^{24}(\sigma)} = (\gamma\sigma+\delta)^{10} \; 2\frac{E_{2}\left(\frac{\alpha\sigma+\beta}{\gamma\sigma+\delta}\right)}{\eta^{24}\left(\frac{\alpha\sigma+\beta}{\gamma\sigma+\delta}\right)}- \frac{12\gamma}{\pi i }\frac{(\gamma\sigma+\delta)^{11}}{\eta^{24}\left(\frac{\alpha\sigma+\beta}{\gamma\sigma+\delta}\right)},
\end{equation}
and reads
\begin{equation}
\label{rad}
	c(n) = \sum_{\gamma=1}^{+\infty} {\rm Kl} (n,-1,\gamma) \left( \frac{24}{n^6}I_{12}\left(\frac{4\pi\sqrt{n}}{\gamma}\right)-\frac{4\pi}{\gamma\, n^{11/2}}I_{11}\left(\frac{4\pi\sqrt{n}}{\gamma}\right) \right) \;.
\end{equation}
This precisely agrees with \eqref{imm-dd-ml0},
\begin{equation}
	(-1)c(n) = d(0,n,0).
\end{equation}

Finally, we note that the extra sign stemming 
from the zeta function regularization \eqref{zet} is crucial for obtaining agreement with the expression of the Rademacher expansion \eqref{rad}.

\section{Conclusions \label{sec:conc}}

By using two distinct ${\rm SL}(2, \mathbb{Z})$ subgroups of ${\rm Sp}(2, \mathbb{Z})$ we have obtained a Rademacher type expansion for the exact degeneracies of $\frac14$ BPS states with unit torsion in heterotic
string theory compactified on a six-torus. This expansion is obtained by summing over the contributions from the quadratic poles 
\eqref{n2p} of $1/\Phi_{10}$.
 The resulting expansion
 is 
 given by
\eqref{crbs} in the case when $\Delta > 0$, and by \eqref{imm-dd-ml0} in the case when  $\Delta = m = \ell =0$.
When $\Delta > 0$, the exact expression for the degeneracies exhibits a fine-grained structure that is tied to 
the presence of the two sets $S_G$ and $S_{\Gamma}$ of  ${\rm SL}(2, \mathbb{Z})$ matrices.   
The result \eqref{crbs} reproduces the expression that was obtained in \cite{Ferrari:2017msn} by a very different approach,
namely by the Rademacher expansion of the Fourier coefficients of 
the mixed Mock Jacobi forms $\psi_m^F$. The latter were shown in \cite{Dabholkar:2012nd}  to 
encode the coefficients $d(m, n, \ell)$ computed in a region \cite{Cheng:2007ch} where there are no linear pole contributions when $n>m$.
These coefficients are called  immortal degeneracies. The Rademacher expansion in this case involved information about
the shadow which repairs the modular behaviour of  $\psi_m^F$. The Mock behaviour manifests itself through the $I_{12}$ and
the $I_{25/2}$ terms in \eqref{crbs}, both of which would be absent in the case of a modular form.
In our approach, the $v$-integral yields error functions, each of which can be split into three distinct types of terms, namely a constant, a term involving
$E_{m,\gamma(\gamma\sigma+\delta)}$ and a term involving $I_{m,\gamma(\gamma\sigma+\delta)}$ (c.f. \eqref{e-function} and \eqref{i-function}). These
respectively yield the  $I_{23/2}$ terms, the $I_{12}$ term and 
the $I_{25/2}$ term. Further, the sum over $S_G$ matrices which have a non-vanishing contribution to the $I_{25/2}$ term yields an integral of a Jacobi theta
function minus a constant term. This, when added to the regularized contribution  from the $E_{m,\gamma(\gamma\sigma+\delta)}$ terms, yields precisely the Eichler integral of the shadow for $\psi_m^F$.
Further, the restriction on the range of $v_1$, which arises by demanding that the poles lie in the $\sigma$ upper half plane, coupled
with the constraint $v_2/\sigma_2 = - \ell/2m$, reveals  the continued fraction structure
underlying the polar coefficients $c^F_m(\tilde{n},\tilde{\ell})$. This structure is not revealed in the approach of \cite{Ferrari:2017msn}.

If the microscopic result \eqref{crbs}
is to be obtained from a suitably defined gravity path integral, such as the quantum entropy function \cite{Sen:2008vm}, then 
the latter has to make use of the two ${\rm SL}(2, \mathbb{Z})$ subsets $S_G$ and $S_{\Gamma}$.
How these two subsets are built into the quantum entropy function is an interesting research question worth exploring \cite{Murthy:2009dq}.
Earlier work  \cite{Banerjee:2008ky,Murthy:2015zzy} on the quantum entropy function for this heterotic model 
did not identify the distinct roles played by $S_G$ and $S_{\Gamma}$. Moreover, inspection of \eqref{crbs} shows that the series associated with
the
Bessel function $I_{23/2}$ is organized in powers of $1/\gamma$, and not in powers of $1/n_2$ as in the semi-classical asymptotic expansion
\cite{LopesCardoso:2004law,David:2006yn,Banerjee:2008ky} (we recall that $n_2 = - a c \gamma$). The dependence 
on $a c $ in the argument of $I_{23/2}$ is encoded in $\tilde{\Delta}$ through the expression
\eqref{tdelMN}. The results in this paper constitute the non-perturbative completion of the semi-classical results that exist in the literature.
Understanding
how the Rademacher expansion can be used to rigorously write down the quantum entropy function for immortal degeneracies
associated with single centre $\frac14$ BPS black holes in this heterotic model 
 is a challenging question.

\vskip 10mm

\subsection*{Acknowledgements}
We would like to thank Sergei Alexandrov, Abhiram Kidambi, Boris Pioline and Valentin Reys for valuable discusssions.
This work was partially
supported by FCT/Portugal through
CAMGSD, IST-ID,
projects UIDB/04459/2020 and UIDP/04459/2020, and 
through the LisMath PhD fellowship PD/BD/135527/2018 (M. Rossell\'o).

\appendix

\section{Error function relations} \label{appendix-error-functions}

We summarise various useful relations involving error functions.

The error function $\text{Erf}$ and the complementary error function $\text{Erfc}$  are defined by
\bea
\text{Erf}(x) &=& \frac{2}{\sqrt{\pi}}\int_{0}^{x }e^{-t^2}dt  \nonumber\\
		\text{Erfc}(x) &=& 1- \text{Erf}(x) = \frac{2}{\sqrt{\pi}}\int_{x}^{+\infty}e^{-t^2}dt\;,
\eea
with $x \in \mathbb{C}$. Note that  $\text{Erf}(-x) = -\text{Erf}(x)$.
When ${\rm Re} \, x > 0$, the complementary error function can be brought to the form
\begin{equation} \label{erfc}
	\text{Erfc}(x) = \frac{1}{\sqrt{\pi}}\frac{e^{-x^2}}{x}-\frac{1}{2\sqrt{\pi}}\int_{x^2}^\infty t^{-3/2}e^{-t}dt, 
	\end{equation}
by performing the change of variables $t' = t^2$ and subsequently integrating by parts.

Using \eqref{gtr} we infer
\begin{equation}
 \text{Im}\left(\frac{1}{\gamma(\gamma\sigma+\delta)}\right) = \text{Im}\left(-\frac{\alpha\sigma+\beta}{\gamma\sigma+\delta}\right)<0,
\end{equation}
since ${\rm Im} \, \sigma' > 0$. Then
\begin{equation}
	\text{Re}\left(\sqrt{\frac{2\pi i m}{\gamma(\gamma\sigma+\delta)}} \right)>0\;,
\end{equation}
where we take the principal branch of the square root. Defining 
\begin{equation} \label{def-capital-x-y}
	X = \frac{b}{a}+\frac{\tilde{\ell}}{2m}, \hspace{6mm}Y = -\frac{b}{a}-\frac{1}{ac}-\frac{\tilde{\ell}}{2m},
\end{equation}
the quantities $x = \sqrt{\frac{2\pi i m}{\gamma(\gamma\sigma+\delta)}} \, X$ and  $x = \sqrt{\frac{2\pi i m}{\gamma(\gamma\sigma+\delta)}} \, Y$
have positive real part when $X, Y >0$.

Let us then set $x = \sqrt{\frac{2\pi i m}{\gamma(\gamma\sigma+\delta)}} X $ with $X >0$, so that ${\rm Re} \, x > 0$.
Performing the change of variables
\begin{equation}
	t =2\pi i m \, X^2 \left( \frac{1}{\gamma(\gamma\sigma+\delta)}-z \right),
	\end{equation}
the complementary error function \eqref{erfc} becomes expressed as
\bea
	\text{Erfc}\left[\sqrt{\frac{2\pi i m}{\gamma(\gamma\sigma+\delta)}}X\right] &=& \frac{1}{\sqrt{\pi}}\frac{1}{\sqrt{\frac{2\pi i m}{\gamma(\gamma\sigma+\delta)}}X}e^{-\frac{2\pi i m}{\gamma(\gamma\sigma+\delta)}X^2} \\
	&& +\frac{1}{2\sqrt{\pi}} \frac{1}{\sqrt{2\pi i m}X}e^{-\frac{2\pi i m}{\gamma(\gamma\sigma+\delta)}X^2}\int_0^{i\infty}\left(\frac{1}{\gamma(\gamma\sigma+\delta)}-z \right)^{-3/2}e^{2\pi i m X^2 z}dz. \nonumber
\eea

On the other hand, when $X<0$, we use the relations
\begin{equation}
	\text{Erf}\left[\sqrt{\frac{2\pi i m}{\gamma(\gamma\sigma+\delta)}}X\right] = -\text{Erf}\left[\sqrt{\frac{2\pi i m}{\gamma(\gamma\sigma+\delta)}}(-X)\right] = -1 + \text{Erfc}\left[\sqrt{\frac{2\pi i m}{\gamma(\gamma\sigma+\delta)}}(-X)\right].
\end{equation}

We define
\begin{equation}
	E_{m,\gamma(\gamma\sigma+\delta)}(X) = - \frac{1}{\sqrt{\pi}}\frac{1}{\sqrt{\frac{2\pi i m}{\gamma(\gamma\sigma+\delta)}}X}e^{-\frac{2\pi i m}{\gamma(\gamma\sigma+\delta)}X^2}
\end{equation}
and
\begin{equation}
	I_{m,\gamma(\gamma\sigma+\delta)}(X) =-\frac{1}{2\sqrt{\pi}} \frac{1}{\sqrt{2\pi i m}X}e^{-\frac{2\pi i m}{\gamma(\gamma\sigma+\delta)}X^2}\int_0^{i\infty}\left(\frac{1}{\gamma(\gamma\sigma+\delta)}-z \right)^{-3/2}e^{2\pi i m X^2 z}dz,
\end{equation}
so that for $X>0$
\begin{equation}
	-\text{Erfc}\left[\sqrt{\frac{2\pi i m}{\gamma(\gamma\sigma+\delta)}}X\right] = E_{m,\gamma(\gamma\sigma+\delta)}(X) + I_{m,\gamma(\gamma\sigma+\delta)}(X).
\end{equation}

\section{Deriving \eqref{fsinh} \label{sec:sinh}}

We briefly describe how to obtain \eqref{fsinh}, following \cite{2007arXiv0708.0692B,Ferrari:2017msn}.

Performing the change of variables $z = iz'$ in \eqref{exp-interm}, we obtain
\begin{equation}
	\int_0^{i\infty}\left(\frac{1}{\tilde{\sigma}}-z \right)^{-3/2}e^{2\pi i m \left(\gamma p +  \frac{g}{2m} \right)^2 z}dz = i^{5/2}\int_0^{+\infty}\left(\frac{i}{\tilde{\sigma}}+z' \right)^{-3/2}e^{-2\pi  \left(\sqrt{m}(\gamma p +  \frac{g}{2m}) \right)^2 z'}dz'.
\end{equation}
Using \eqref{Bringmann-Lovejoy-1} and setting
\begin{equation}
	t = \frac{i}{\tilde{\sigma}}, \;\;\; s = \sqrt{m}\left(\gamma p +\frac{g}{2m} \right),
\end{equation}
we infer
\begin{equation}
	i^{5/2}\int_0^{+\infty}\left(\frac{i}{\tilde{\sigma}}+z' \right)^{-3/2}e^{-2\pi  \left(\sqrt{m}(\gamma p +  \frac{g}{2m}) \right)^2 z'}dz' = -\frac{i^{3/2}}{\sqrt{2}\pi}\tilde{\sigma}\int_\mathbb{R} \frac{e^{-2\pi i \frac{1}{\tilde{\sigma}}x^2}}{\left( x-i \sqrt{m} \left(\gamma p +\frac{g}{2m} \right)\right)^2}dx.
\end{equation}
Performing another change of variables, $x = \sqrt{m}x'$, this becomes
\begin{equation}
\label{exp-p-den}
	-\frac{i^{3/2}}{\sqrt{2m}\pi}\tilde{\sigma}\int_\mathbb{R} \frac{e^{-2\pi i m\frac{1}{\tilde{\sigma}}x'^2}}{\left( x'-i  \gamma p -i\frac{g}{2m} \right)^2}dx'.
\end{equation}
Next, we perform the sum over $p \in \mathbb{Z}$ in \eqref{exp-interm}. Noting that $p$ only enters in the denominator of \eqref{exp-p-den},
we evaluate
\begin{equation} 
	 \sum_{\substack{p\in\mathbb{Z}\\2m\gamma p+ g\neq 0}}	 \frac{1}{\left( x'-i  \gamma p -i\frac{g}{2m} \right)^2}.
\end{equation}
Using the Mittag-Leffler formula 
\begin{equation}
	\frac{1}{\sin^2(z)} = \sum_{n\in\mathbb{Z}} \frac{1}{(z-n\pi)^2}, 
\end{equation}
we infer, in the case when $g  \neq 0 \text{ mod } 2m\gamma$, 
\begin{equation}
	\sum_{\substack{p\in\mathbb{Z}\\2m\gamma p+ g\neq 0}}	
		\frac{1}{\left( x'-i  \gamma p -i\frac{g}{2m} \right)^2} 
		= \frac{1}{\gamma^2}\frac{\pi^2}{\sinh^2\left( \frac{\pi x'}{\gamma} - \frac{\pi i g}{2m \gamma} \right)},
\end{equation}
while when $g = 0 \text{ mod } 2m\gamma$,
there is one term in the sum that gets excluded and hence 
\begin{equation}
	\frac{\pi^2}{\gamma^2\sinh^2\left( \frac{\pi x'}{\gamma} \right)} - \frac{1}{x'^2}.
\end{equation}

\section{Rademacher expansion for $2E_2(\sigma)/\eta^{24}(\sigma)$ \label{sec:rade}}

The Fourier coefficients $c_0^F(n)$ of 
\begin{equation}
\psi_0^F(\sigma) = 
	2\frac{E_2(\sigma)}{\eta^{24}(\sigma)} = \sum_{n = -1}^{+\infty} c_0^F(n)q^n
	\label{modfe2}
\end{equation} 
have the following well-known exact expression. Namely, using 
\begin{equation}
	\frac{E_2(\sigma)}{\eta^{24}(\sigma)} = -q\frac{d}{dq}\frac{1}{\eta^{24}(\sigma)},
\end{equation}
the Fourier coefficients $c_0^F(n)$ can be written as 
\begin{equation}
	c_0^F(n) = -2 n \, d(n) ,
\end{equation}
where $d(n)$ are the Fourier coefficients of $\eta^{-24}(\sigma)$. The latter admit the following Rademacher expansion in terms
of the modified Bessel function $I_{13}$,
\begin{equation}
	d(n) = \sum_{\gamma = 1}^{+\infty} \frac{2\pi}{\gamma \,n^{13/2}} {\rm Kl} (n,-1,\gamma) I_{13}\left( \frac{4\pi \sqrt{n}}{\gamma} \right),
\end{equation}
and hence
\begin{equation}
\label{rade1}
	-c_0^F(n) =  2n \, \sum_{\gamma = 1}^{+\infty} \frac{2\pi}{\gamma \,n^{13/2}} 
	{\rm Kl}(n,-1,\gamma) I_{13}\left( \frac{4\pi \sqrt{n}}{\gamma} \right).
\end{equation}
Using the recurrence relation for the modified Bessel functions 
\begin{equation}
	I_{n-1}(x)-I_{n+1}(x) = \frac{2n}{x}I_n (x),
\end{equation}
we infer
\begin{equation}
	n\, I_{13}\left( \frac{4\pi \sqrt{n}}{\gamma} \right) = n I_{11}\left( \frac{4\pi \sqrt{n}}{\gamma} \right) -\frac{6}{\pi}\sqrt{n}\gamma I_{12}\left( \frac{4\pi \sqrt{n}}{\gamma} \right),
\end{equation}
and hence, \eqref{rade1} may also be written as 
\begin{align}
\label{eeexp}
	 -c_0^F(n) = 
		 \sum_{\gamma = 1}^{+\infty}  {\rm Kl}(n,-1,\gamma) \left( \frac{4\pi}{\gamma \,n^{11/2}} I_{11}\left( \frac{4\pi \sqrt{n}}{\gamma} \right) -\frac{24}{n^6} I_{12}\left( \frac{4\pi \sqrt{n}}{\gamma} \right) \right).
\end{align}

The Rademacher expression \eqref{eeexp} can also be obtained by deforming the integration contour along Ford circles and subsequently
subjecting \eqref{modfe2} to a modular transformation to re-express the integrand on each Ford circle.
In the following, we will use this approach to obtain the expression \eqref{eeexp}.

\subsection{Ford circles and Farey fractions: the Rademacher contour \label{sec:forad}}
As is well known, the function 
$E_2(\sigma)$ transforms as a quasi-modular form under ${\rm SL}(2, \mathbb{Z})$ transformations,
\begin{equation}
	E_2\left(\frac{\alpha\sigma+\beta}{\gamma\sigma+\delta} \right) = (\gamma\sigma+\delta)^2E_2(\sigma) + \frac{6\gamma(\gamma\sigma+\delta)}{\pi i}.
\end{equation}
Therefore,
\begin{equation}
	2\frac{E_2(\sigma)}{\eta^{24}(\sigma)} = (\gamma\sigma+\delta)^{10}2\frac{E_2\left(\frac{\alpha\sigma+\beta}{\gamma\sigma+\delta} \right)}{\eta^{24}\left(\frac{\alpha\sigma+\beta}{\gamma\sigma+\delta} \right)}-\frac{12\gamma(\gamma\sigma+\delta)^{11}}{\eta^{24}\left(\frac{\alpha\sigma+\beta}{\gamma\sigma+\delta} \right)\pi i}.
	\label{qme2}
\end{equation}
To obtain the Rademacher expansion for its Fourier coefficients, 
\begin{equation}
	c_0^F(n) = \int_{\mathcal{C}} d\sigma e^{-2\pi i n \sigma} \; 2\frac{E_2(\sigma)}{\eta^{24}(\sigma)} , 
		\label{initial-integral}
\end{equation}
we will apply Rademacher's method 
(see \cite{rademacher, apostol,Dijkgraaf:2000fq}),
which was originally developed for modular forms, to this quasi-modular form, as follows.
The contour $\mathcal{C}$ is any contour in the $\sigma$ upper half plane that starts at some point, $\sigma_0$, and ends at $\sigma_0+1$. 
Rademacher's approach consists in deforming the contour $\mathcal{C}$ to a new contour that is the union of upper arcs 
along Ford circles defined by the Farey sequence of order $N$, and subsequently use the modular properties of the integrand to re-express it
on each of these arcs. Then, in the limit $N\to +\infty$, the integral becomes expressed as an infinite sum of integrals over Ford circles, yielding a convergent series called the Rademacher expansion. We now review this construction following
 \cite{rademacher, apostol,Dijkgraaf:2000fq}.\\

\noindent
{\bf Definition:}  
The set of \emph{Farey fractions of order $N$}, denoted by $\mathcal{F}_N$, is the set of reduced fractions in the interval $[0,1]$ with denominators $\leq N$, listed in increasing order of magnitude. \\

\noindent
{\bf Definition:}   For a given rational number $\delta/\gamma$ with $\gcd(\delta,\gamma)=1$, the \emph{Ford circle} $\mathcal{C}(\delta,\gamma)$ is defined as the circle in the complex upper half plane with radius $1/(2\gamma^2)$ that is tangent to the point $\delta/\gamma$. \\

\begin{figure}
\begin{center}
		\includegraphics*[width = 6cm]{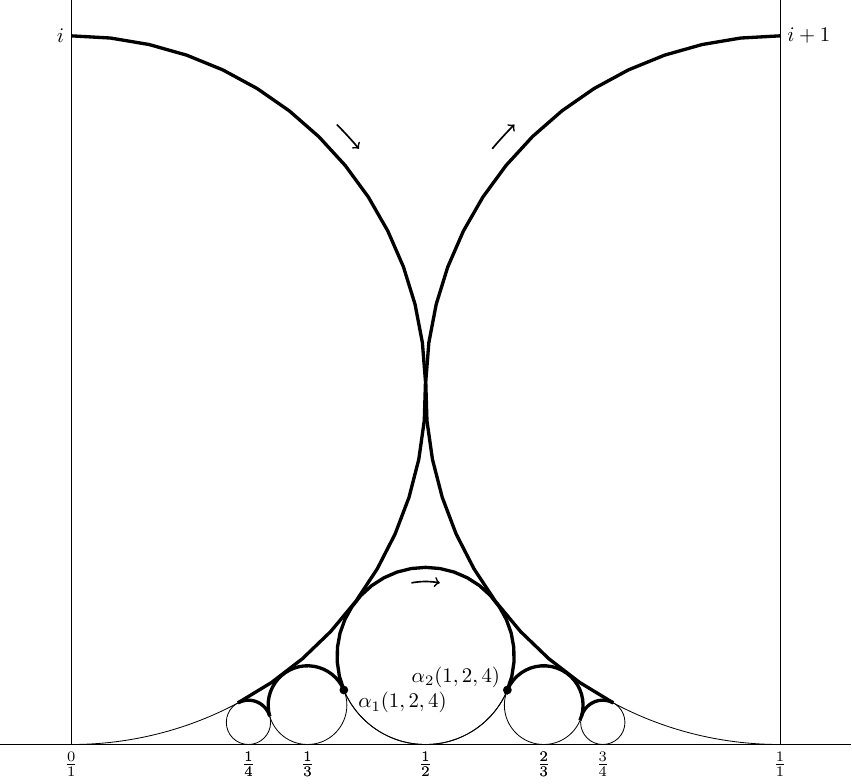}
		\includegraphics*[width = 6cm]{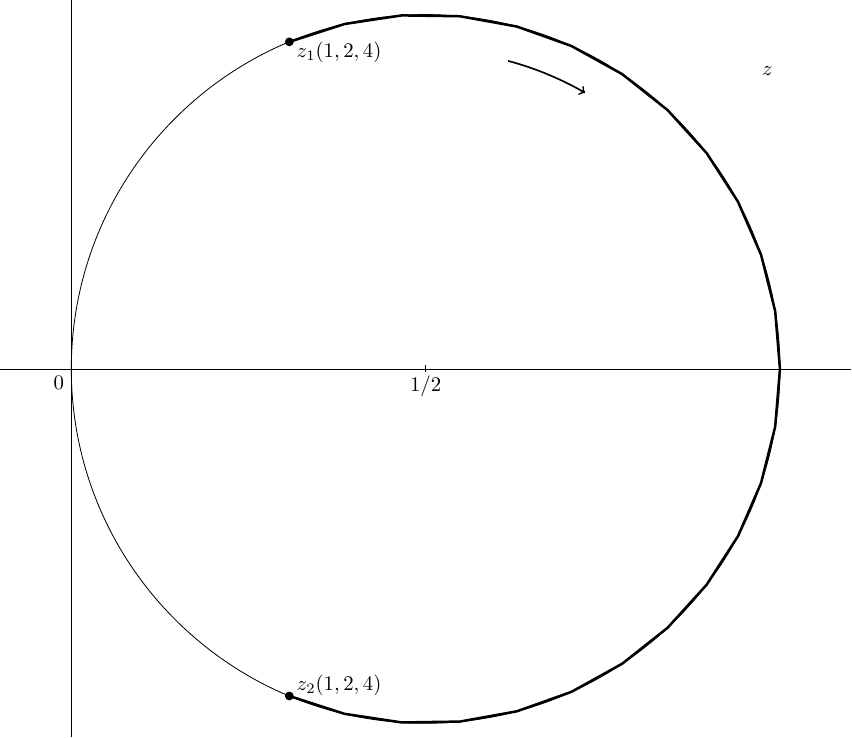}
		\caption{On the left: Rademacher contour for $P(4)$. On the right: the circle $C_{1/2}$ with $z_1(1,2,4)$ and $z_2(1,2,4)$. \label{fig1}}
\end{center}
\end{figure}

Two Ford circles $\mathcal{C}(\delta,\gamma)$ and $\mathcal{C}(\delta',\gamma')$ are either tangent to each other or they do not intersect. They are tangent if, and only if, $\gamma \delta'-\gamma'\delta = \pm 1$, which is the same as saying that they are the Ford circles associated to two consecutive Farey fractions. If $\delta_1/\gamma_1< \delta/\gamma < \delta_2/\gamma_2$ are three consecutive Farey fractions in $\mathcal{F}_N$, the points of tangency of $\mathcal{C}(\delta,\gamma)$ with $\mathcal{C}(\delta_1,\gamma_2)$ and $\mathcal{C}(\delta_2,\gamma_2)$ are
\begin{equation}
	\alpha_1(\delta,\gamma;N) = \frac{\delta}{\gamma}- \frac{\gamma_1}{\gamma(\gamma^2+\gamma_1^2)}+\frac{i}{\gamma^2+\gamma_1^2}
\end{equation}
and
\begin{equation}
	\alpha_2(\delta,\gamma;N) = \frac{\delta}{\gamma}+ \frac{\gamma_2}{\gamma(\gamma^2+\gamma_2^2)}+\frac{i}{\gamma^2+\gamma_2^2}.
\end{equation}
Then, for a given $N\in\mathbb{N}$, 
the contour $\mathcal{C}$ in \eqref{initial-integral} that goes horizontally from $i$ to $i+1$ can be deformed to a contour $P(N)$ which also  goes from $i$ to $i+1$, but follows a different path. This new contour $P(N)$, the Rademacher contour, consists in taking the upper arcs of the Ford circles defined by the Farey sequence $\mathcal{F}_N$. If $\delta_1/\gamma_1<\delta/\gamma<\delta_2/\gamma_2$ are three consecutive Farey fractions, the upper arc of $\mathcal{C}(\delta,\gamma)$, denoted by $\mathcal{C}_{\delta,\gamma,N}$, is the path that joins the two points of tangency $\alpha_1(\delta,\gamma;N)$ and $\alpha_2(\delta,\gamma;N)$ following the upper arc of $\mathcal{C}(\delta,\gamma)$. For the fractions $0/1$ and $1/1$ we only take the upper arcs lying in the interval $[0,1]$. The union of all the upper arcs in $\mathcal{F}_N$ is $P(N)$ (Fig. \ref{fig1}).

Therefore, we can write the integral over $\mathcal{C}$ as a sum over integrals over the upper arcs
\begin{equation}
	\int_{\mathcal{C}} = \sum_{\gamma=1}^{N}\sum_{\substack{0\leq -\delta<\gamma\\ (\delta,\gamma)=1}}\int_{\mathcal{C}_{-\delta,\gamma,N}}  \equiv \sum_{\gamma,\delta}^N\int_{\mathcal{C}_{-\delta,\gamma,N}},
\end{equation}
where from now on we take $\delta$ to be negative, for later convenience. We introduce the following transformation
\begin{equation}
	z = -i\gamma(\gamma\sigma+\delta),
	\label{z-change-of-variables}
\end{equation}
which maps the Ford circle $\mathcal{C}(-\delta,\gamma)$ to the circle with centre $1/2$ and with radius $1/2$, and hence tangent to zero, denoted by $C_{1/2}$. The points of tangency $\alpha_1(-\delta,\gamma;N),\alpha_2(-\delta,\gamma;N)$ of $\mathcal{C}(-\delta,\gamma)$ with $\mathcal{C}(-\delta_1,\gamma_1)$ and $\mathcal{C}(-\delta_2,\gamma_2)$ with $-\delta_1/\gamma_1 < -\delta/\gamma < -\delta_2/\gamma_2$ become
\begin{equation}
	z_1(-\delta,\gamma;N) = \frac{\gamma^2}{\gamma^2+\gamma_1^2}+i\frac{\gamma \gamma_1}{\gamma^2+\gamma_1^2}
\end{equation}
and
\begin{equation}
	z_2(-\delta,\gamma;N) =\frac{\gamma^2}{\gamma^2+\gamma_2^2}-i\frac{\gamma \gamma_2}{\gamma^2+\gamma_2^2}.
\end{equation}
The upper arc joining $\alpha_1(-\delta,\gamma;N)$ and $\alpha_2(-\delta,\gamma;N)$ becomes the arc on the right, namely the one that joins $z_1(-\delta,\gamma;N)$ and $z_2(-\delta,\gamma;N)$ without touching the imaginary axis. If $z$ is on the chord joining the points $z_1(-\delta,\gamma;N)$ and $z_2(-\delta,\gamma;N)$, then $|z|\leq \max(|z_1|,|z_2|)$. Using the property that successive Farey fractions $-\delta/\gamma,-\delta'/\gamma'$ in ${\cal F}_N$ satisfy the inequalities
\begin{equation}
	\max(\gamma,\gamma') \leq N \leq \gamma+\gamma'-1,
\end{equation}
one obtains the bound $|z_i|< \frac{\sqrt{2}\gamma}{N}$, and hence
\begin{equation}
	\vert z \vert < \frac{\sqrt{2}\gamma}{N}
	\label{z-on-chord}
\end{equation}
for any $z$ on the chord.
Moreover, using the triangle inequality, the length of the chord has to be $\leq |z_1|+|z_2|$, and hence it may not exceed
\begin{equation}
	\frac{2\sqrt{2}\gamma}{N}.
	\label{length-chord}
\end{equation}
Now we perform the Rademacher expansion. We deform the initial contour in \eqref{initial-integral} to the Rademacher contour $P(N)$, for some $N\in\mathbb{N}$,
\begin{equation}
	c_0^F(n) =  \int_{\mathcal{C}} d\sigma e^{-2\pi i n \sigma} \, 2\frac{E_2(\sigma)}{\eta^{24}(\sigma)} = 
	 \sum_{\gamma=1}^{N}\sum_{\substack{0\leq -\delta<\gamma\\ (\delta,\gamma)=1}}\int_{\mathcal{C}_{-\delta,\gamma,N}} d\sigma e^{-2\pi i n \sigma}\,2\frac{E_2(\sigma)}{\eta^{24}(\sigma)} .
\end{equation}
Next, we apply the 
modular transformation
\begin{equation}
	\begin{pmatrix}
		\alpha & \beta \\
		\gamma & \delta
	\end{pmatrix}\in {\rm SL}(2,\mathbb{Z}) \;\;\;,\;\;\; \alpha\in\mathbb{Z}/\gamma\mathbb{Z}
\end{equation}
to the integrand.
Note that this maps the cusp $-\delta/\gamma$ to $+i\infty$.
 
 Using \eqref{qme2}, we obtain
 \begin{align}
	c_0^F(n) = &  \sum_{\gamma,\delta}^N\int_{\mathcal{C}_{-\delta,\gamma,N}} d\sigma e^{-2\pi i n \sigma}
	2\left( (\gamma\sigma+\delta)^{10}\frac{E_2\left(\frac{\alpha\sigma+\beta}{\gamma\sigma+\delta} \right)}{\eta^{24}\left(\frac{\alpha\sigma+\beta}{\gamma\sigma+\delta} \right)}-\frac{6\gamma(\gamma\sigma+\delta)^{11}}{\eta^{24}\left(\frac{\alpha\sigma+\beta}{\gamma\sigma+\delta} \right)\pi i} \right)
	\nonumber\\
	= &\sum_{\gamma,\delta}^N\int_{\mathcal{C}_{-\delta,\gamma,N}}d\sigma e^{-2\pi i n \sigma}\sum_{m=-1}^{+\infty}\left( (\gamma\sigma+\delta)^{10}c_0^F(m) - \frac{12}{\pi i }\gamma(\gamma\sigma+\delta)^{11} d(m)  \right)e^{2\pi i m \frac{\alpha\sigma+\beta}{\gamma\sigma+\delta}}.
\end{align}
Next, on each of the arcs we perform the $z$-change of variables \eqref{z-change-of-variables}, using
\begin{equation}
	\frac{\alpha\sigma+\beta}{\gamma\sigma+\delta} = \frac{\alpha}{\gamma} - \frac{1}{\gamma(\gamma\sigma+\delta)}, \hspace{5mm}     \gamma\sigma+\delta = \frac{iz}{\gamma}, 
	\end{equation}
and obtain
\begin{align}
	c_0^F(n) = & \sum_{\gamma,\delta}^N\int_{z_1(-\delta,\gamma;N)}^{z_2(-\delta,\gamma;N)} dz \frac{i}{\gamma^2} e^{\frac{2\pi n}{\gamma^2} z+2\pi i n\frac{\delta}{\gamma} }2\left( \left(\frac{iz}{\gamma}\right)^{10}\frac{E_2\left(\frac{\alpha}{\gamma}+\frac{i}{z} \right)}{\eta^{24}\left(\frac{\alpha}{\gamma}+\frac{i}{z} \right)}-\frac{6\gamma\left(\frac{iz}{\gamma}\right)^{11}}{\eta^{24}\left(\frac{\alpha}{\gamma}+\frac{i}{z}\right)\pi i} \right)
	\nonumber\\
	= & \sum_{\gamma,\delta}^N\int_{z_1(-\delta,\gamma;N)}^{z_2(-\delta,\gamma;N)} dz \frac{i}{\gamma^2} e^{\frac{2\pi n}{\gamma^2} z+2\pi i n\frac{\delta}{\gamma} }\sum_{m=-1}^{+\infty}\left( \left(\frac{iz}{\gamma}\right)^{10}c_0^F(m) - \frac{12}{\pi i }\gamma\left(\frac{iz}{\gamma}\right)^{11} d(m)  \right)
	\nonumber\\
	& e^{2\pi i m\frac{\alpha}{\gamma} -2\pi \frac{m}{z}}.
\end{align}
Now we split the sum in the integral into two pieces, namely the polar part, which corresponds to $m=-1$, 
\begin{equation}
	I^-(-\delta,\gamma;N) = \int_{z_1(-\delta,\gamma;N)}^{z_2(-\delta,\gamma;N)} dz \frac{i}{\gamma^{12}} e^{\frac{2\pi n}{\gamma^2} z+2\pi i n\frac{\delta}{\gamma} }\left( -z^{10}c_0^F(-1) + \frac{12}{\pi  }z^{11} d(-1)  \right)e^{-2\pi i \frac{\alpha}{\gamma} +\frac{2\pi}{z}},
\end{equation}
and the non-polar part,
\begin{equation}
	I^+(-\delta,\gamma;N) = \int_{z_1(-\delta,\gamma;N)}^{z_2(-\delta,\gamma;N)} dz \frac{i}{\gamma^{12}} e^{\frac{2\pi n}{\gamma^2} z+2\pi i n\frac{\delta}{\gamma} }\sum_{m=0}^{+\infty}\left( -z^{10}c_0^F(m) + \frac{12}{\pi }z^{11} d(m)  \right)e^{2\pi i m\frac{\alpha}{\gamma}-\frac{2\pi m}{z}}.
\end{equation}
Using the analysis given in \cite{apostol}, it can be shown that the non-polar part does not contribute in the limit  $N\to +\infty$.
The polar part, on the other hand, is non-vanishing in this limit and given by\footnote{The integral should be understood as an improper integral with both endpoints tending to $z=0$. The integrand has an essential singularity at $z=0$. However, the remarkable property of the Rademacher contour is that the integrand is bounded on this path of integration, and therefore the integral over the circle $C_{1/2}$ is well defined.}
\begin{equation}
	c_0^F(n) = \sum_{\gamma=1}^{+\infty}\sum_{\substack{0\leq-\delta<\gamma\\ (\delta,\gamma)=1}}e^{2\pi i\left(- \frac{\alpha}{\gamma}+ n\frac{\delta}{\gamma} \right)}\int_{C_{1/2}}dz \frac{i}{\gamma^{12}} \left( -2z^{10} + \frac{12}{\pi  }z^{11}  \right)e^{\frac{2\pi n}{\gamma^2} z+\frac{2\pi}{z}}.
\end{equation}
The sum over $\delta$ can be written as a classical Kloosterman sum,
\begin{equation}
	\text{Kl}(n,-1,\gamma) = \sum_{\substack{0\leq-\delta<\gamma\\ \alpha\delta = 1 \text{ mod } \gamma}}e^{2\pi i\left(- \frac{\alpha}{\gamma}+ n\frac{\delta}{\gamma} \right)},
\end{equation}
where $\alpha\in\mathbb{Z}/\gamma\mathbb{Z}$ is the modular inverse of $\delta$, and hence uniquely specified. By performing the change of variables
\begin{equation}
	t = \frac{2\pi}{z},
\end{equation}
the $z$ integral over the circle $C_{1/2}$ becomes the $t$ integral over the vertical line defined by $\text{Re}(t) = 2\pi$,
\bea
	&& \int_{C_{1/2}}dz \frac{i}{\gamma^{12}} \left( -2z^{10} + \frac{12}{\pi  }z^{11}  \right)e^{\frac{2\pi n}{\gamma^2} z+\frac{2\pi}{z}}  \nonumber\\
	&& = \frac{1}{2\pi i }\int_{2\pi -i\infty}^{2\pi +i\infty} dt \frac{1}{\gamma^{12}}\left(-2\left( \frac{2\pi}{t}\right)^{12} +\frac{12}{\pi}\left( \frac{2\pi}{t}\right)^{13} \right)
	e^{\frac{4\pi^2n}{\gamma^2 t}+ t},
\eea
which equals a sum of modified Bessel functions of the first kind,
\begin{equation}
	I_\nu(z) = \frac{(\frac{1}{2}z)^\nu}{2\pi i }\int_{\epsilon-i\infty}^{\epsilon+i \infty} dt  \,\,t^{-\nu-1} e^{t + \frac{z^2}{4t}},
	\label{bessel-definition}
\end{equation}
and which results in
\begin{equation}
	c_0^F(n) = \sum_{\gamma=1}^{+\infty}\text{Kl}(n,-1,\gamma)
	\left(-\frac{4\pi}{n^{11/2}\gamma}I_{11}\left(\frac{4\pi \sqrt{n}}{\gamma} \right)  + \frac{24}{n^6}I_{12} \left(\frac{4\pi \sqrt{n}}{\gamma} \right)   \right).
\end{equation}

\providecommand{\href}[2]{#2}\begingroup\raggedright\endgroup

\end{document}